\documentclass[twocolumn]{aastex631}

\hypersetup{linkcolor=red,citecolor=blue,filecolor=cyan,urlcolor=magenta}

\usepackage{amsmath}
\received{January 12, 2022}
\revised{March 22, 2022}
\accepted{March 25, 2022}
\shorttitle{SN~2020jfo: A short plateau SN}
\shortauthors{Teja et al.}

\graphicspath{{./}}

\begin{document}

\title{SN~2020jfo: A short plateau Type II supernova from a low mass progenitor}

\author[0000-0002-0525-0872]{Rishabh Singh Teja}
\affiliation{Indian Institute of Astrophysics, II Block, Koramangala, Bengaluru-560034, Karnataka, India}
\affiliation{Pondicherry University, R.V. Nagar, Kalapet, Pondicherry-605014, UT of Puducherry, India}
\correspondingauthor{Rishabh Singh Teja}
\email{rishabh.teja@iiap.res.in, rsteja001@gmail.com}

\author[0000-0003-2091-622X]{Avinash Singh}
\affiliation{Hiroshima Astrophysical Science Center, Hiroshima University, Higashi-Hiroshima, Hiroshima 739-8526, Japan}

\author[0000-0002-6688-0800]{D.K. Sahu}
\affiliation{Indian Institute of Astrophysics, II Block, Koramangala, Bengaluru-560034, Karnataka, India}

\author[0000-0003-3533-7183]{G.C. Anupama}
\affiliation{Indian Institute of Astrophysics, II Block, Koramangala, Bengaluru-560034, Karnataka, India}
\author[0000-0001-7225-2475]{Brajesh Kumar}
\affiliation{Aryabhatta Research Institute of Observational Sciences, Manora Peak, Nainital-263001, Uttarakhand, India}

\author[0000-0002-8070-5400]{Nayana A.J.}
\affiliation{Indian Institute of Astrophysics, II Block, Koramangala, Bengaluru-560034, Karnataka, India}

\begin{abstract}

We present spectroscopic and photometric observations of the Type IIP supernova, SN~2020jfo, in ultraviolet and optical wavelengths. SN~2020jfo occurred in the spiral galaxy M61 (NGC\,4303), with eight observed supernovae in the past 100 years. SN~2020jfo exhibited a short plateau lasting $<\,65$\,d, and achieved a maximum brightness in \emph{V}-band of $\rm M_V=-17.4\pm0.4$\,mag at about $\rm 8.0\pm0.5$\,d since explosion. From the bolometric light curve, we have estimated the mass of $\rm ^{56}Ni$ synthesised in the explosion to be $\rm 0.033\pm0.006\,M_\odot$. The observed spectral features are typical for a type IIP supernova except for shallow H$\alpha$ absorption throughout the evolution and the presence of stable $\rm ^{58}$Ni feature at 7378\,\AA, in the nebular phase. Using hydrodynamical modelling in the \texttt{MESA\,+\,STELLA} framework, an ejecta mass of $\rm \sim 5\,M_\odot$ is estimated. Models also indicate SN~2020jfo could be the result of a Red Super Giant progenitor with $\rm M_{ZAMS}\,\sim\,12\,M_\odot$. Bolometric light curve modelling revealed the presence of a secondary radiation source for initial $\rm \sim 20$\,d, which has been attributed to interaction with a circumstellar material of mass $\rm \sim0.2\,M_\odot$, which most likely was ejected due to enhanced mass loss about 20 years prior to the supernova explosion.

\end{abstract}


\section{Introduction} \label{sec:intro}
Core-Collapse Supernovae (CCSNe) are a diverse and heterogeneous class, mainly due to a vast variety and complexity of their possible progenitors, surroundings, and the physics associated with these. Stars exceeding the threshold of $\rm \sim 8\,M_\odot$, terminate their lives in these violent explosions injecting recently synthesised elements, which in turn enrich the interstellar medium. During the progenitor's evolution, they develop a degenerate core with an outer envelope surrounding it, and when this core reaches its ``Chandrasekhar mass" ($\rm \sim 1.5\ M_\odot$), they implode due to gravitational instability \citep{BurrowsINtro}. They leave neutron stars or black holes as a remnant post collapse. Various routes have been proposed for such events, viz. Electron-Capture SNe, Fe-Core SNe, $\rm \gamma$-Ray Burst SNe, and Pair-Instability SNe (\citealp{2012Janka} and references therein).

The primary classification of supernovae is based on the presence of spectral features around optical maximum \citep{1997AVFillippenko}. SNe with an absence of hydrogen features in their spectra are termed as Type I events, whereas Type II events show prominent hydrogen Balmer features in their spectra \citep{minkwoski1941}. Type II events are further classified based on their light curves, as a Type IIP if a plateau phase of constant luminosity is observed and a Type IIL when the decline from the peak is linear before falling to the tail powered by the radioactive decay. Type IIP SNe are more commonly observed with a definitive plateau of around 100 days, but with varied luminosity, before falling to the radioactive-decay powered tail. In certain events, narrow and strong emission features in the spectra arise due to the strong interaction of ejecta with the circumstellar material (CSM), hence terming such events as Type IIn. In addition to these, some events, labelled type IIb, show hydrogen lines initially in their spectra, which weaken in the later phases when strong helium features are visible. 

Direct detection of a progenitor (\citealp{2017VanDyk} and references therein) have tightly constrained the mass range of  Red Super Giant (RSG) progenitors of Type IIP supernovae. Type IIP SNe, such as the SN~2012aw \citep{2012awVanDyk, 2012awFraser} and SN~2012ec \citep{2012ecMaund} have a directly detected progenitor in archival images that puts a mass limit of $\rm 8 \lesssim M/M_\odot \lesssim 17$ \citep{Branch2017TypeIIP, HydrogenRich} for the Type IIP progenitors. However, theoretical models predict a much higher upper limit for the progenitor mass.

Many statistical studies, including both observational \citep{2014Anderson, 2015Sanders, 2016Valenti, 2017Claudia} and synthetic/hydrodynamical modelling \citep{2018Eldridge, 2021Hiramatsu} have shown that Type IIP, IIL, and IIb form a continuous sequence of objects with the plateau or decline depending on the outer hydrogen envelope mass. It has been seen statistically in both types of attempts that the probability of events occurring with a characteristic small plateau of tens of days is relatively low.  

Apart from the above-mentioned properties, detecting the presence of CSM around these events has become quite usual. Conspicuous signatures of CSM have been observed in many CCSNe events both in spectra and light curves \citep{2013fsBullivant, 2017eawRui, 2016gfy, 2018hfmZhang, 2020tlfJacobson}. RSGs have a significant mass-loss history, and it is intuitive for the CSM to be present \citep{mauron}. However, in many cases, the CSM evades from being detected directly \citep{2017gmrAndrews, 2018cufDong}. Nonetheless, the heating effects of such hidden CSM are at times observed in the form of enhanced luminosity \citep{discCSM, PS1-13arp}.

Even with a robust and ever-increasing classification scheme for these phenomena, some do not fit very well in either one of the classified designations. With the mushrooming supernova numbers discovered by the various existing regular night-sky surveys and those expected from forthcoming survey projects, there would be more such discrete events that would not be considered standard. Hence, studying individual events in detail with an endeavour to learn about their origins would help significantly in understanding of the classes as a whole.  

SN~2020jfo (also known as ZTF20aaynrrh) was discovered by Zwicky Transient Facility (ZTF, \citealp{2019bellm}), using the Palomar 1.2m Oschin telescope on 2020 May 06 (JD\,=\,2458975.70) in the galaxy M61 (NGC~4303) at $\alpha=12^h21^m50^s.479,\ \delta=+04\degr28\arcmin54.14\arcsec\ (J2000)$. It was discovered at an AB magnitude of 16.0 mag in ZTF $r$-band. Merely within a day after the discovery, spectroscopic classification of SN~2020jfo was performed by the ZTF group \citep{2020TNSCR1259....1P} using spectra obtained with LT/SPRAT, NOT/ALFOSC, and P60/SEDM. Cross-correlation of the observed spectra with the \texttt{SNID} \citep{2007ApJ...666.1024B} library showed a good match to Type IIP supernova SN~1999gi,  about 7 days before maximum light. SN~2020jfo was suggested to be a young Type II supernova. Based on nebular spectra, \cite{2021Sollerman} suggest the SN to have a low mass progenitor.

\begin{figure}[htb!]
    \centering
    \resizebox{\hsize}{!}{0.9\includegraphics{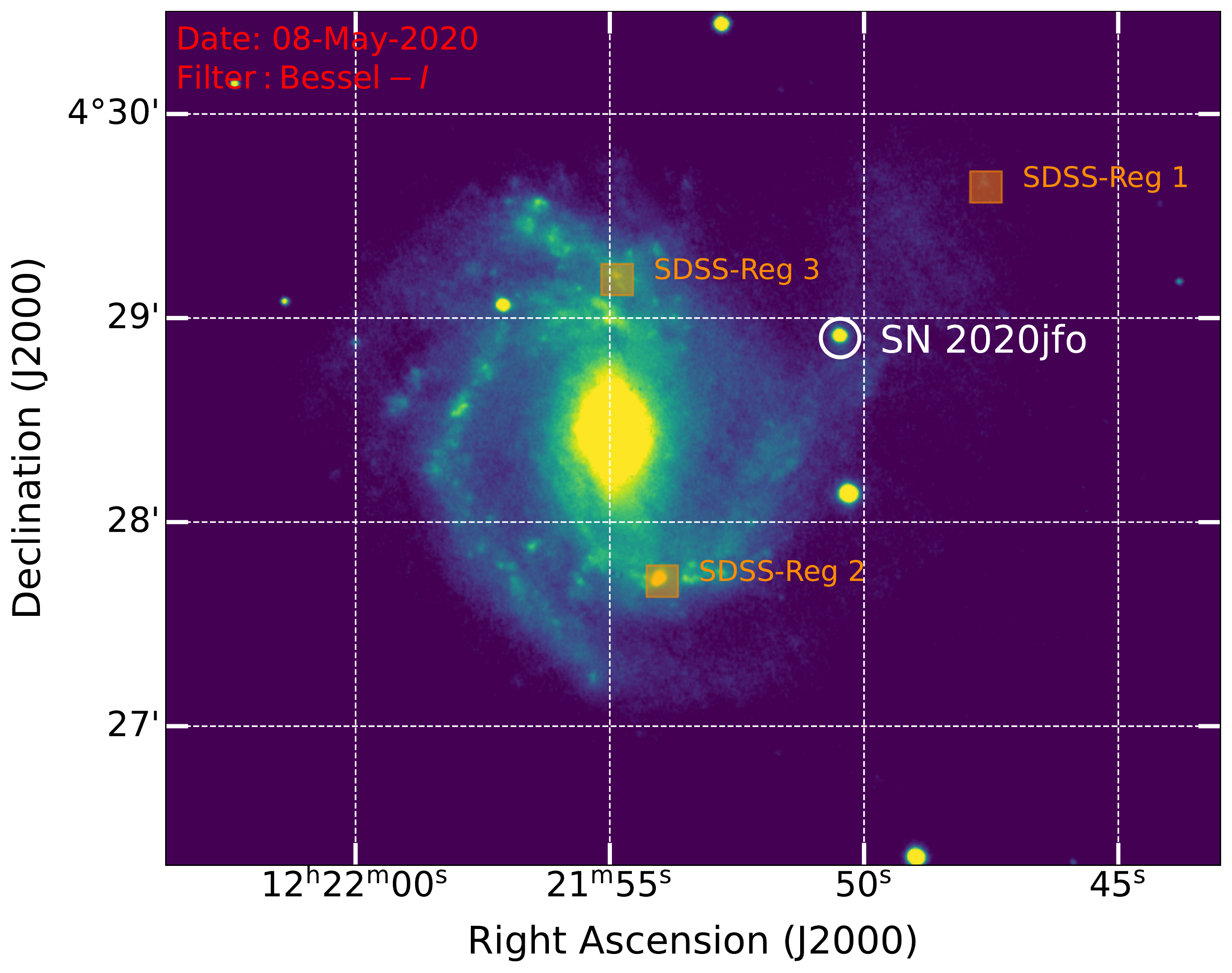}}
    \caption{The $I$ band image of SN~2020jfo in M61 obtained on 2020 May 08. The positions of the three regions of the archival SDSS-spectra [dark orange squares] along with SN~2020jfo [white circle] have been marked.}
    \label{fig:sdss_region}
\end{figure}

In this paper, we present a comprehensive photometric and spectroscopic analysis along with hydrodynamical modelling of SN~2020jfo. Subsequent sections are divided in the following manner. Section~\ref{sec:obsdata} describes optical observations of SN~2020jfo along with a brief outline of the data reduction procedure. Host galaxy properties, light curve analysis and various physical parameters of SN~2020jfo are presented in Section~\ref{sec:analysis}. Section~\ref{sec:spectroscopyevolution} deals with the spectroscopic properties, while possible progenitor and its properties estimated using various methods viz. semi-analytical modelling, nebular phase spectrum, and hydrodynamical modelling are discussed in Section~\ref{sec:possibleprogenitor}. Section~\ref{sec:summary} provides a brief summary of the results presented.

\section{Observations and Data Reduction}
\label{sec:obsdata}

\begin{table*}[htb!]
 \centering
\caption{Optical Photometry of SN~2020jfo from HCT}
    \begin{tabular}{|c | c | c | c | c | c | c |}
    \hline
    \hline
\bf{JD} 	&	 \bf{Phase}  	&	 \bf{\emph{U}}   	&	 \bf{\emph{B}} 	&	   \bf{\emph{V}}   	&	 \bf{\emph{R}}   	&	 \bf{\emph{I}} 	\\
\emph{(2458900+)} 	&	 \emph{(d)} 	&	 \emph{(mag)} 	&	 \emph{(mag)} 	&	 \emph{(mag)} 	&	 \emph{(mag)} 	&	 \emph{(mag)} 	\\
\hline
77.2 	&	3.2	&	 13.89 $\pm$ 0.17 	&	 14.63 $\pm$ 0.05 	&	 14.85 $\pm$ 0.05 	&	 -  	&	 14.84 $\pm$ 0.01 	\\
	78.3 	&	4.3	&	 - 	&	 14.57 $\pm$ 0.02 	&	 14.80 $\pm$ 0.03 	&	 14.74 $\pm$ 0.07 	&	 14.68 $\pm$ 0.02 	\\
	79.2 	&	5.2	&	 -  	&	 14.51 $\pm$ 0.03 	&	 14.61 $\pm$ 0.04 	&	 14.57 $\pm$ 0.04 	&	 14.48 $\pm$ 0.05 	\\
	80.2 	&	6.2	&	 13.84 $\pm$ 0.08 	&	 14.55 $\pm$ 0.03 	&	 14.61 $\pm$ 0.03 	&	 - 	&	 14.34 $\pm$ 0.05 	\\
	81.1 	&	7.1	&	 13.85 $\pm$ 0.02 	&	 14.57 $\pm$ 0.02 	&	 - 	&	 14.32 $\pm$ 0.05 	&	 -  	\\
	83.1 	&	9.1	&	 -  	&	 14.57 $\pm$ 0.01 	&	 14.59 $\pm$ 0.01 	&	 14.38 $\pm$ 0.02 	&	 14.36 $\pm$ 0.03 	\\
	85.1 	&	11.1	&	 14.09 $\pm$ 0.05 	&	 14.63 $\pm$ 0.01 	&	 14.57 $\pm$ 0.02 	&	 - 	&	 14.39 $\pm$ 0.02 	\\
	86.1 	&	12.1	&	 -  	&	 14.55 $\pm$ 0.03 	&	 - 	&	 -  	&	 14.30 $\pm$ 0.03 	\\
	87.1 	&	13.1	&	 -  	&	 14.72 $\pm$ 0.02 	&	 -  	&	 14.46 $\pm$ 0.04 	&	 - 	\\
	89.3 	&	15.3	&	 14.41 $\pm$ 0.07 	&	 14.80 $\pm$ 0.02 	&	 14.70 $\pm$ 0.03 	&	 14.42 $\pm$ 0.02 	&	 14.36 $\pm$ 0.03 	\\
	90.2 	&	16.2	&	 14.54 $\pm$ 0.06 	&	 14.84 $\pm$ 0.02 	&	 - 	&	 14.47 $\pm$ 0.04 	&	 - 	\\
	100.2 	&	26.2	&	 15.64 $\pm$ 0.11 	&	 15.35 $\pm$ 0.03 	&	 14.80 $\pm$ 0.03 	&	 14.47 $\pm$ 0.03 	&	 -  	\\
	102.3 	&	28.3	&	 15.66 $\pm$ 0.13 	&	 15.47 $\pm$ 0.05 	&	 14.80 $\pm$ 0.05 	&	 14.50 $\pm$ 0.02 	&	 14.35 $\pm$ 0.03 	\\
	104.3 	&	30.3	&	 -  	&	 15.57 $\pm$ 0.04 	&	 14.82 $\pm$ 0.02 	&	 14.51 $\pm$ 0.02 	&	 14.33 $\pm$ 0.14 	\\
	109.2 	&	35.2	&	 16.28 $\pm$ 0.11 	&	 15.69 $\pm$ 0.02 	&	 14.86 $\pm$ 0.02 	&	 14.54 $\pm$ 0.02 	&	 14.35 $\pm$ 0.02 	\\
	111.1 	&	37.1	&	 16.46 $\pm$ 0.06 	&	 15.70 $\pm$ 0.01 	&	 14.91 $\pm$ 0.02 	&	 14.53 $\pm$ 0.03 	&	 14.36 $\pm$ 0.04 	\\
	119.2 & 45.2 & 16.77 $\pm$ 0.13 & 15.96 $\pm$ 0.03 & 15.01 $\pm$ 0.02 & 14.62 $\pm$ 0.05 & 14.40 $\pm$ 0.03 \\
	130.2 	&	56.2	&	 -  	&	 16.36 $\pm$ 0.04 	&	 15.20 $\pm$ 0.03 	&	 14.79 $\pm$ 0.03 	&	 14.51 $\pm$ 0.02 	\\
	144.2 & 70.2 & 19.04 $\pm$	0.14 & 18.15 $\pm$	0.03 & 16.90 $\pm$	0.01 & 16.14 $\pm$	0.02 & 15.89 $\pm$	0.04 \\
	267.5 	&	193.5	&	 - 	&	 19.50 $\pm$ 0.07 	&	 18.67 $\pm$ 0.03 	&	 17.71 $\pm$ 0.06 	&	 17.54 $\pm$ 0.04 	\\
	280.5 	&	206.5	&	 -  	&	 19.51 $\pm$ 0.02 	&	 18.71 $\pm$ 0.04 	&	 17.92 $\pm$ 0.02 	&	 17.85 $\pm$ 0.05 	\\
	309.5 	&	235.5	&	 -  	&	 19.75 $\pm$ 0.04 	&	 19.21 $\pm$ 0.03 	&	 18.40 $\pm$ 0.04 	&	 18.36 $\pm$ 0.01 	\\
	316.4 	&	242.4	&	 -  	&	 19.50 $\pm$ 0.49 	&	 19.48 $\pm$ 0.29 	&	 18.46 $\pm$ 0.17 	&	 18.55 $\pm$ 0.26 	\\
	341.4 	&	267.4	&	 -  	&	 -  	&	 19.71 $\pm$ 0.05 	&	 18.93 $\pm$ 0.04 	&	 - 	\\

\hline
\hline
    \end{tabular}
    \label{table:photometryHCT}
\end{table*}

\subsection{Optical Photometry with 2m HCT}
\label{sec:HCTphotometry}


A quick follow up of SN~2020jfo began on 2020 May 07 (JD 2458977.2), i.e., $\sim$ 2 days after discovery, with the Himalayan Faint Object Spectrograph Camera (HFOSC, \citealp{2014Prabhu}) mounted on the 2-m Himalayan Chandra Telescope (HCT), situated at the Indian Astronomical Observatory (IAO), Hanle, India. It was monitored in two phases. In the first phase, it was observed until 2020 July 14 (JD 2459044.1), after which it went into Solar conjunction. When it reappeared in the night sky, the second phase of observations were carried out from 2020 November 14 (JD 2459167.5) to 2021 January 26 (JD 2459241.5). Broad-band photometric observations in Bessell $UBVRI$ filters were carried out for a total of 23 epochs. The HCT optical data presented here are supplemented with data from the ZTF in $g$ and $r$ bands, obtained through ALeRCE \citep{2021Alerce}.

HFOSC is equipped with an E2V CCD chip having a dimension of $2048 \times 4096$ pixels. The readout noise and gain of the camera are $\rm 5.75\,e^-$ and $\rm 0.28\,e^-\,ADU^{-1}$, respectively. The central $\rm 2K\times 2K$ pixels used for imaging, covers a field of view (FOV) of $\rm 10\arcmin\times10\arcmin$ at an image scale of $\rm 0.296\arcsec\,pixel^{-1}$. Object frames were obtained in multiple filters. In addition to the object frames, several bias and sky flat frames were observed at each epoch. The data were pre-processed by performing the standard tasks of bias subtraction, flat fielding, and cosmic ray removal through packages available in IRAF implemented through pyRAF as given in \citet{2021redpipe}. At certain epochs, especially during the late phase, multiple frames in the same band were observed, which were later aligned and combined to improve the signal-to-noise ratio (SNR) in the resultant object frame. The zero points used to calibrate the secondary standards in the SN field were determined using the average colour terms for the telescope detector system and field stars calibrated from SN~2014dt field \citep{2014dt}.

\begin{figure*}[htb!]
    \centering
     \resizebox{\hsize}{!}{ \includegraphics{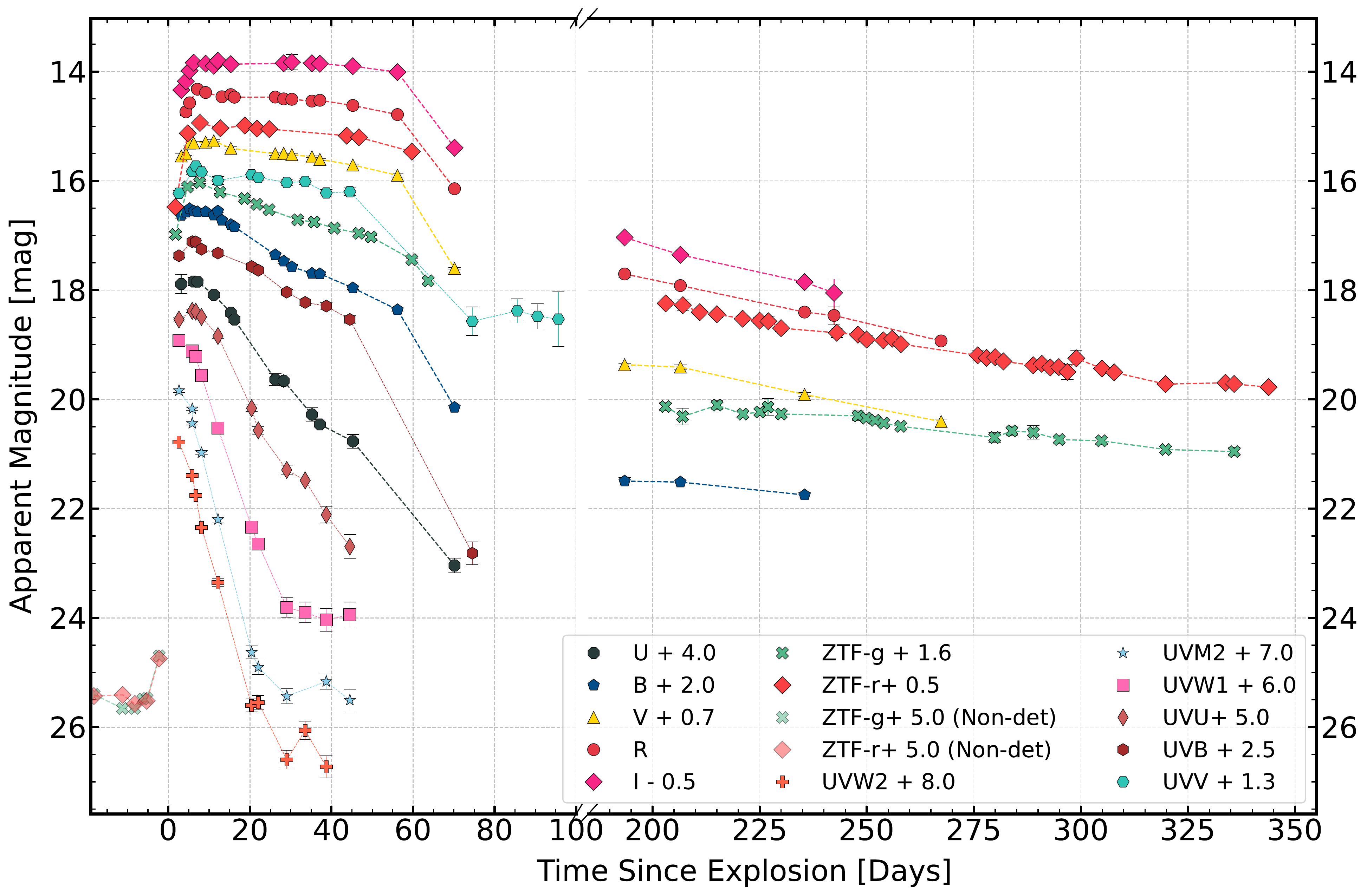}}
    \caption{Panchromatic light curves for SN~2020jfo with photometry from HCT, Swift/$UVOT$ and ZTF. The time period for which SN~2020jfo went behind the Sun has been obliterated from the plot and is marked by the discontinuity in the abscissa. Offsets in the apparent magnitudes are for visual clarity.}
    \label{fig:optical_lc}
\end{figure*}

Since SN~2020jfo is situated in an outer spiral arm of M61, the host brightness could significantly affect the supernova luminosity, especially during the late phase. Hence, we used template images of the host, obtained as a part of our monitoring programme of SN~2008in, to remove the contribution from the host. The template images were aligned with the field of SN~2020jfo, background subtracted, PSF-matched, and scaled. The scaled templates were then subtracted, leaving only the SN in the resultant images. Aperture photometry of the SN was then performed, and the SN magnitudes were calibrated using the nightly zero-points obtained from the original images. The template subtraction procedure adopted is given in \citet{2016gfy}.
The estimated magnitudes are listed in Table~\ref{table:photometryHCT} and plotted in Figure~\ref{fig:optical_lc}.

\subsection{UV-optical photometry with {\emph Swift}/UVOT}
\label{sec:Swiftphotometry}

\begin{table*}[htb!]
\caption{UV-Optical Photometry of SN~2020jfo from \emph{Swift}/UVOT}
    \label{table:photometrySwift}
    \centering
    \begin{tabular}{| c | c | c | c | c | c | c | c |}
    \hline
    \hline
    \bf{JD}	&  \bf{Phase} &	\bf{\emph{UVW2}}			&	\bf{\emph{UVM2}}			&	\bf{\emph{UVW1}}			&	\bf{\emph{UVU}}			&	\bf{\emph{UVB}}			&	\bf{\emph{UVV}}		\\
\emph{(2458900+)}& \emph{(d)} & \emph{(mag)} & \emph{(mag)} & \emph{(mag)} & \emph{(mag)} & \emph{(mag)} & \emph{(mag)} \\
\hline
76.6 &	2.6		&	12.78	$\pm$	0.02	&	12.84	$\pm$	0.03	&	12.92	$\pm$	0.03	&	13.54	$\pm$	0.03	&	14.87	$\pm$	0.03	&	14.93	$\pm$	0.05	\\
79.9 &	5.9		&	13.39	$\pm$	0.03	&	13.17	$\pm$	0.03	&	13.11	$\pm$	0.03	&	13.38	$\pm$	0.03	&	14.61	$\pm$	0.03	&	14.53	$\pm$	0.05\\
80.7	&	6.7		&	13.76	$\pm$	0.03	&	13.44	$\pm$	0.03	&	13.22	$\pm$	0.03	&	13.39	$\pm$	0.03	&	14.62	$\pm$	0.03	&	14.43	$\pm$	0.04\\
82.1	&	8.1		&	14.34	$\pm$	0.04	&	13.98	$\pm$	0.04	&	13.56	$\pm$	0.04	&	13.49	$\pm$	0.04	&	14.75	$\pm$	0.04	&	14.54	$\pm$	0.07\\
86.2	&	12.2		&	15.35	$\pm$	0.07	&	15.20	$\pm$	0.06	&	14.52	$\pm$	0.05	&	13.84	$\pm$	0.04	&	14.82	$\pm$	0.05	&	14.69	$\pm$	0.07\\
94.4	&	20.4		&	17.60	$\pm$	0.13	&	17.63	$\pm$	0.12	&	16.34	$\pm$	0.09	&	15.16	$\pm$	0.06	&	15.06	$\pm$	0.04	&	14.59	$\pm$	0.05\\
96.0	&	22.0		&	17.55	$\pm$	0.13	&	17.90	$\pm$	0.13	&	16.64	$\pm$	0.11	&	15.57	$\pm$	0.07	&	15.14	$\pm$	0.04	&	14.63	$\pm$	0.06\\
103.0	&	29.0		&	18.59	$\pm$	0.17	&	17.43	$\pm$	0.14	&	17.81	$\pm$	0.18	&	16.29	$\pm$	0.09	&	15.54	$\pm$	0.05	&	14.73	$\pm$	0.06\\
107.5		&	33.5		&	18.05	$\pm$	0.17	&	-	&	17.89	$\pm$	0.19	&	16.48	$\pm$	0.10	&	15.72	$\pm$	0.06	&	14.71	$\pm$	0.06	\\
112.7	&	38.7		&	18.72	$\pm$	0.20	&	18.16	$\pm$	0.14	&	18.04	$\pm$	0.21	&	17.11	$\pm$	0.15	&	15.79	$\pm$	0.06	&	14.92	$\pm$	0.06\\
118.5	&	44.5		&	-	&	18.50	$\pm$	0.20	&	17.94	$\pm$	0.23	&	17.70	$\pm$	0.22	&	16.04	$\pm$	0.07	&	14.90	$\pm$	0.07\\
148.5   & 74.5 & - & -& - & - & 20.31 $\pm$ 0.21 & 17.27 $\pm$ 0.26 \\
159.5   & 85.5 & - & -& - & - & - & 17.08 $\pm$ 0.22 \\
164.5   & 90.5 & - & -& - & - & - & 17.18 $\pm$ 0.23 \\
169.6   & 95.6 & - & -& - & - & - & 17.23 $\pm$ 0.50 \\

    \hline
    \hline
    \end{tabular}
    \end{table*}

The \emph{Neil Gehrels Swift Observatory} \citep{2004gehrels} database indicated that SN~2020jfo was observed with the Ultra Violet Optical Telescope (UVOT, \citealp{2005roming}) onboard \emph{Neil Gehrels Swift Observatory} \citep{2004gehrels} in the \emph{UVW2}, \emph{UVM2}, \emph{UVW1}, \emph{UVU}, \emph{UVB} and \emph{UVV} bands starting from 2020 May 07 (JD 2458976.6) and continued till 2020 August 08 (JD 2459069.6). The openly accessible archival images\footnote{\href{https://www.swift.ac.uk/}{https://www.swift.ac.uk/}} were reduced using packages available via High Energy Astrophysics Software (\texttt{HEASOFT, v6.27}) and with the latest calibration database for the UVOT instrument, following the methods as described in \cite{poole2008MNRAS.383..627P}, and \cite{brownAJ....137.4517B}. The SN magnitude was extracted using \texttt{UVOTSOURCE} task with an aperture size of 5 arcsec for the source and a similar aperture size to extract the background counts. The final UVOT magnitudes (see Figure~\ref{fig:optical_lc}) were obtained in the Vega system and are tabulated in Table~\ref{table:photometrySwift}. Template subtraction for UVOT images was performed using the mean background flux estimated at the SN~2020jfo location from the archival images of M61 obtained during follow-up of SN~2014dt. A similar flux was also obtained at the SN location as the light curve in the UV filters ($UVW2$, $UVM2$ and $UVW1$) flattened out during the post-plateau phase ($\rm\gtrsim 60\,d$).



\subsection{Optical spectroscopy}
\label{sec:spectroscopy}

\begin{table}[htb!]
\centering
\caption{Log of Spectroscopic observations of SN~2020jfo}
    \label{table:Spectroscopy}
    \begin{tabular}{| c | c | c |}
    \hline
    \hline
  \bf{ JD} 	&	 \bf{Phase}	&	 \bf{Wavelength} 	\\
  \emph{(2458900+)} 	&	 \emph{(d)}	&	 \emph{(\AA)} 	\\
\hline
77.1 	&	3	&	 4000–8000; 5200–9000 	\\
78.3 	&	4	&	 4000–8000; 5200–9000 	\\
79.2 	&	5	&	 4000–8000; 5200–9000 	\\
81.2 	&	7	&	 4000–8000; 5200–9000 	\\
85.1 	&	11	&	 4000–8000; 5200–9000 	\\
86.2 	&	12	&	 4000–8000; 5200–9000 	\\
89.3 	&	15	&	 4000–8000; 5200–9000 	\\
102.2 	&	28	&	 4000–8000; 5200–9000 	\\
110.1 	&	36	&	 4000–8000; 5200–9000 	\\
112.1 	&	38	&	 4000–8000; 5200–9000 	\\
119.2 	&	45	&	 4000–8000; 5200–9000 	\\
129.2 	&	55	&	 4000–8000; 5200–9000 	\\
144.1 	&	70	&	 4000–8000 	\\
270.5 	&	196	&	 4000–8000 	\\
276.5 	&	202	&	 4000–8000; 5200–9000 	\\
309.4 	&	235	&	 4000–8000; 5200–9000 	\\
341.4 	&	267	&	 4000–8000; 5200–9000 	\\
366.4   &  292 & 4000-9000 [DOT] \\

\hline
\hline
    \end{tabular}
    \end{table}

Spectroscopic observations of SN~2020jfo were primarily carried out with the HCT starting from 2020 May 07 (JD 2458977.1) to 2021 January 26 (JD 2459241.4), using HFOSC with grisms Gr7 and Gr8. One spectrum at nebular phase was obtained on 2021 February 21 (JD 2459266.5) with the ADFOSC instrument mounted at the 3.6\,m Devasthal Optical Telescope \citep[DOT,][]{2019DOTOmar, Sagar2019}. Standard \texttt{IRAF} packages were used to extract, reduce and calibrate the spectra obtained with both the instruments. Details of the reduction procedure are mentioned in \citet{2018avinash}. The spectra of SN~2020jfo were corrected for the host redshift using $z = 0.00502$ \citep{2020TNSCR1259....1P}. Log of spectroscopic observations is provided in Table~\ref{table:Spectroscopy}.

\section{Analysis}
\label{sec:analysis}

\subsection{Reddening, Distance and Metallicity}
\label{subsec:host}

The Milky Way line-of-sight reddening for M61 is $ E(B-V)_{\rm {MW}} =0.0194\pm0.0001$ mag which is obtained from IRSA\footnote{ \href{https://irsa.ipac.caltech.edu/applications/DUST/}{NASA/IPAC Infrared Space Archive}} Galactic Dust Reddening and Extinction map \citep{2011ApJ...737..103S}. We also noted a prominent host \ion{Na}{1}\,D absorption with an equivalent width (EW) of $1.14\pm 0.04$~\AA, in the co-added spectrum obtained from three early phase spectra, spanning 11 to 15 d from the date of explosion (see Section~\ref{sec:lc_color} for explosion epoch). Here, we have adopted two independent methods to estimate the host galaxy reddening ($E(B-V)_{\rm{host}}$). The empirical relations between $E(B-V)$ and equivalent width of \ion{Na}{1}\,D absorption lines provided by \citet{1990A&A...237...79B} and \citet{2012MNRAS.426.1465P} were used to infer $E(B-V)_{\rm{host}}$ of $0.29\pm0.01$ mag and $0.30\pm0.08$ mag, respectively. Secondly, reddening was also estimated using Balmer decrements, with host galaxy spectra from three regions, marked in Figure~\ref{fig:sdss_region}, obtained from the SDSS archive \citep{2020ApJS..249....3A}. The ratio of $\rm H\alpha$ and $\rm H\beta$ line fluxes were measured and the colour excess was estimated using the relation given by \cite{2013ApJ...763..145D}, which resulted in an $E(B-V)_{\rm { host}}$ = $0.25\pm0.02$ mag. The reddening estimated using the two independent methods agree within errors. A weighted mean from the above estimates results in $E(B-V)_{\rm{host}} = 0.27\pm0.08$ mag. A total reddening of $E(B-V) = 0.29$\,mag is adopted throughout this work.

A plethora of distance estimates to the host galaxy M61 are available on the NASA/IPAC Extragalactic Database (NED)\footnote{\href{http://ned.ipac.caltech.edu/}{http://ned.ipac.caltech.edu}}, ranging from 7.59 Mpc \citep{1984A&AS...56..381B} to 35.50 Mpc \citep{1994ApJ...433...19S} including both redshift-dependent and redshift-independent measurements. Recent redshift-independent measurements based on SN~2008in constrain the distance from 12-20 Mpc \citep[see][]{2014AJ....148..107R, 2014ApJ...782...98B}. A simple mean of all these estimates could not be adopted as the values are not continuous but at extremes. \cite{Steer_2020} has defined a robust method to get enhanced mean estimate distances (MED) using weighted mean for the distances from various primary and secondary sources. From the various means, we have estimated MED 7, which is a combination of the unweighted (MED 2), error-weighted (MED 3), and date-weighted (MED 4) means with weights of 1:2:4, respectively. The distance obtained is, $\rm D_L\,=\,16.45\pm2.69$\,Mpc ($\rm \mu\,=\,31.08\pm0.36$\,mag).

\begin{table}[htb!]
    \centering
     \caption{O3N2-index and E(B-V) estimated from the SDSS spectra of 3 regions in M~61. The regions have been marked in the Figure~\ref{fig:sdss_region}.}
    \begin{tabular}{|c|c|c|c|}
    \hline
    \hline
    \bf{Regions} &\bf{O3N2-index}&\bf{12+log[O/H]}&\bf{\emph{E(B-V)}} \\
    \hline
SDSS-Reg 1 &  1.16 & 8.36 & 0.28 \\
SDSS-Reg 2 & -0.32 & 8.83 & 0.22 \\
SDSS-Reg 3 & -0.48 & 8.88 & 0.26 \\
\hline
\hline
    \end{tabular}
    \label{tab:sdss_table}
\end{table}

To estimate the host environment properties, we used archival SDSS spectra of the three regions in M61, as indicated earlier (refer Figure~\ref{fig:sdss_region}). Fluxes of the strong emission lines viz. H$\alpha$, H$\beta$, [\ion{N}{2}] 6584\,\AA\ and [\ion{O}{3}] 5007\,\AA\ were measured and the O3N2 index as prescribed by \citet{o3n2} was estimated. The gas-phase oxygen abundance (12+log[O/H]) was computed using the relation given in \citet{o3n2}. The metallicity estimates for all the three regions are listed in Table~\ref{tab:sdss_table}. We find that towards the outer edge of the galaxy, the metallicity is sub-solar with an oxygen abundance of $\sim$\,8.36\,dex($\rm \sim 0.5\,Z_\odot$)\footnote{Solar value for 12+log[O/H] is taken from \citet{asplund} which is $\rm 8.66\pm0.05$\,dex}, whereas in the regions on the spiral arms (Regions 2 and 3), the metallicity is 8.83\,dex and 8.88\,dex ($\rm \sim 1.6\,Z_\odot$), respectively.

\begin{table}
    \centering
        \caption{Peak magnitudes from UV/Optical light curves }
    \label{tab:exp_time}
    \renewcommand{\arraystretch}{1.2}
    \begin{tabular}{|c|c|c|}
    \hline
    \hline
 \bf{Band}  & \bf{t(m$_{max}$)}  & \bf{ m$_{max}$}\\
  & $(MJD)$ & $(mag)$ \\
 \hline
 $U$  & 58978.71$\pm$0.70 & 13.83$\pm$0.08  \\
 $B$ & 58979.19$\pm$0.28 & 14.53$\pm$0.03  \\
 $V$  & 58981.54$\pm$0.35 & 14.55$\pm$0.03  \\
 $R$ & 58982.58$\pm$0.25 & 14.38$\pm$0.02  \\
 $I$ & 58981.31$\pm$0.56 & 14.30$\pm$0.05  \\
 \hline
 ZTF-$g$& 58980.63$\pm$0.21 & 14.46$\pm$0.02 	\\
  ZTF-$r$  & 58982.35$\pm$0.37  &14.47$\pm$0.03  \\
\hline
 $UVU$  & 58979.23$\pm$0.28 & 13.37$\pm$0.03 \\
 $UVB$& 58980.23$\pm$0.31 & 14.61$\pm$0.03  \\
 $UVV$  & 58981.54$\pm$0.58  & 14.45$\pm$0.05 \\
 
\hline
\hline
    \end{tabular}
\end{table}


\subsection{Light and colour curves evolution}
\label{sec:lc_color}

The last non-detection of SN~2020jfo was on UT 2020 May 02.27 (JD 2458971.8) with a limiting AB magnitude of 19.7 mag in the $g$-band ZTF filter \citep{2020TNSTR1248....1N}. The supernova was discovered on UT 2020 May 06.26 (JD 2458975.7). The mid epoch between the last non-detection and the first detection is JD 2458973.75. Hence, JD 2458974$\pm$2 is taken as date of explosion and has been used for defining phase throughout this work.

The light curve evolution of SN~2020jfo in Bessell $U, B, V, R, I$, the ZTF $g, r$ bands and in \emph{Swift} $UVOT$-bands is shown in Figure~\ref{fig:optical_lc}. Optical light curves show a relatively fast rise to the maximum in all bands. To estimate peak magnitudes and rise times to the peak in various bands, we fitted a cubic spline to the observed photometric data. The estimated peak magnitude and date of maximum in different bands are given in Table~\ref{tab:exp_time}. The rise time ranges from 5.2\,d in $U$ to 9.1\,d in $R$ band, with a similar trend seen in \emph{Swift}-UVOT bands from 5.7\,d in $UVU$ to 8.0\,d in $UVV$, and 7.1\,d and 8.9\,d in ZTF $g$ and $r$-bands, respectively. Early phase light curves show a bump around the maximum which is prominent in the redder bands ($R$ and ZTF-$r$). Post peak, the light curves vary very slowly in the redder bands and settle onto a plateau, that appears to be short.

To have a better estimate of the plateau length, observations during transition from plateau to late declining phase are required. Unfortunately, only one observation could be made during this phase due to observational constraints, however, we notice a steep decline in ZTF-$g$ band at +60 d. Moreover, observations around +70\,d and beyond (in $UVV$) indicate that the SN has already entered into the radioactive decay tail. This puts an upper limit on the length of the plateau to be 70\,d. Also, we do not see any change in slope in the $V$ band light curve until +56\,d. With this the lower limit of the plateau length is constrained as 56\,d. With these limits the plateau (OPTd, \citealp{2014Anderson}) length is estimated as $\simeq$63$\pm$7 d.

\begin{figure}
    \centering
    \resizebox{\hsize}{!}{\includegraphics{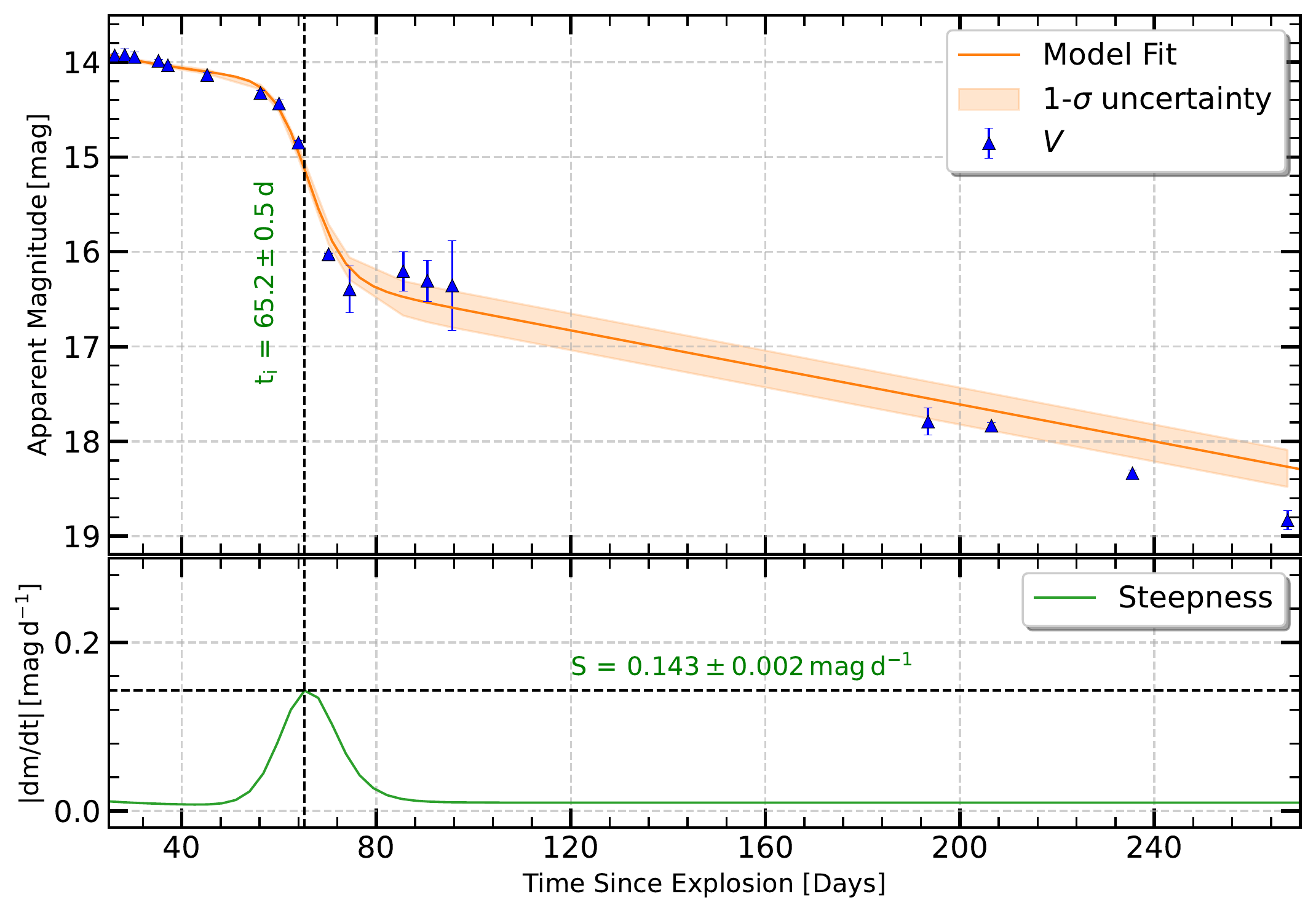}}
    \caption{Estimated Steepness of SN~2020jfo using the functional form from \citet{ElmhamdiIR2003MNRAS.338..939E}.}
    \label{fig:steepness}
\end{figure}

Another way to estimate the upper limit of the plateau length is by estimating the date of inflection during the transition phase, which is defined as the point of maximum steepness/slope. We use the formulation from \citet{ElmhamdiIR2003MNRAS.338..939E} to fit the late-plateau and radioactive decay phase of the $V$ band light curve (Figure~\ref{fig:steepness}). We could include some more points in $V$ band during transition from plateau to nebular phase using ZTF-$g$ and $UVV$ magnitudes. The $g$-band magnitudes were transformed to $V$ magnitudes using the transformation relations given by \citet{Jester2005}. Fit to the better sampled $V$-band light curve yields a steepness parameter of 0.143$\pm$0.002 $\rm mag\,d^{-1}$ and day of inflection as 65.2$\pm$0.5\,d. This is in concurrence with our plateau length estimate.
\begin{figure}[htb!]
    \centering
    \resizebox{\hsize}{!}{\includegraphics{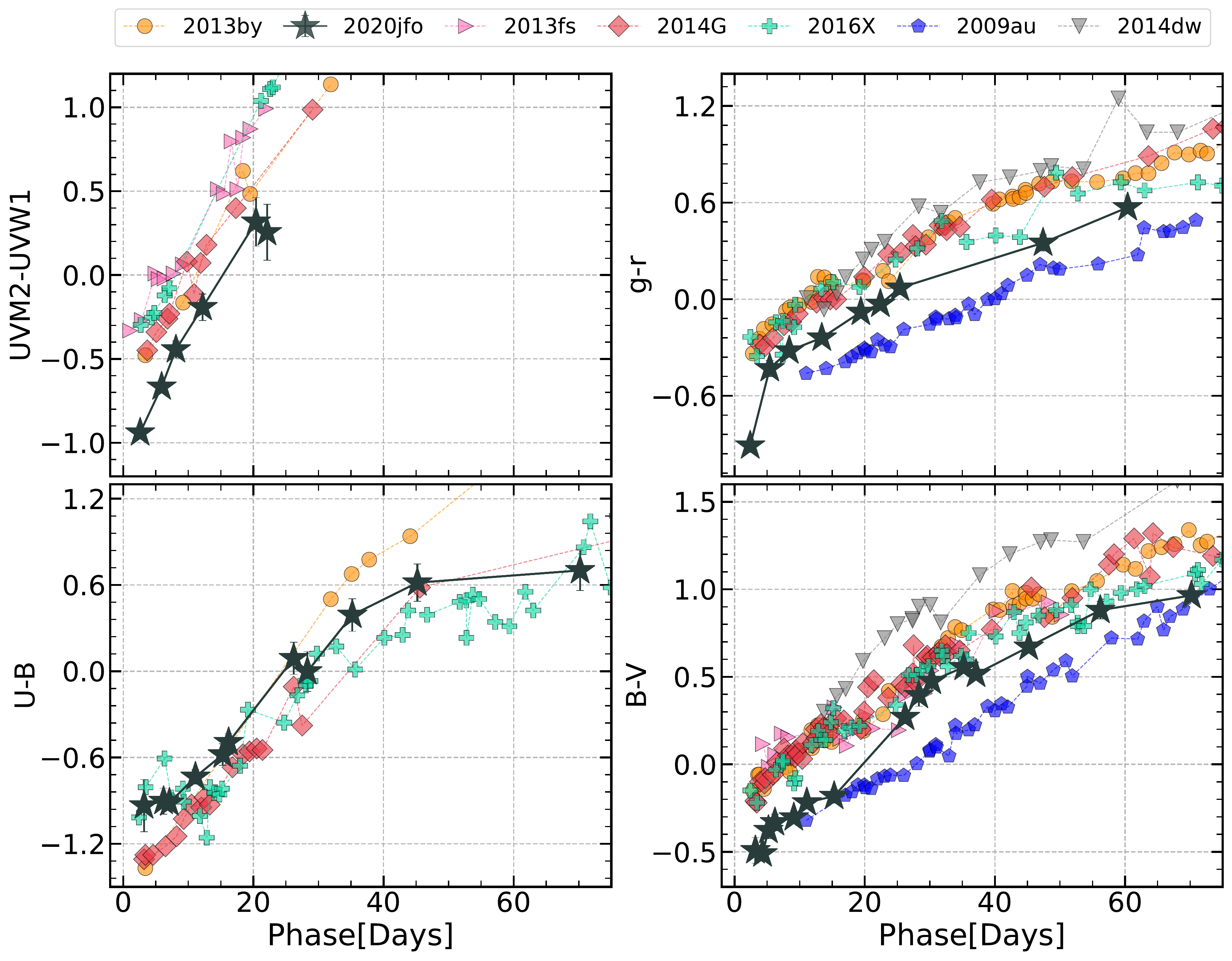}}
    \caption{Colour evolution of SN~2020jfo from the early rise up to the plateau phase is plotted along with some other Type II SNe.}
    \label{fig:color_Curves}
\end{figure}

The mean plateau length for a large sample of type IIP SNe was found to be $\sim$ 100\,d \citep{2014Anderson}, while the estimated plateau length is much shorter, $\sim$\,63\,d for SN~2020jfo. Only a handful of such objects have been discovered till now, namely, SN~2006Y, SN~2006ai, SN~2008bp, SN~2008bu \citep{2014Anderson}, SN~2014G \citep{2014GTerraran}, and SN~2016egz \citep{2021Hiramatsu}. In a recent study, \cite{2021Hiramatsu} estimated that only a small fraction ($\rm \sim 4\,\%$, (3/78)) of  short plateau objects were there in a large sample of Type II SNe studied by \cite{2018guiterez}. The rarity of short plateau objects could also be seen in the supernova lightCURVE POPulation Synthesis by \citet[CURVEPOPS]{2018Eldridge} for Type II SN, as they obtained a mere $\rm 4.7\%$ short-plateau Type IIP events out of their 637 models.

Temporal evolution of $UVW2-UVM2$, $g-r$, $U-B$ and $B-V$ colors for SN~2020jfo, during early phase is shown in Figure~\ref{fig:color_Curves}. The colours have been corrected for reddening estimated in Section~\ref{subsec:host}. The colour evolution of some other well studied objects is also plotted in the same figure for comparison. The colour evolution of SN~2020jfo follows the blue to red trend, indicating cooling of the ejecta as the supernova evolves. SN~2020jfo shows overall bluer colour, the $UVM2-UVW1$, $B-V$, and $g-r$ colour of SN~2020jfo is bluer than all other supernovae used for comparison, with the exception of SN~2009au.

\subsection{Absolute V-band light curve}
\label{subsec:M_V_lc}

\begin{figure*}[htb!]
\resizebox{\hsize}{!}{\includegraphics{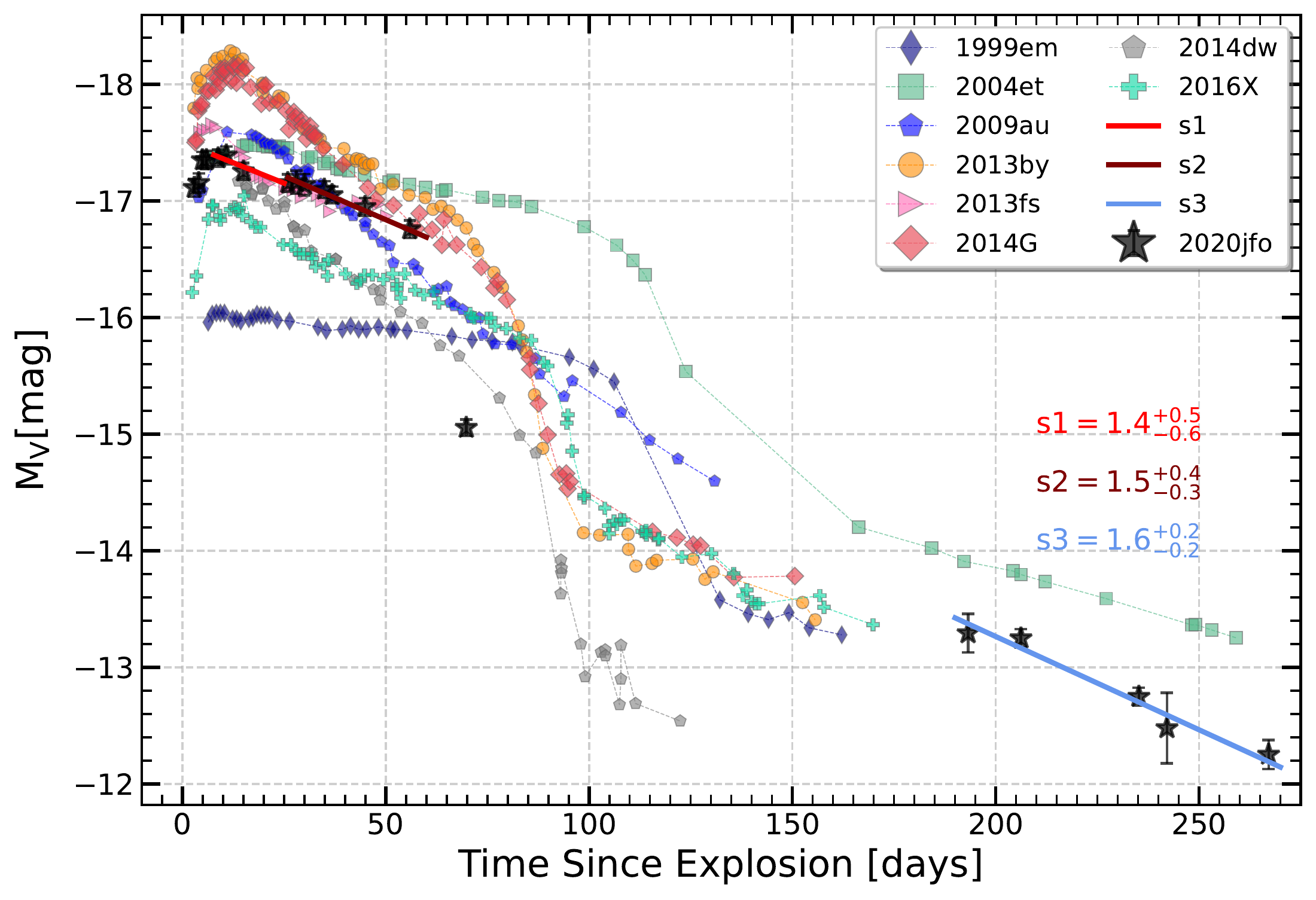}}
    \caption{Absolute V-band light curve of SN~2020jfo is plotted with some other Type II SNe. Distance and extinction correction for individual objects are obtained from their references as provided in Section~\ref{subsec:M_V_lc}. The decline rates during early plateau (s1), late-plateau (s2) and the nebular (s3) phases determined using linear fit are also mentioned.} 
    \label{fig:absolute_comp}
\end{figure*}

Absolute $V$ band light curve of SN~2020jfo is obtained after correcting the observed $V$ band magnitude for extinction and distance estimated in Section~\ref{subsec:host} and $\rm R_V = 3.1$ \citep{Cardelli}. The $V$ band light curve peaked on $\sim \rm 8.0$\,d (JD 2458982.04$\pm$0.35) after explosion with an absolute magnitude, $M_V=-17.40\pm0.37$\,mag. This puts it under the category of luminous Type IIP events. We estimated the light curve slopes during different phases $s1$, $s2$ and $s3$, \citep{2014Anderson} for SN~2020jfo as $1.4^{+0.5}_{-0.6}$, $1.5^{+0.4}_{-0.3}$ and $1.6^{+0.2}_{-0.2}$ mag per 100 days, respectively. Based on a large sample of Type II SNe light curves, \cite{2014Anderson} estimate mean values of 2.65 ($s1$), 1.23 ($s2$) and 1.47 ($s3$) mag per 100 days, indicating a clear transition from the early decline to the plateau phase. The estimated values of $s1$ and $s2$ in the case of SN 2020jfo indicate the absence of such a clear transition, although the rise to maximum is similar to other Type II SNe. It thus appears that either the $s1$ phase lasted for a very short period, or is missing entirely.


The comparison of $V$-band absolute magnitude light curve of SN~2020jfo with other Type II SNe including short plateau events is shown in Figure~\ref{fig:absolute_comp}. As the number of short plateau objects studied in detail so far is small, we compared the light curve of SN~2020jfo with a sample of objects including archetypal Type IIP SNe, SN~1999em \citep{1999emElmhamdi} and SN~2004et \citep{2004etSahu}, Type II SNe with CSM-signatures, SN~2009au \citep{2009auRodriguez}, SN~2013fs \citep{2013fsBullivant}, and SN~2014G \citep{2014GTerraran}, and faster declining or short plateau Type II SNe, SN~2013by \citep{2013byValenti}, SN~2014dw \citep{2016Valenti} and SN~2016X \citep{2016XHuang}. Although SN~2014G and SN~2013by are brighter than SN~2020jfo during the pre-maximum, early decline and plateau phase, whereas in the nebular phase their light curve merges with that of SN~2020jfo. In case of SN~2013fs, the early post-maximum decline is faster in comparison to SN~2020jfo but the plateau brightness is similar.

\begin{figure*}[htb!]
    \centering
\resizebox{0.98\hsize}{!}{\includegraphics{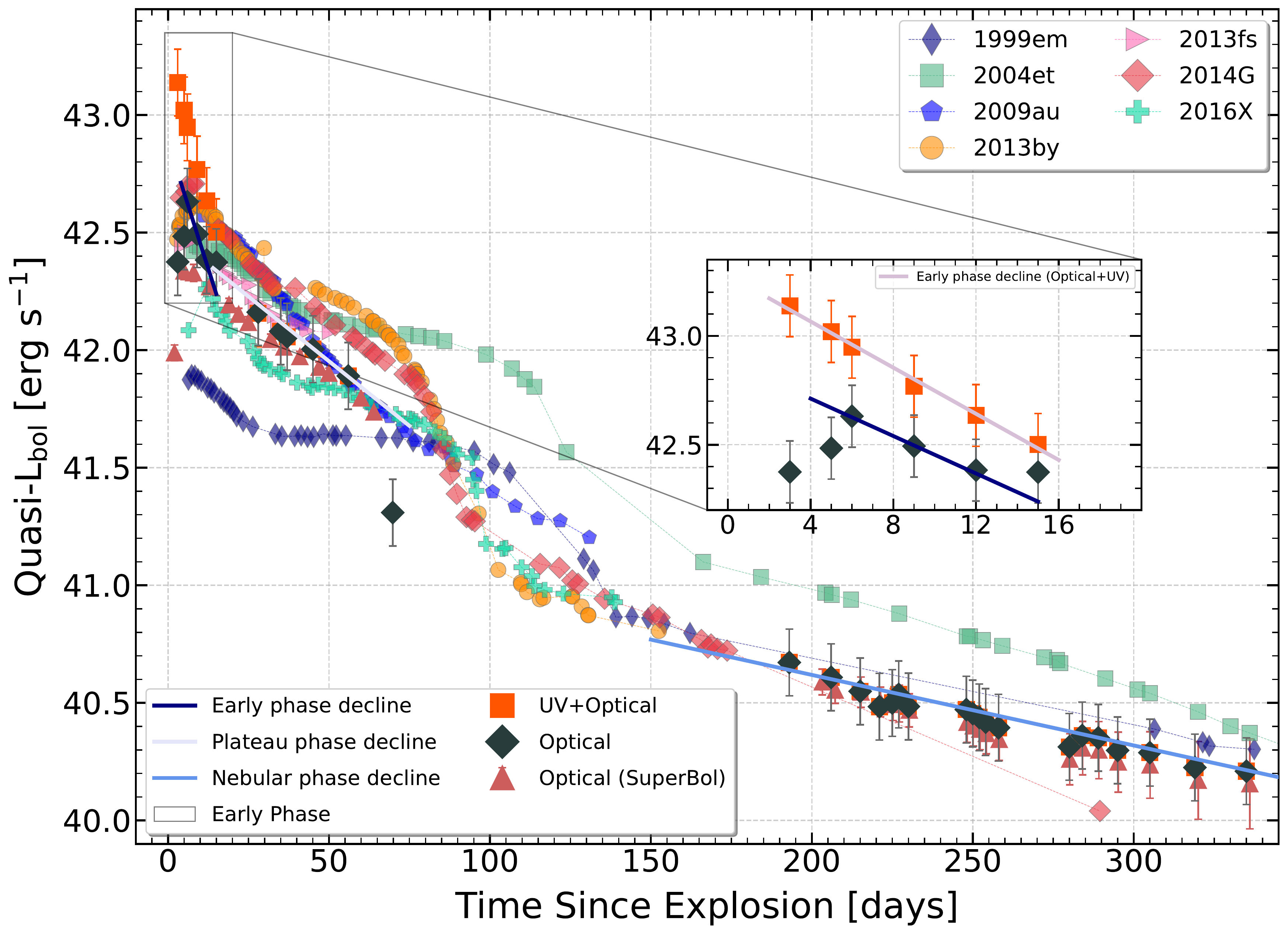}}
    \caption{Quasi-bolometric light curve (Q-bol) of SN~2020jfo along with other Type II SNe. Q-bol with contribution from UV fluxes and from \texttt{SuperBol} (without BB-corrections) are also plotted. Inset shows Optical and UV+Optical Q-bol during early phase.} 
    \label{fig:bolom1}
\end{figure*}
\subsection{Quasi-Bolometric LC}

The quasi-bolometric light curve (Q-bol) of SN~2020jfo is estimated using the observed magnitudes in $UVW2, UVM2, UVW1, U, B,$ ZTF$-g, V, R$ and $I$ filters, corrected for reddening due to the Milky Way and the host galaxy. Extinction corrections in individual photometric bands is applied using the relations by \citet[$R_V=3.1$]{Cardelli}. The extinction-corrected apparent magnitudes were converted to monochromatic fluxes at the effective filter wavelength, using the magnitude-to-flux conversion zero points listed in \citep{bessell}. Zero-point for bands other than Bessell $U,\,B,\,V,\,R,\,I$ filters were taken from the SVO Filter Profile Service\footnote{\href{http://svo2.cab.inta-csic.es/theory/fps/}{http://svo2.cab.inta-csic.es/theory/fps/}}. The spectral energy distribution (SED) curve for each epoch was estimated by interpolating the estimated flux in different bands using a Cubic spline. Finally, the quasi-bolometric flux was estimated by integrating the SED through the initial wavelength of the first band to the upper cut-off wavelength of the last band. On the nights when magnitudes were not available for some bands, we used linear interpolation to estimate them.

The quasi-bolometric luminosity for initial epochs, {\it i.e.}, up to $+$28 d, includes UV fluxes obtained from {\it Swift-UVOT} and, beyond that, the contribution is computed only using $UBgVRI$ filters. Figure~\ref{fig:bolom1} shows the quasi-bolometric light curve with and without UV contribution. It is evident from Figure~\ref{fig:bolom1} (and its inset) that during the first $\sim$ 15 days, contribution from UV bands to the bolometric flux is significant, and beyond this, it becomes very small in comparison to the optical flux. During the late nebular phase, where only ZTF data are available, bolometric correction ($BC_g$) was derived using the last few points in the LC for which the $BgVRI$ bolometric luminosity could be obtained. The estimated $BC_g$ was applied to ZTF-$g$ band magnitudes to obtain the bolometric luminosity till the very late phase.

For comparing our bolometric estimates we also use \texttt{SuperBol} \citep{SuperBol}, with ZTF-$g$ as the reference band. \texttt{SuperBol} fits a polynomial to bands with missing data and integrates those at epochs of the reference band. It seems to slightly underestimate the luminosity at the earlier epochs, where we see some signs of enhanced flux in individual optical light curves. It might be due to the smoothing of the data with a polynomial approximation. At other phases, quasi-bolometric light flux estimated in two different ways, match quite well. The contribution from optical flux to the UV+Optical bolometric flux is $\approx 20\%$ at +3\,d, which increases to  $\approx 80\%$ at $\sim$\,$\rm +15\,d$ and almost in entirety at $\sim$\,$\rm +28\,d$.

Clearly, discrete decline trends are visible in the Q-bol light curve where the initial decline from $\rm+6\,d$ to $\rm+15\,d$ is significantly steeper than other supernovae. Each decline phase is linearly fitted using Python's \texttt{emcee} routine. For comparison, the slopes for other objects during similar phases are computed and tabulated in Table~\ref{tab:slopes}.

\begin{table*}[hbt!]
\begin{center}
\caption{Best fit slopes for various phases where decline is conspicuous in the Q-bol light curve of SN~2020jfo. Slopes for other Type II SNe have also been estimated for comparison wherever possible.}
\label{tab:slopes}
\renewcommand{\arraystretch}{1.4}
\begin{tabular}{|c|c|c|c|c|c|c|c|c|}
\hline
\hline
    \bf{ SNe$\Rightarrow$} & 2020jfo & 1999em & 2004et & 2009au & 2013by & 2013fs & 2014G & 2016X  \\
     \hline
     \bf{Phases $\Downarrow$} &\multicolumn{8}{c|}{Slopes$\rm (dex \{log[L(erg\ s^{-1}]\}\ 100\ d^{-1})$}\\
     \hline
     Early &
     $-4.00^{+1.02}_{-1.09}$&$-1.27^{+0.06}_{-0.05}$&$-1.33^{+0.02}_{-0.02}$&$-1.84^{+0.01}_{-0.01}$&$-1.60^{+0.01}_{-0.01}$&$-1.33^{+0.01}_{-0.01}$&$-1.32^{+0.02}_{-0.02}$&$-1.56^{+0.02}_{-0.02}$\\
     Plateau  & $-0.91^{+0.19}_{-0.20}$&$-0.08^{+0.01}_{-0.01}$&$-0.22^{+0.01}_{-0.01}$&-&$-0.89^{+0.01}_{-0.01}$&-&$-0.98^{+0.03}_{-0.04}$&$-0.52^{+0.01}_{-0.01}$\\
     Nebular &$-0.32^{+0.03}_{-0.04}$&$-0.30^{+0.01}_{-0.01}$&$-0.44^{+0.01}_{-0.01}$&-&$-0.45^{+0.05}_{-0.04}$&-&$-0.68^{+0.04}_{-0.03}$&$-0.77^{+0.02}_{-0.02}$\\
     Early UV+Optical & $-5.33^{+0.42}_{-0.41}$&-&-&-&-&-&-&-\\
    \hline
    \hline
\end{tabular}
\end{center}
\end{table*}

Q-bol light curve of SN~2020jfo peaks at $\sim\,4.3\pm1.4\times 10^{42}$ $\rm erg\ s^{-1}$ in optical bands around $\rm +6\,d$, whereas we missed the peak in the UV+Optical data. During the very early phase, Q-bol declines at a rate of $\rm 4.00^{+1.02}_{-1.09}\, dex\,100\,d^{-1}$ and $\rm 5.33^{+0.42}_{-0.41}\,dex\,100\,d^{-1}$ in Optical and UV+Optical respectively, whereas for the other SNe this early phase decline is less steeper. For the Type IIP events SN~1999em and SN~2004et, and SN~2013fs we estimated an early phase decline of 1.27\,$\rm dex\,100\,d^{-1}$, 1.33\,$\rm dex\,100\,d^{-1}$, and 1.33\,$\rm dex\,100\,d^{-1}$, respectively. For SN~2009au (1.84\,$\rm dex\,100\,d^{-1}$), SN~2013by (1.60\,$\rm dex\,100\,d^{-1}$) and SN~2016X (1.56\,$\rm dex\,100\,d^{-1}$), we find the decline to be steeper than normal Type II SNe, but significantly lower than SN~2020jfo (see Table~\ref{tab:slopes}). During the plateau phase and nebular phase, we find decline rates for SN~2020jfo to be $\rm0.91^{+0.19}_{-0.20}\, dex\,100\,d^{-1}$ and $\rm 0.32^{+0.03}_{-0.04}\,dex\,100\,d^{-1}$, respectively. Plateau phase decline is found to be similar to SN~2013by (0.89\,$\rm dex\,100\,d^{-1}$) and SN~2014G (0.98\,$\rm dex\,100\,d^{-1}$). In terms of magnitude, the slope in the nebular phase is found to be $0.80^{+0.08}_{-0.10}\rm\,mag\,100\,d^{-1}$. 


\subsection{\texorpdfstring{$^{56}$}{}Ni Mass}
\label{subsec:nickel_mass}

In addition to ionising and heating the outer envelope, the shock assists in the synthesis of heavy radioactive nuclei that decay and radiate \citep{NiArnett}. Out of all these, $\rm ^{56}Ni$ is the most significant contributor whose decay results in the daughter nuclei $\rm ^{56}Co$ with a half-life of $\sim$\,6.1\,d. These daughter nuclei then decay to $\rm ^{56}Fe$ with a half-life of $\sim$\,77.3\,d.

To calculate the synthesised mass of $\rm ^{56}Ni$, we employed two independent methods. Firstly, we used the following relations by \cite{hamuy2003ApJ...582..905H}:

\begin{equation} 
\label{eq:hamuay1}
\frac{M(^{56}Ni)}{M_\odot} = \frac{L_t}{L_*}\exp\left[\frac{(t_t-t_{exp})/(1+z)-t_{1/2} (^{56}Ni)}{t_{e-folding} (^{56}Co)}\right]
\end{equation}

where $\rm L_* = 1.271\times 10^{43}\,erg\,s^{-1}$, $\rm t_{1/2} (^{56}Ni)$ is 6.1\,d and the $e$-folding time of $\rm ^{56}Co$ decay used is 111.26\,d. Using the quasi-bolometric luminosity from $\sim$ 192\,d onward as tail luminosity, $L_t$, the mass of synthesised $\rm ^{56}Ni$ is estimated as $\rm 0.019\pm0.005\,M_\odot$. It is to be noted that IR contribution to the bolometric luminosity is not included, and hence, it could be considered as a lower limit on $\rm ^{56}Ni$ mass.


Secondly, we compared the late phase quasi-bolometric luminosity of SN~2020jfo with that of SN 1987A. For SN~1987A the bolometric luminosity and the mass of $\rm ^{56}Ni$ synthesised in the explosion is estimated with significant accuracy \citep[0.075 $\rm M_\odot$,][]{56NIMass1987A}. Assuming that the $\gamma$-ray deposition in SN~2020jfo is similar to SN~1987A, mass of $\rm ^{56}Ni$ in SN~2020jfo was estimated using, $\rm M_{Ni} = M_{Ni}(1987A)\times L_{bol}(2020jfo)/L_{bol}(1987A)\,M_\odot$. If a constant fraction of about $35\%$ (as estimated by \citealp{patat2001ApJ...555..900P} and \citealp{ElmhamdiIR2003MNRAS.338..939E}) is added to the quasi-bolometric flux to account for missing NIR flux, the mass of $\rm ^{56}Ni$ synthesised in SN~2020jfo becomes  $\rm 0.033\pm0.004\,M_\odot$, which is consistent with our earlier estimate if a similar IR correction is used. This value is also typical of Type II SNe as it is similar to the mean value of $\rm ^{56}Ni$ mass ($\rm =0.033\,M_\odot$) obtained by \cite{2019AndersonNi} for a sample consisting of more than 40 Type II supernovae.

It was empirically shown by \cite{ElmhamdiIR2003MNRAS.338..939E} that the $\rm ^{56}Ni$ mass anti-correlates with the maximum of the steepness parameter ($\rm S=dM_V/dt$) during the transition from the plateau phase to the nebular phase. This relation was further refined by \cite{2018avinash} by incorporating a larger sample of Type IIP SNe including low-luminosity events. Mass of $\rm ^{56}Ni$ estimated using steepness parameters of 0.143\,$\rm mag\,d^{-1}$ (refer Section~\ref{sec:lc_color}) is 0.030$\pm$0.002\,$\rm M_{\odot}$, which is similar to earlier estimates.

\citet{maguire2012} showed that the mass of $\rm ^{56}Ni$ is correlated with full-width at half-maximum (FWHM) of H$\alpha$ feature during the late nebular phase. The FWHM  of H$\alpha$ line was measured in the spectrum obtained at +292\,d by fitting a Gaussian profile. The observed FWHM was corrected for instrumental broadening using the width of the night sky emission lines present in the spectrum. Mass of $\rm ^{56}Ni$ estimated using this method is found to be $\rm 0.047^{+0.005}_{-0.004}\,M_\odot$, which is higher than our earlier estimates. It clearly signifies a broadened line emission profile in SN~2020jfo, implying a larger velocity dispersion in the line forming region, whereas in a typical Type IIP SN, the dispersion would have been lower due to a massive hydrogen envelope.



\section{Spectroscopic Evolution}
\label{sec:spectroscopyevolution}

\begin{figure*}[htb!]
\centering
\resizebox{0.8\hsize}{!}{\includegraphics{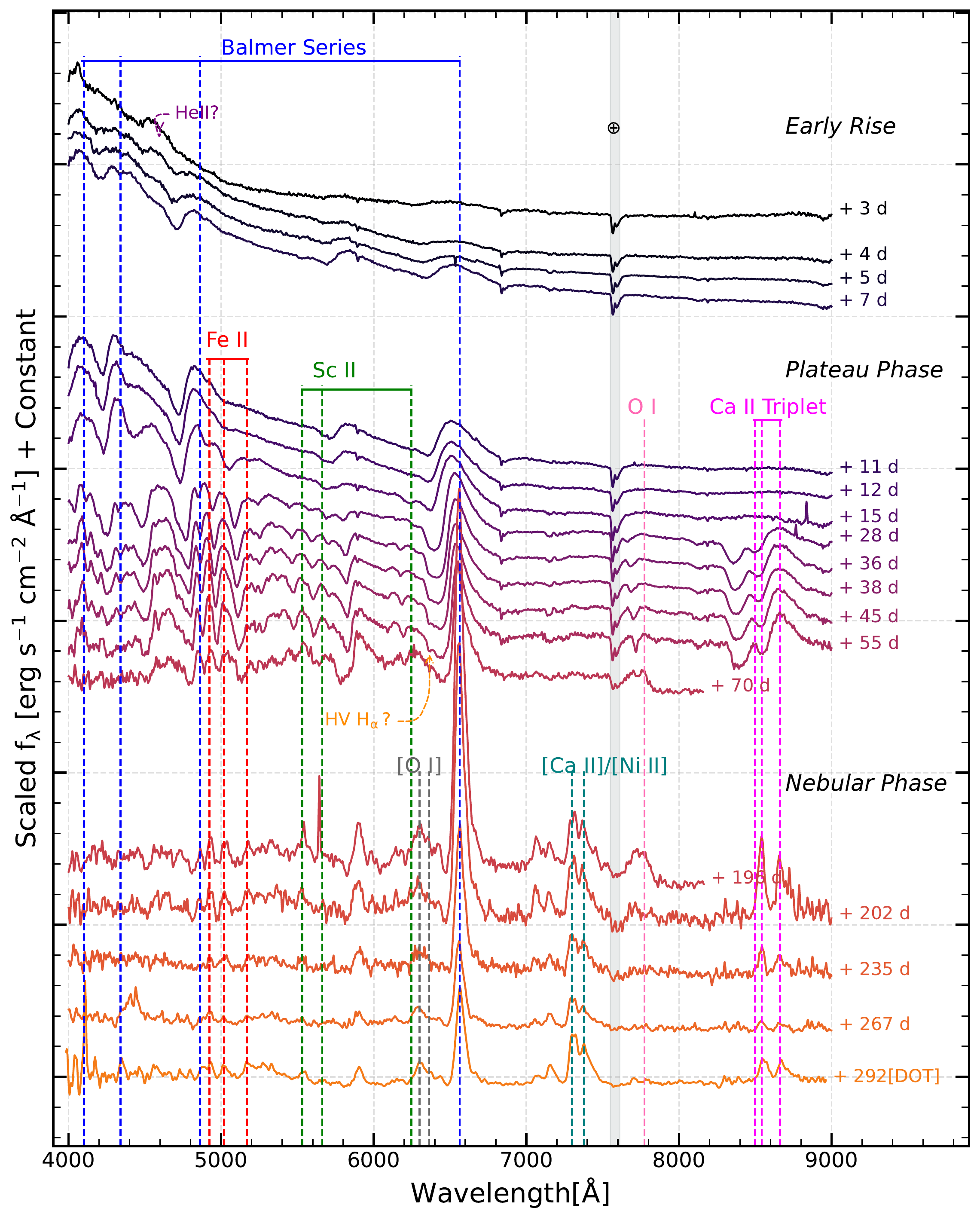}}
    \caption{Spectral evolution of SN~2020jfo from 3\,d until 292\,d post explosion. Lines have been identified following \citealp{2017gutierrez}, some of the prominent lines are marked.  All spectra are flux calibrated and corrected for reddening and redshift.} 
    \label{fig:spectra_sequence}.
\end{figure*}

The optical spectral sequence of SN~2020jfo spanning from +3\,d to around +292\,d is shown in Figure~\ref{fig:spectra_sequence}. The spectral evolution at various phases together with a comparison with other type II supernovae is discussed in this Section.

\subsection{Pre-Maximum Spectral Evolution}

\begin{figure}[htb!]
\resizebox{\hsize}{!}{\includegraphics{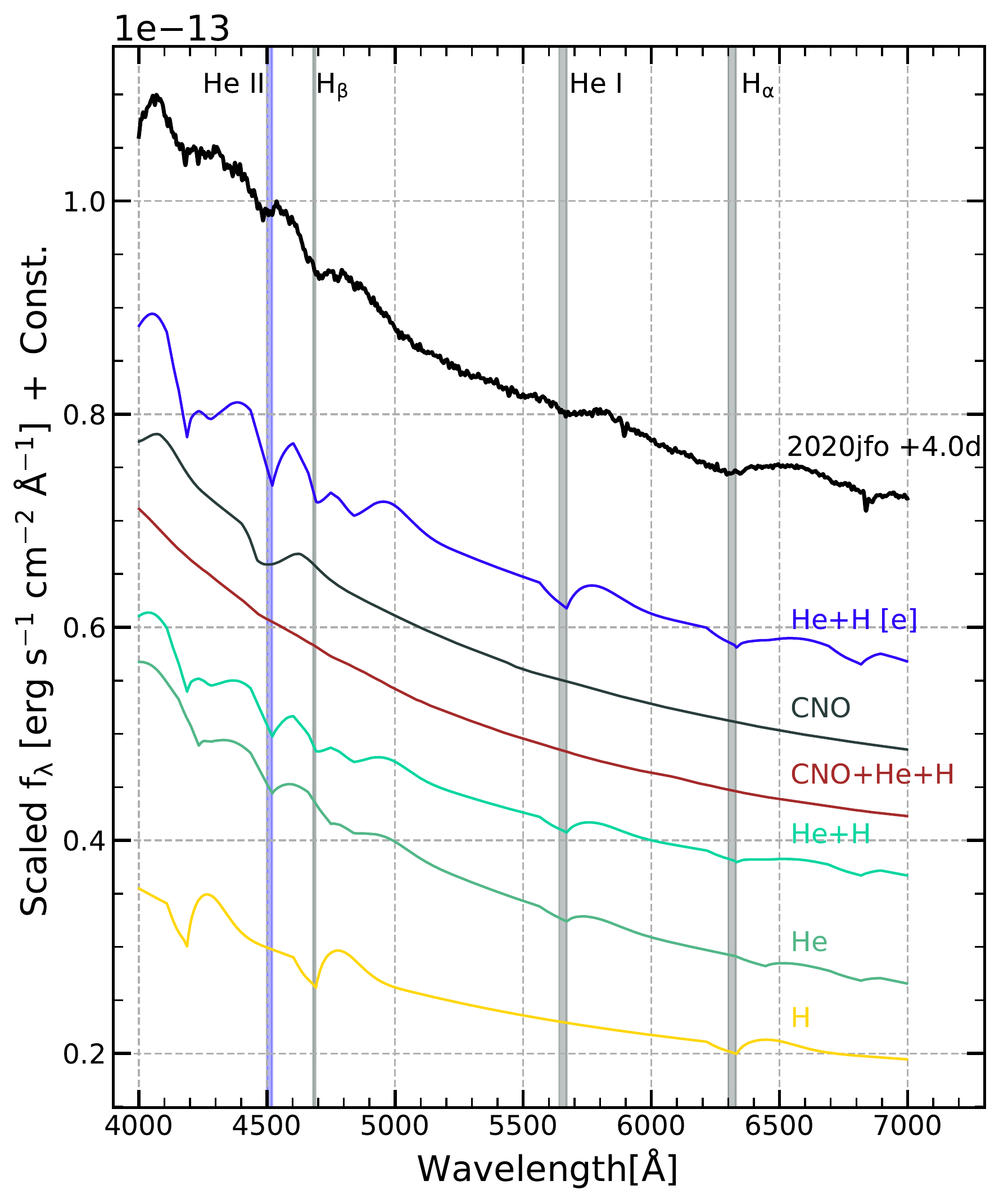}}
    \caption{Early phase (+4.0\,d) spectrum of the SN~2020jfo is compared with synthetic spectra generated using \textsc{tardis}, indicating the presence of ionised Helium. The relative contributions due to different elemental compositions are plotted. In the square brackets 'e' implies the enhanced abundance for He in the composition.} 
    \label{fig:tardis_comp}
\end{figure}

In the first spectrum obtained on +3\,d, we detect a broad absorption trough at 6266\,\AA, which is likely due to $\rm H\alpha$ and yields a line velocity of around 13,500 $\rm km\,s^{-1}$ (see Figure~\ref{fig:spectra_sequence}, \ref{fig:tardis_comp}). If we look for an $\rm H\beta$ counterpart at a similar velocity, we should detect an absorption dip at 4650\,\AA. Instead, we observe a broad P-Cygni feature with emission at around 4686\,\AA\ and its absorption counterpart at roughly 4466\,\AA. The feature is likely a broad feature of \ion{He}{2} 4686\,\AA\ at roughly 14,000\,$\rm km\,s^{-1}$, consistent with the line velocity of $\rm H\alpha$ feature. This feature faded after $+$4\,d, and a feature redward of this started appearing, which was identified as $\rm H\beta$ owing to a similar velocity with the $\rm H\alpha$ feature. Broad \ion{He}{2} 4686\,\AA\ was also seen in SN~2013fs \citep{2018bullivant, 2020chugai} and is indicative of the presence of a cold dense shell (CDS) above the photosphere. The \ion{He}{2} feature has a blue-skewed boxy profile which suggests a geometrically thin and unfragmented CDS \citep{2020chugai}. The presence of \ion{He}{2} in the early spectrum typically arises from the rapid recombination resulting from the interaction of the SN ejecta with extended supergiant atmosphere \citep{2021bruch}, however, these would lead to the existence of narrow emission lines in the spectrum. The presence of a broad P-Cygni feature indicates that the line originated in the SN ejecta. This would require that the ejecta and the nearby CSM is highly ionised by the passage of the shock, which was also seen in SN~2006bp \citep{Quimby_2007}.  

To ascertain the identification of the \ion{He}{2} feature in the early spectra, we used rapid spectral modelling code \textsc{tardis} \citep{2014TARDIS}. Incorporating modifications from \cite{2019Vogl}, \textsc{tardis} is now capable of synthesising spectra for Type II events as well. For our initial setup, we used uniform density configuration with a density profile in the form of power law \citep{2019Vogl}. Hydrogen was treated in the non-local thermodynamic equilibrium (NLTE) approximation. We used different compositions for the outer layers, including CNO, H only, He only, H+He only and H+He+CNO. We fixed luminosity parameters for +4\,d calculations and used temperature as a free parameter. The observed velocities ($\rm \sim 14000-16000\,km\,s^{-1}$) in the spectrum are used as velocities of the envelope layers. The resulting spectral luminosity was scaled with distance to obtain the observed flux values. The synthesised spectra along with the SN~2020jfo spectrum at +4\,d are shown in Figure~\ref{fig:tardis_comp}. It was noticed that, reproducing the ionised Helium feature required a temperature range of $\sim$9000\,K to $\sim$18000\,K along with a higher Helium abundance than Solar values. On the other hand, for this feature to be a blend of Nitrogen and Carbon, the modelling required much higher CNO abundances which are almost order of magnitudes higher than the Solar values and is rather non-physical. However, some amount of blending along with the Helium could not be ruled out altogether. This strengthens the case that the observed broad absorption feature is likely a \ion{He}{2} feature.

\begin{figure}
\resizebox{\hsize}{!}{\includegraphics{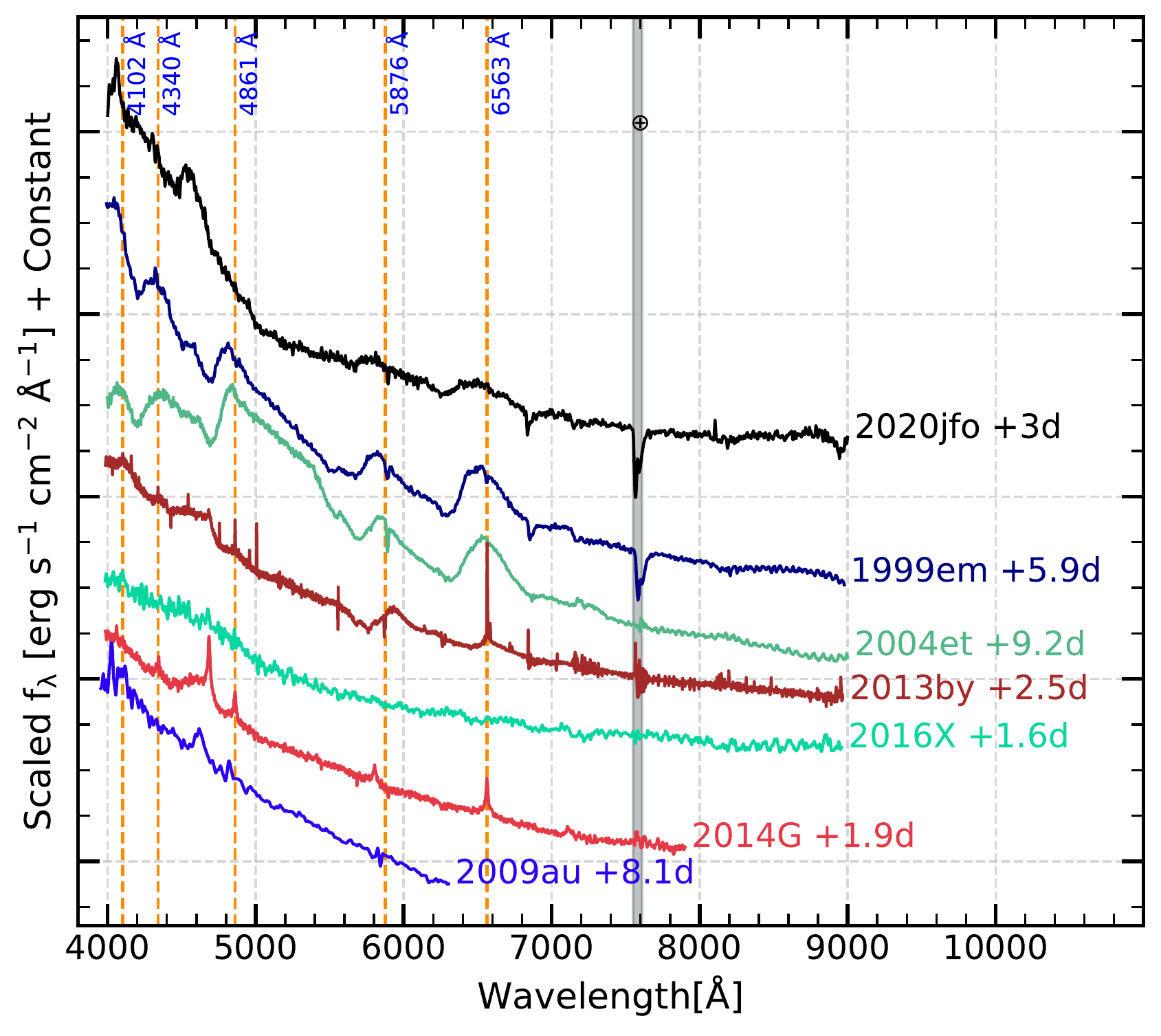}}
    \caption{The pre-maximum (+2.8\,d) spectrum of SN~2020jfo compared with the spectra of other Type II SNe at similar early phase.} 
    \label{fig:early_spec_comp}
\end{figure}

The spectrum obtained during $+$4\,d to $+$7\,d shows gradual development of Balmer spectral features. Absorption trough around 5600\AA\ is seen in the spectrum obtained on $+$4\,d which is likely due to \ion{He}{1} 5876\,\AA\ which evolved into a fully developed P-Cygni profile on $+$7\,d. As the SN evolves, the continuum becomes redder. The spectrum of SN~2020jfo obtained on $+$3\,d is compared with the spectrum of some other objects at comparable epochs and shown in Figure~\ref{fig:early_spec_comp}. The early phase spectrum of SN~1999em and SN~2004et shows a blue continuum with broad absorption due to hydrogen Balmer lines, while the early spectrum of SN~2009au, SN~2013by and SN~2014G show narrow flash-ionised lines. The spectrum of SN~2020jfo appears different than the other objects with shallow absorption due to H$\alpha$ and the presence of broad absorption due to \ion{He}{2}.


 \subsection{Plateau Phase Spectral Evolution}

\begin{figure}
\resizebox{\hsize}{!}{\includegraphics{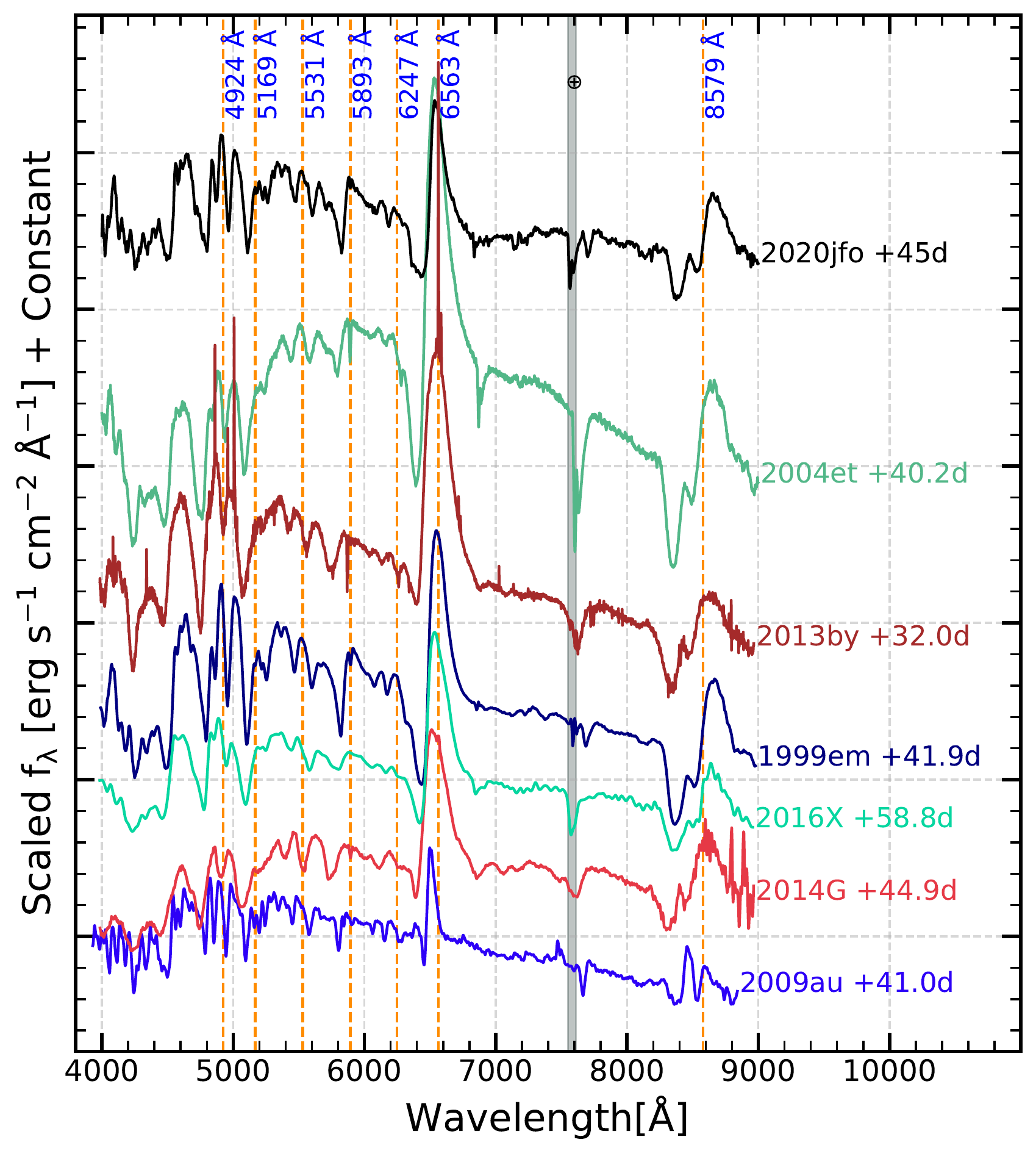}}
    \caption{Spectrum of SN~2020jfo during the plateau phase (+44.9\,d) compared with other Type II SNe at a similar epoch.} 
    \label{fig:plateau_spec_comp}
\end{figure}

As the SN enters the plateau phase, the photosphere cools to the recombination temperature and stays in the hydrogen envelope leading to the development of various metallic lines of Iron, Scandium, Oxygen and Calcium in the spectra. \ion{He}{1} feature in the early phase slowly vanishes by +15\,d and the \ion{Na}{1}\,D feature from the SN appears at its place. The shallow absorption feature seen at 5000\,\AA\ in the spectrum obtained on $+$12 d is due to the \ion{Fe}{2} (multiplet 42) features at 4924\,\AA, 5018\,\AA, and 5169\,\AA). This feature strengthens as the photosphere moves deep inside the hydrogen envelope. Hydrogen Balmer lines become stronger, and other metallic lines such as, \ion{Sc}{2} (5663\,\AA), \ion{Sc}{2}/\ion{Fe}{2} (5531\,\AA), \ion{He}{1}/\ion{Na}{1}D, \ion{Ba}{1} and Calcium NIR triplet develop in the spectra. We also detect \ion{O}{1} 7774\,\AA\ absorption feature in the spectrum of +28\,d. 

The spectrum of SN~2020jfo obtained around +45\,d is compared with the spectra of some other well studied objects in Figure~\ref{fig:plateau_spec_comp}. Most spectral features are identical in all the Type II SNe used for comparison. The H$\alpha$ absorption feature in SN~2020jfo is shallower compared to the normal Type II SNe like SN~1999em and SN~2004et, whereas it is similar to fast-declining Type II SNe such as SN~2009au, SN~2013by and SN~2014G. Further, in SN~2020jfo, the H$\alpha$ absorption trough is broader than other objects.
\begin{figure*}[htb!]
    \centering
 \resizebox{\hsize}{!}{\includegraphics{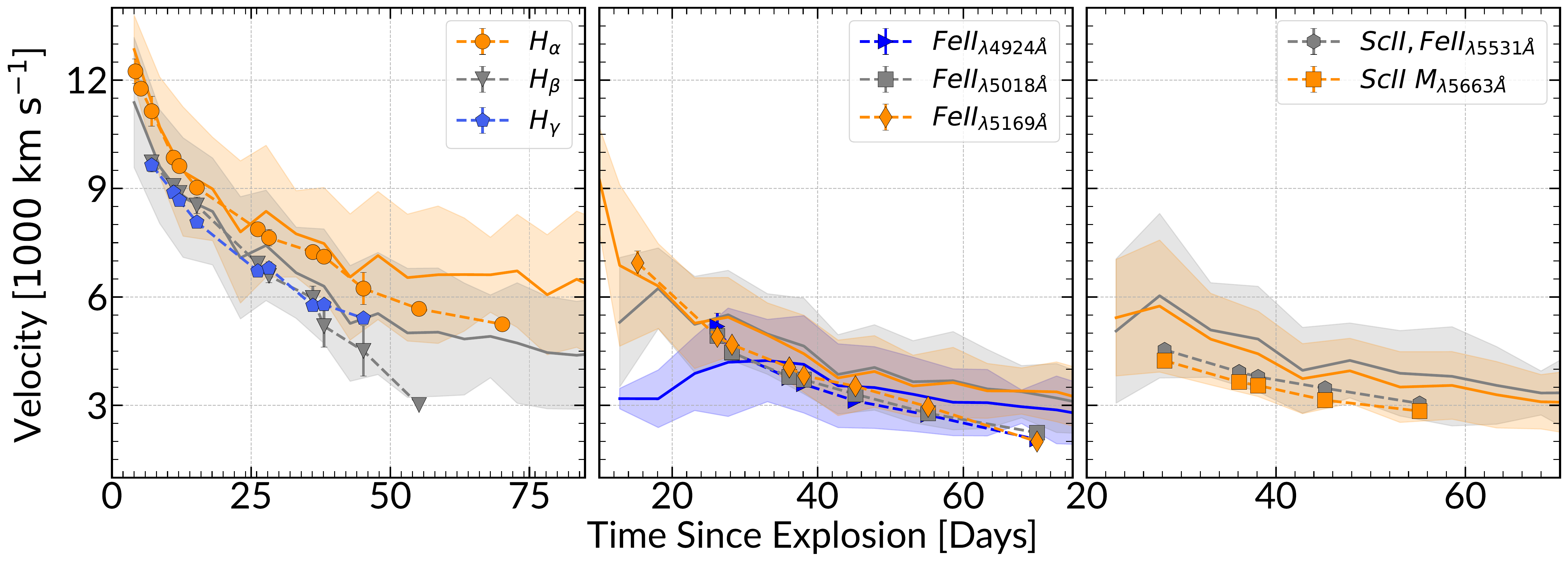}}
   \caption{Line velocity evolution of Balmer, \ion{Fe}{2} and \ion{Sc}{2} features obtained using their absorption minima are shown here. A comparison with mean Type II SNe velocities from \citet{Guti_rrez_2014} is also shown. The solid line represents mean value while the shaded region displays the 1-$\rm \sigma$ scatter from the mean.}
    \label{fig:spectra1c}
\end{figure*}
The velocities inferred from the metallic lines are similar and they fall in the range $\rm \sim 5000\,km\,s^{-1}$ at $+$28 d to $\rm \sim 2000\,km\,s^{-1}$ at $+$70 d. The expansion velocities obtained using various species is compared with velocity estimates for a larger sample of Type II supernovae \citep{Guti_rrez_2014} and is shown in Figure~\ref{fig:spectra1c}. Except for the early phase ($<+25\,\rm d$), where a steep decline in the H$\alpha$ velocity is observed, the velocities measured using  H$\alpha$, H$\beta$ and H$\gamma$ lines are similar to the average velocity for Type II sample and closely follow the observed trend in Type II supernovae throughout the photospheric phase. The velocities calculated using Fe and Sc lines are found to be marginally lower than the mean velocities of the Type II sample. This might probably indicate that SN~2020jfo was an explosion with lower energy, albeit the higher luminosity indicates otherwise. The steep decline observed during the early phase in the H$\alpha$ velocity could possibly hint towards slowing down of outer layers while encountering circumstellar matter around the progenitor. Nevertheless, if we look at the Figure~\ref{fig:bolom1}, we could see that the higher luminosity, in comparison to other Type II events, is only visible initially and reaches a moderate value later achieved through a faster decline, again indicating a short-lived source of secondary radiation, likely CSM.

From +36\,d onward, the H$\alpha$ absorption feature starts to broaden up and a deep and narrow absorption feature (Cachito) starts to develop blue-wards of H$\alpha$. This feature is prominently visible on the spectrum of +55\,d at a wavelength of 6365\,\AA. The possibility of this feature arising due to \ion{Si}{2}\,6355\,\AA\ \citep{2014Valenti} or \ion{Ba}{2}\,6497\,\AA\ was explored. The Cachito absorption lies redwards of the rest-wavelength of \ion{Si}{2} and is hence unlikely to be related to it. Assuming the feature originated due to the \ion{Ba}{2} line, the line velocity inferred is $\rm \sim 6000\,km\,s^{-1}$ which is almost twice the velocity obtained from other metal lines ($\rm \sim 3000\,km\,s^{-1}$ for \ion{Fe}{2}) at the same epoch. 

\citet{2007Chugai} proposed the emergence of high velocity absorption features of hydrogen as a result of the interaction between the RSG wind and SN ejecta. We also explored the possibility of this feature being a high velocity feature of hydrogen. The measured velocity of this HV-H$\alpha$ feature is $\rm \sim 9000\,km\,s^{-1}$ and this velocity is similar to the post-maximum expansion velocity of H$\alpha$. However, we did not observe a clear H$\beta$ counterpart, likely due to the blending of several metallic lines in that region. This HV feature is similar to the ``narrow and deeper" Cachito, seen in \citet[][e.g. SN~2003hl]{2017gutierrez} Type II SN sample study, and is similar to the case of low-velocity/low-luminosity SNe, where no H$\beta$ counterpart is seen. The likely presence of the HV feature of H$\alpha$ in the photospheric spectra favours the case of circumstellar interaction \citep{2017gutierrez}.


\subsection{Nebular Phase Spectral Evolution}

Nebular spectra of SN~2020jfo during $+$196 d to $+$292 d is plotted at  the lower panel of Figure~\ref{fig:spectra_sequence}. As the recombination phase ends, the photosphere recedes into the innermost part of the ejecta. The luminosity during this phase varies in direct proportion to the $\rm ^{56}Ni$, which was synthesised during explosion \citep{2020qmp_nebular}. The light curve in this phase is mainly powered by the radioactive decay of $\rm ^{56}Co$ to $\rm ^{56}Fe$. The nebular phase spectrum of SN~2020jfo is dominated by prominent emission lines of \ion{Na}{1}D, [\ion{O}{1}], H$\alpha$ and [\ion{Ca}{2}].

\begin{figure*}[htb!]
\centering
\resizebox{\hsize}{!}{\includegraphics{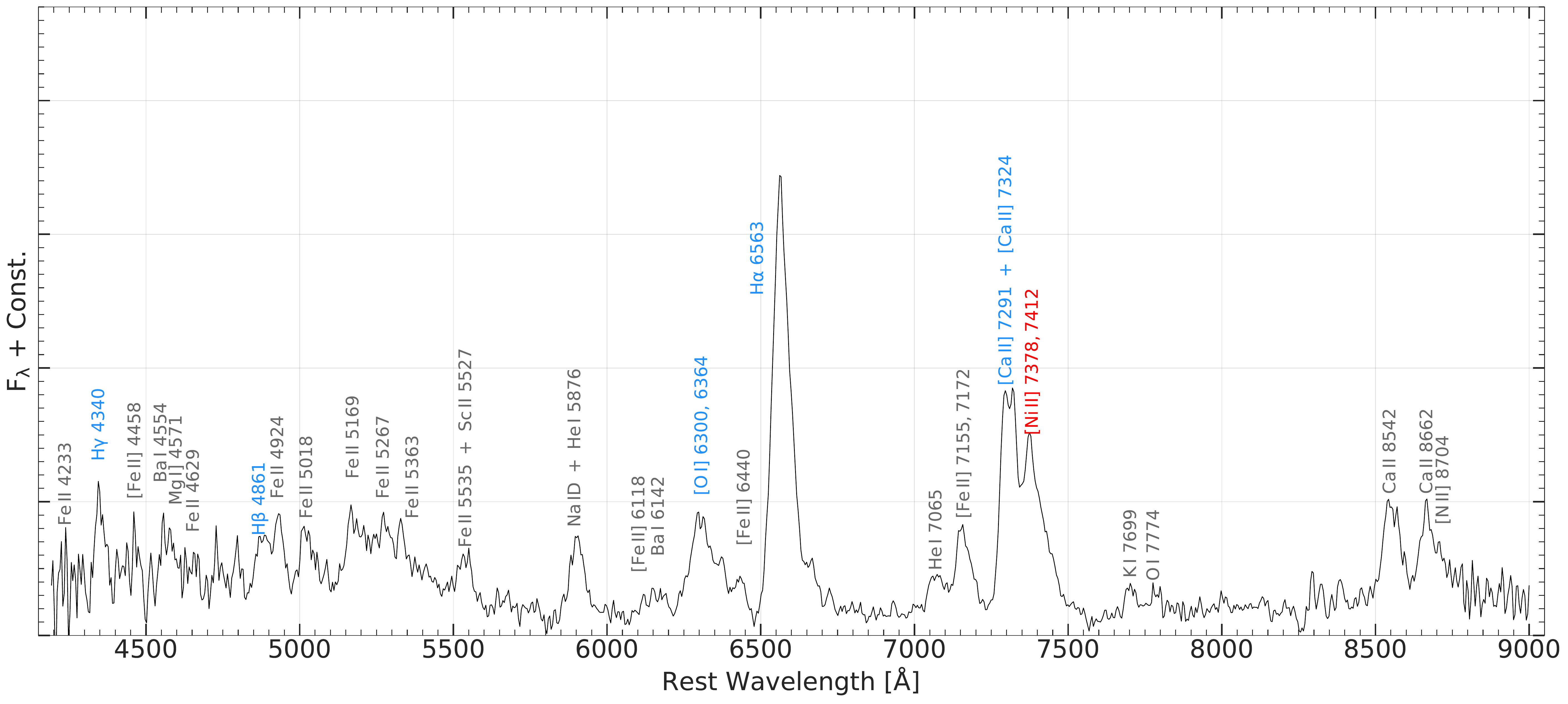}}
\caption{Identification of lines in the nebular spectrum (+292\,d) of SN~2020jfo.}
\label{fig:specline}
\end{figure*}

Narrow emission lines from metals are also seen, which become progressively more prominent as the supernova evolves into the late nebular phase. The bluer part of the spectrum is dominated by lines due to Fe, Ba, Sc, Mg etc. Hydrogen Balmer lines are seen with decreased absorption strength. As the medium becomes more rarefied forbidden lines of Fe, Ca, and O appear in the spectrum. The prominent lines seen in the nebular phase spectrum are identified and marked in Figure~\ref{fig:specline}. The spectra taken during this phase could be used to estimate the Zero-Age-Main-Sequence (ZAMS) mass of the progenitor when compared with line strengths of model spectra at similar phases. This method has been deployed in many cases in the literature to constrain progenitor mass of Type II SNe (e.g., \cite{VanDyk2019, 2019sazlai, 2018zdHiramatsu}.

\begin{figure}[htb!]
\resizebox{\hsize}{!}{\includegraphics{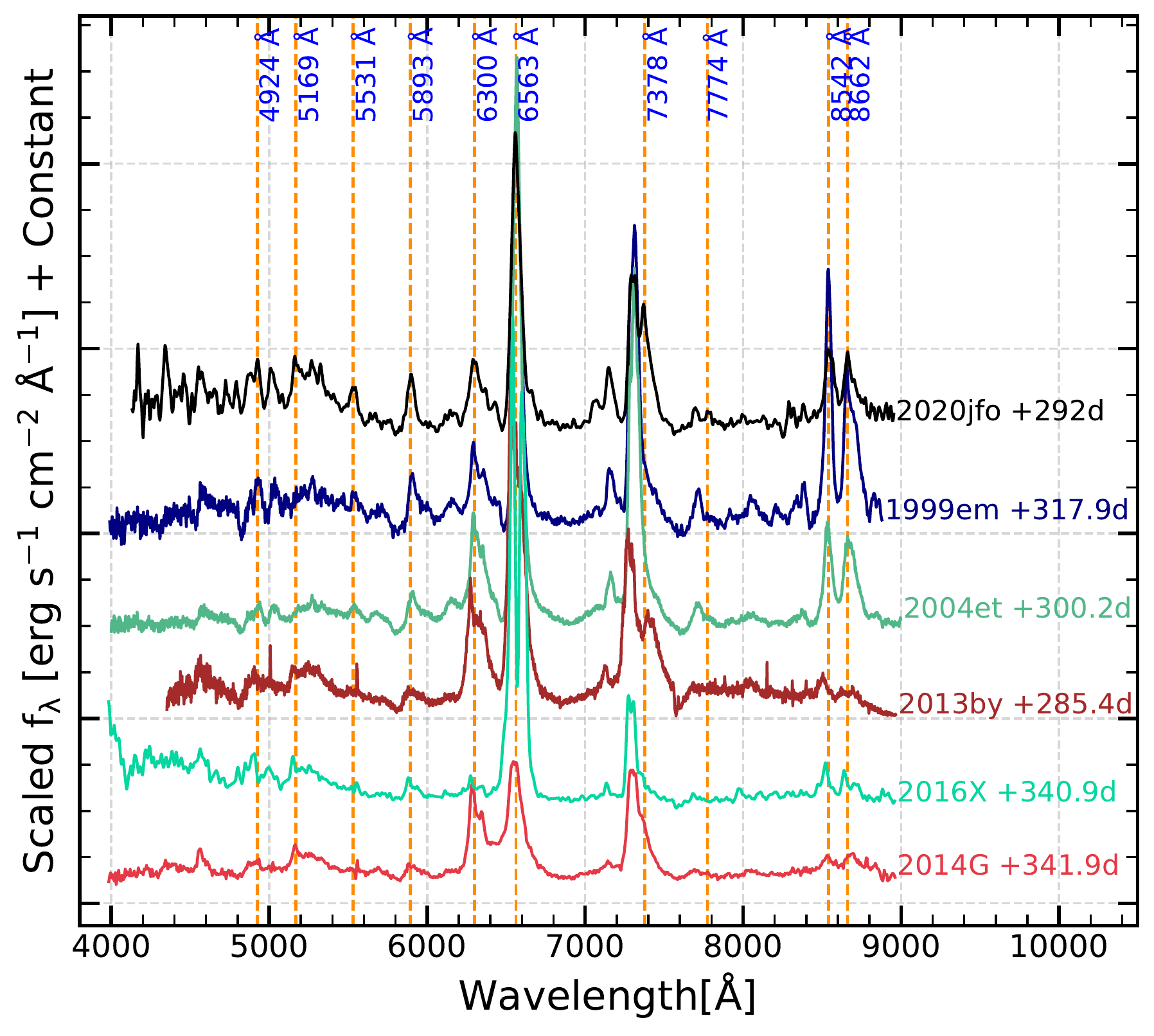}}
    \caption{Spectrum of SN~2020jfo during the nebular phase (+292.2\,d) is compared with other Type II SNe at similar epochs.} 
    \label{fig:late_spec_comp}
\end{figure}

Nebular spectrum of SN~2020jfo at $+$292\,d is compared with nebular phase spectra of other Type II SNe at similar epochs (see Figure~\ref{fig:late_spec_comp}). We find that the prominent emission lines of [\ion{O}{1}], H$\alpha$, [\ion{Ca}{2}] and \ion{Ca}{2} NIR triplet is similar to that of other normal and fast-declining Type II SNe. However, if we look closely, the spectrum of SN~2020jfo shows a clear blue excess and a forest of features due to lines of [\ion{Fe}{2}] and \ion{Fe}{2}. The \ion{Na}{1}D feature in SN~2020jfo is similar to that of normal type II SN~1999em but is more pronounced in comparison to fast-declining events SN~2013by and SN~2014G. The red-wing of [\ion{Ca}{2}] shows a clear secondary peak due to [\ion{Ni}{2}] which is not seen in normal Type II SNe like SN~1999em and SN~2004et, but seen in the fast-declining Type II SN~2013by.


\section{Characteristics of the possible progenitor}
\label{sec:possibleprogenitor}
In this section, we perform observational analysis along with the semi-analytical and hydrodynamical modelling to discuss possible progenitor scenarios for SN~2020jfo.  
\subsection{Semi-Analytical Modelling}
\label{subsec:semi}
\begin{figure}[htb!]
\resizebox{\hsize}{!}{\includegraphics{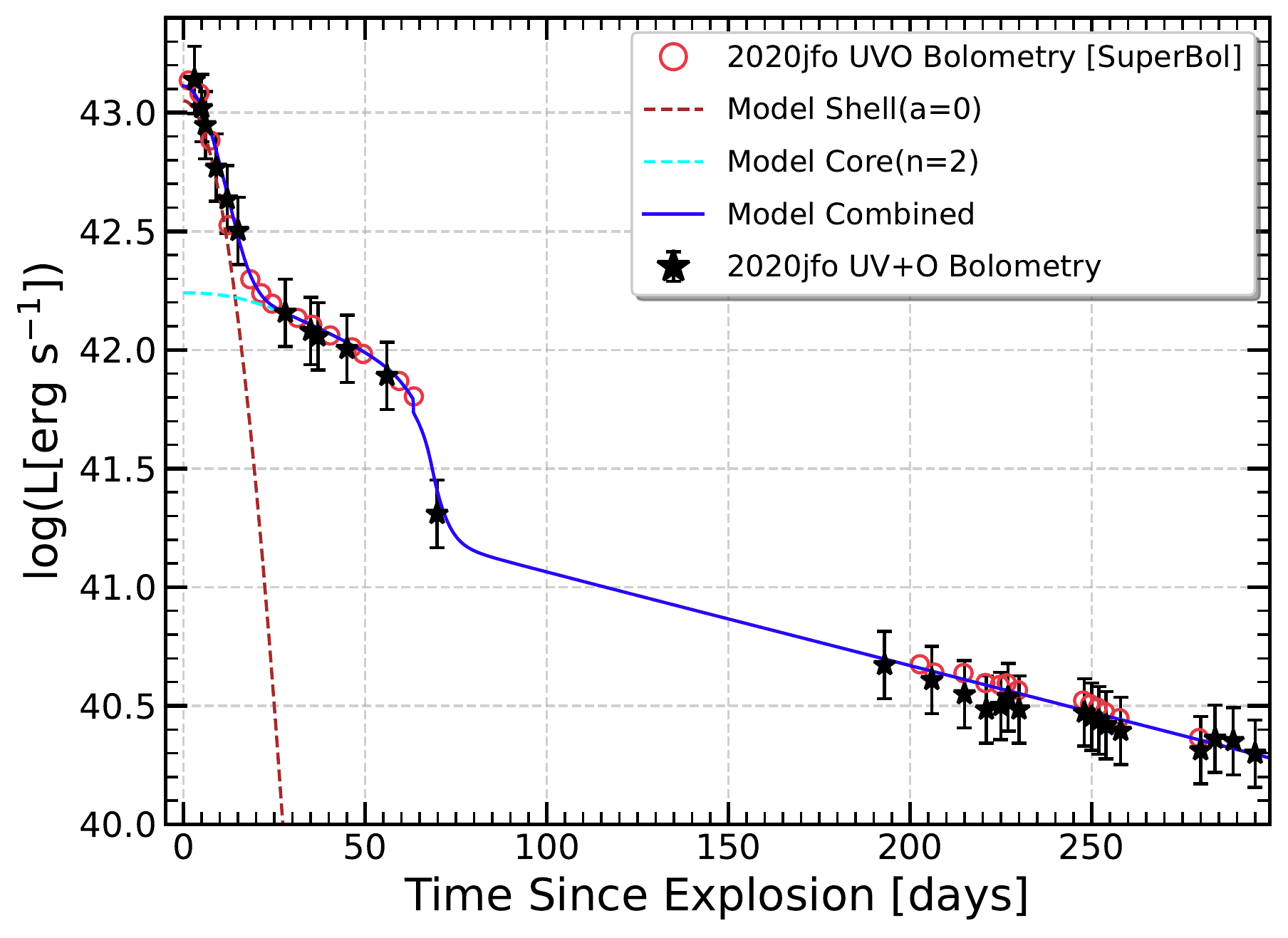}}
    \caption{Best fitting model curves to the bolometric light curve using two-component model from \citet{NagyandVinko}. Individual contributions from the shell and core component are shown with the combined bolometric luminosity.} 
    \label{fig:Nagy_fit}
\end{figure}
To obtain estimates on progenitor properties, we used a semi-analytical model described in \cite{NagyandVinko}, which was initially described by \cite{ArnettAndFu}. It models the supernova as a two-component system consisting of a core region that is dense and a shell region with a low mass extended envelope. It assumes that the SN ejecta is spherically symmetric and expanding homologously. For the density profile of the ejecta, the core region is assumed with a flat or constant density profile with a constant Thompson-scattering opacity of $\rm \kappa=0.4\,cm^{2}\,g^{-1}$ whereas the shell region has density profile which decreases as a power-law function ($\rm n=2$) or as an exponential ($\rm a=0$) with an opacity of $\rm \kappa=0.2\,cm^{2}\,g^{-1}$ \citep{NagyandVinko}. We obtained an ejecta mass of $\sim$ 7.5 $\rm M_\odot$ (core+shell), an RSG radius ranging from 310-340 $\rm R_\odot$, and a total energy (thermal and kinetic) of $\sim$\,3\,foe based on the best fitting model shown in Figure~\ref{fig:Nagy_fit}. The mass of $\rm ^{56}Ni$ obtained from this semi-analytical model, with the amount of gamma-leakage added to the model to follow the nebular light curve evolution is $\rm 0.03\pm 0.01\,M_\odot$, which is in corroboration with our earlier estimates.

\subsection{Estimate from nebular spectrum}
\label{subsec:jerkstrand}

\begin{figure}[htb!]
    \centering
  \resizebox{\hsize}{!}{\includegraphics{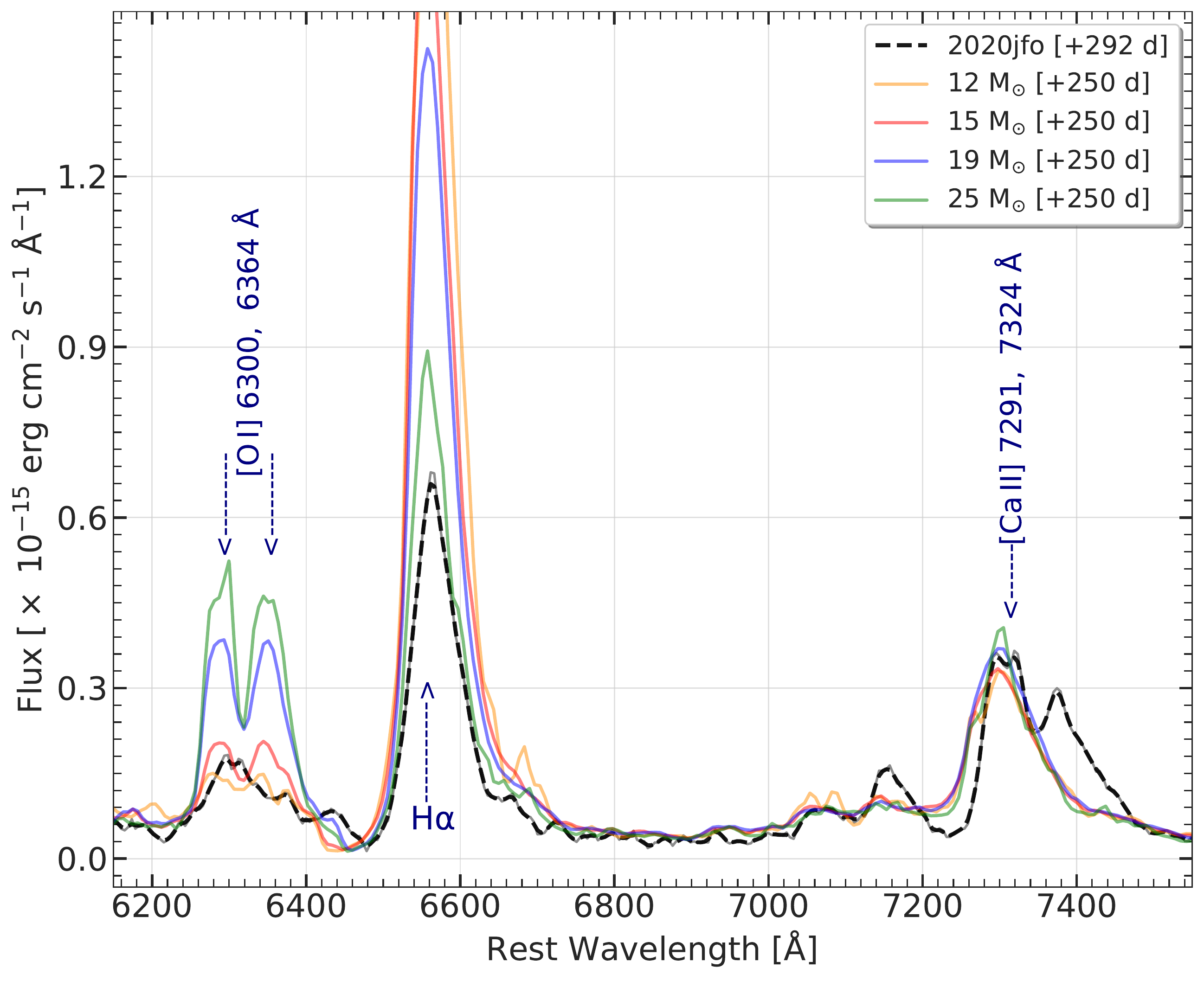}}
    \caption{Nebular spectrum (+292\,d) of SN~2020jfo is compared with the model spectra for 12, 15, 19 and 25 $\rm M_\odot$ models, at 250\,d since explosion. The model spectra obtained from \citet{2014jerkstrand} are scaled for distance and nickel mass, and corrected for phase mismatch using the characteristic decay time corresponding to SN~2020jfo.}
    \label{fig:specjerk}
\end{figure}

To constrain the progenitor mass, we compared the nebular phase spectrum of $+$292\,d with model spectra from \citet{2014jerkstrand} (Figure~\ref{fig:specjerk}). The model spectra for different progenitor masses viz. 12, 15, 19, and 25\,$\rm M_\odot$ have been scaled with respect to $\rm ^{56}Ni$ mass and the distance of SN~2020jfo (in contrast to 5.5\,Mpc for model spectra). In order to account for the difference in phase between the model spectra and the observed spectrum, the observed spectrum was scaled by the amount determined from the characteristic time scale of $\rm ^{56}Ni$-decay chain and the dissimilarity in phases. The comparison of [\ion{O}{1}]\,6300\,\AA, 6364\,\AA\ line fluxes of the observed spectra with the spectral models suggest a lower mass progenitor of $\rm \sim$\,12\,$\rm M_{\odot}$. However, the flux of $\rm H\alpha$ is quite weak compared to the 12\,$\rm M_{\odot}$ progenitor, indicating a stripped hydrogen envelope in SN~2020jfo. In a core-collapse SN, the mass of calcium synthesised is insensitive to the ZAMS mass of the progenitor, whereas the mass of oxygen synthesised depends on it, and the  [\ion{Ca}{2}]\,/\,[\ion{O}{1}] flux ratio is an indicator of the progenitor mass \citep{1989fransson}. The observed [\ion{Ca}{2}]\,/\,[\ion{O}{1}] flux ratio of $\sim$\,1.5 in the spectrum of $\sim$\,292\,d is also suggestive of a low-mass progenitor for SN~2020jfo.

\begin{figure}
\centering

\resizebox{\hsize}{!}{\includegraphics{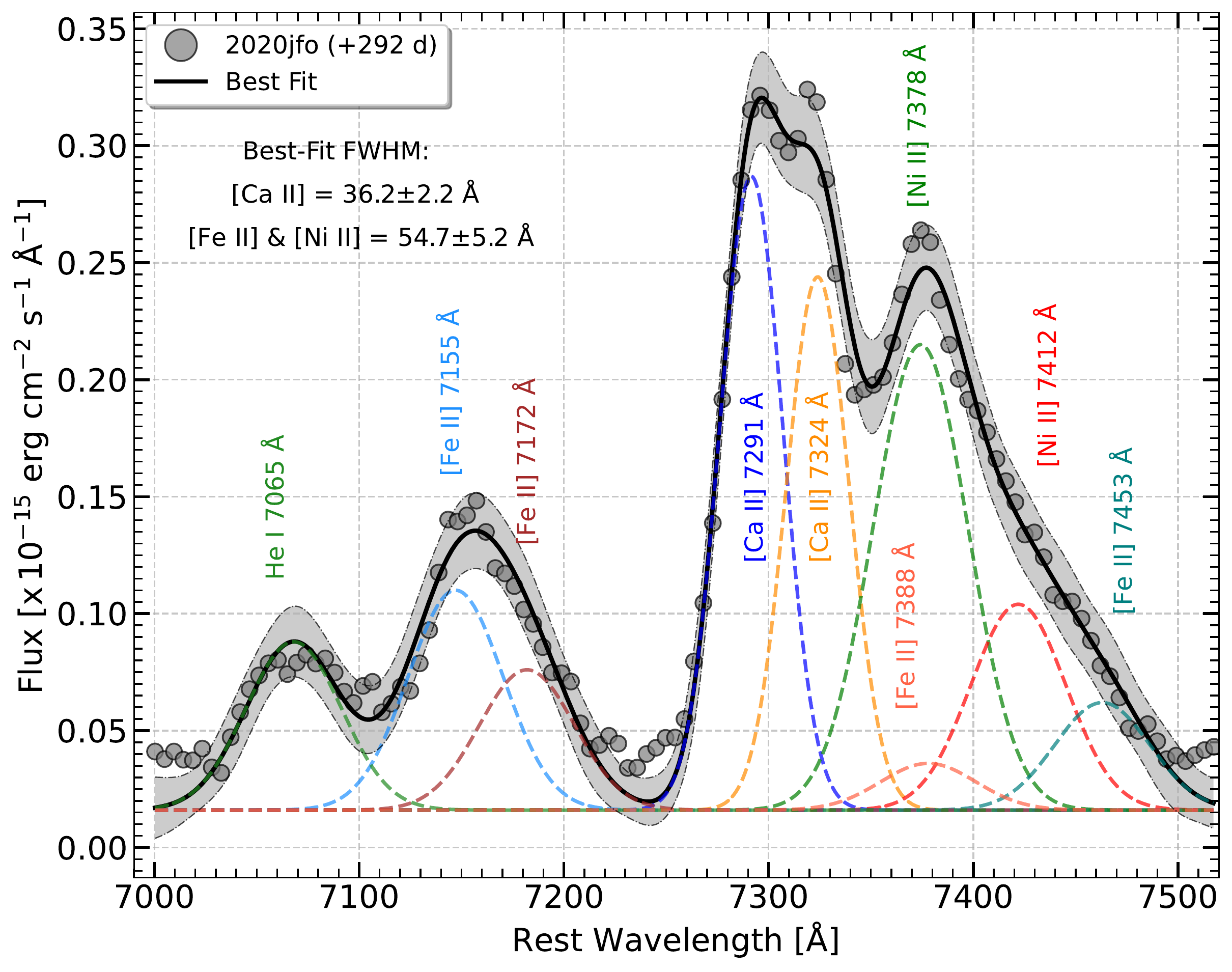}}

\caption{Multi-component Gaussian fit to the nebular spectrum of SN~2020jfo (+292\,d). The fit was performed by having different values of line broadening for [\ion{Ca}{2}] and [\ion{Fe}{2}]/[\ion{Ni}{2}] owing to their differing origins in the ejecta. The FWHM values obtained are mentioned in the figure.}
\label{fig:NiFeRatio}
\end{figure}

SN~2020jfo is also one of the few hydrogen-rich SNe where a clear, distinct spectral feature of [\ion{Ni}{2}]\,7378\,\AA\ is seen adjacent to the [\ion{Ca}{2}] feature in nebular phase spectral evolution. The feature has its origins from stable $\rm ^{58}Ni$, synthesised during explosive nucleosynthesis \citep{2015NiFe}. Following the methodology described in \cite{2015FeNiMethodology}, we computed a Ni/Fe luminosity ratio for SN~2020jfo (see Figure~\ref{fig:NiFeRatio}) as $2.10\pm0.43$ (similar to the value obtained in \citealp{2021Sollerman}). This translates to a Ni/Fe ratio by mass as $0.18\pm0.04$, which is roughly $3.0\pm0.6$ times the Solar value. This could be achieved either by a neutron excess usually found in the Silicon layer or due to a very high progenitor metallicity ($\rm >5\,Z_\odot$) that increases the neutron excess in Oxygen shell \citep{2015NiFe}. The estimated metallicity close to the SN site is $\rm \sim 1.5\,Z_\odot$ which is not high enough to produce such a Ni/Fe ratio in the ejecta. The only plausible scenario for such excess is seen in spherically symmetric models of $\rm M_{ZAMS}\lesssim 13\,M_\odot$ \citep{2015NiFe} that house a thick Si layer with a neutron excess. This is concurrent with our estimates of a lower mass progenitor.

\subsection{Hydrodynamical Modelling}
\label{subsec:mesahydro}

We resort to detailed hydrodynamical modelling for better constraints about the progenitor, its evolution, mass loss history and its immediate environment. We used the publicly available 1-D stellar evolution code \texttt{MESA r-15140} \citep{Paxton2011, Paxton2013, Paxton2015, Paxton2018, Paxton2019} and a simplified version of \texttt{STELLA} \citep{Blinnikov2004, Baklanov2005, Blinnikov2006} included with MESA to simulate light curves and photospheric velocities of SN~2020jfo. \texttt{MESA\,+\,STELLA} has been successfully used in many studies to investigate properties of Type IIP SNe progenitors \citep{2011Moriya, 2019Goldberg, 2021Hiramatsu}. We also try this framework to get more insights about the progenitor of SN~2020jfo. Some of the aspects regarding various hydrodynamical parameters are as follows:

\begin{enumerate}
\item The built-in nuclear reactions rates were taken from `approx21\_cr60\_plus\_co56.net'. Nuclear reaction rates are mostly from the Nuclear Astrophysics Compilation of Reaction rates, \citep[NACRE,][]{1999nacre} and the Joint Institute for Nuclear Astrophysics, JINA reaction rates \citep[REACLIB,][]{2010JINA}.

\item Cool and hot wind schemes for the Red Giant Branch or Asymptotic Giant Branch phase are taken as `Dutch', as described in MESA IV. This wind scheme for massive stars is a combination of results from work by various Dutch authors. The particular combination chosen is based on the work by \citet{dutch1}. Typically, if the surface hydrogen has a mass fraction less than 0.4 and an effective temperature greater than $\rm 10^4\,K$, the prescription used is from \cite{vink1}, otherwise it is taken from \cite{nugis1}.

\item The mixing length parameter (\texttt{MLT\_option}) is set to $Henyey$, which is based on the work by  \cite{henyey1965}, with $\rm \alpha_{MLT} = 1.5$, where, $\rm \alpha_{MLT}$ is the ratio of mixing length to the pressure scale height ($\rm = P/g\rho$). 

\item To determine the position of the convective boundaries, the default $Ledoux$ criterion is used.

\end{enumerate}

From the evolution of pre-main-sequence star to finally retrieving the optical light curve post-explosion, the modelling process was completed in three steps inside the \texttt{MESA} framework. First, a pre--MS star was evolved till the Fe-core developed, and there was an onset of rapid infall of the iron core. It was accomplished using \texttt{`make\_pre\_ccsnIIp'} test suite provided in MESA. Default values of the controls in \texttt{inlists} were used with slight variations for convergence with the help from \citet{Farmer}. We fixed our metallicity to $\rm Z=0.024$ as estimated in Section~\ref{subsec:host} for all the simulations.

\begin{figure}
\resizebox{\hsize}{!}{\includegraphics{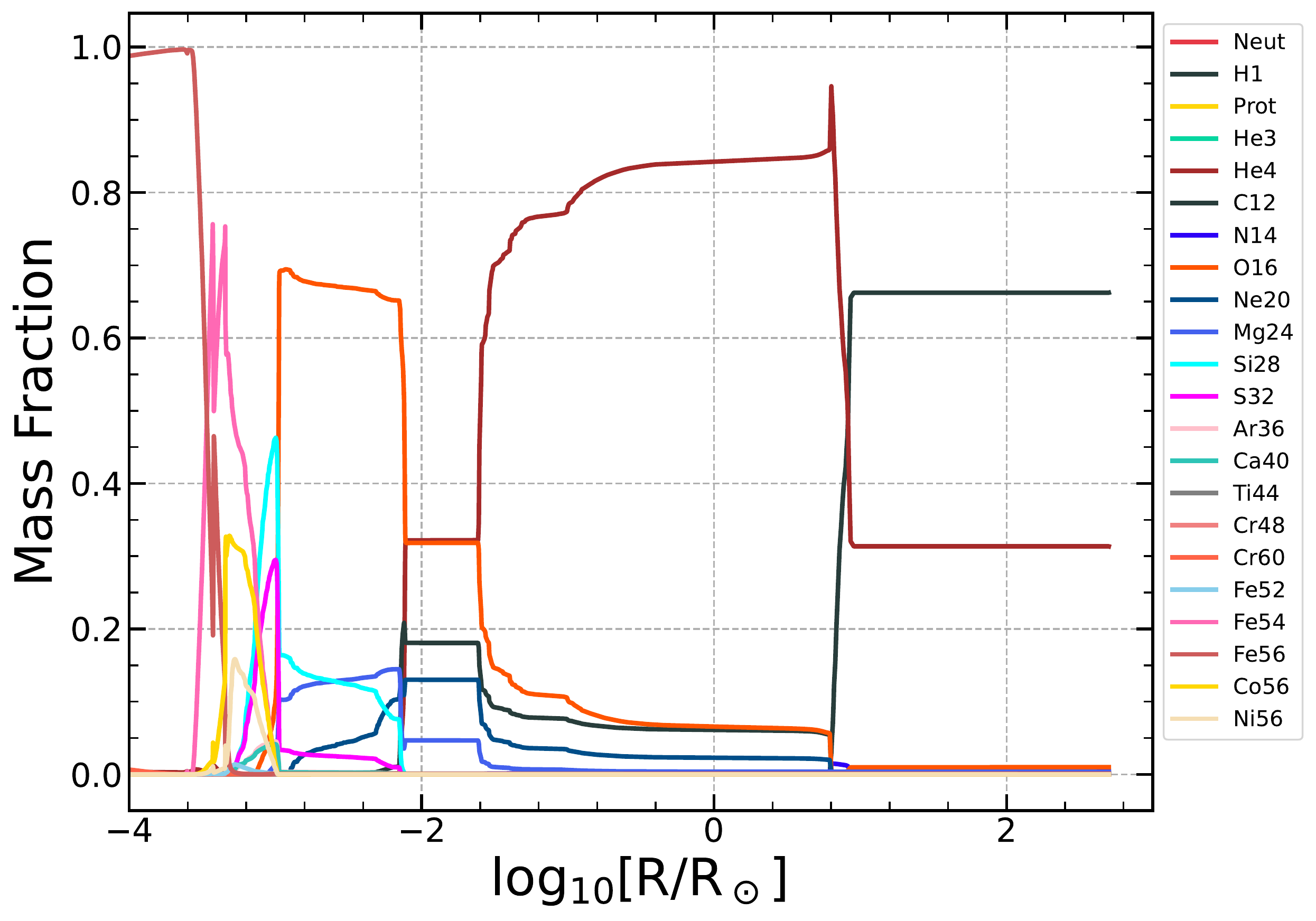}}
    \caption{Pre-supernova mass fractions for an evolved 12\,$\rm M_\odot$ model for different species present in the `approx21\_cr60\_plus\_co56.net' (approx21) network used in \texttt{MESA} modelling.} 
    \label{fig:MesaFractions}
\end{figure}

Since the explosion could not be achieved directly by MESA, we proceeded to the second step, which closely followed the \texttt{`ccsn\_IIp'} test suite. In this step, a section of the core was removed, which would have eventually collapsed onto a proto--NS. This centre section was removed from the model at the location where $\rm entropy/baryon = 4k_B$. Later, the explosion was induced by the synthetic injection of energy into a thin layer of $\rm \approx 0.01\,M_\odot$ at the inner boundary (IB) for 5\,ms, and the rate was scaled such that the $\rm E_{exp}$ reached the desired input value. Shock then proceeded through the various steps until it reached just below the surface where the hand-off was performed from MESA to STELLA \citep{Paxton2018}. STELLA then dealt with the shock-breakout and post-explosion evolution. For the second step, we did not vary $\rm ^{56}Ni$ mass estimate of 0.033\,M$_\odot$ obtained in Section~\ref{subsec:nickel_mass} in order to reduce the parameter space.


A recent study by \citet{2021Hiramatsu} discusses the possibility of obtaining shorter plateaus in Type II SN light curves from progenitors with ZAMS mass of 18-25 $\rm M_\odot$ with an enhanced mass loss ($\dot{M}\rm \simeq10^{-2}\,M_\odot\,yr^{-1}$) in the decades prior to collapse. The objects of interest in their study were more luminous events ($\rm Peak\,M_V<-18\,mag$), with a higher $\rm ^{56}Ni$ yield and higher expansion velocities in contrast to SN~2020jfo, which has an average $\rm ^{56}Ni$ mass and lower expansion velocities in comparison to a typical Type II SNe. In addition, a number of other numerical modelling works \citep{dessart2010, sukhbold} showed that low mass progenitors of $\rm M \leq 18\,M_\odot$ were not able to produce a shorter plateau duration of around 60 days. In all, none of the simulations for masses $\rm M \leq 15\,M_\odot$ under standard conditions were able to produce light curves with short plateaus \citep{NSP1, 2021Hiramatsu, dessart2010, sukhbold}. Instead of exploring the parameter space favoured by other works for short plateau Type II SNe, we took a different approach, where the inputs were driven from the results of semi-analytical modelling and nebular phase spectra. Hence, we went ahead with the evolution of a ZAMS model of $\rm 12 \,M_\odot$ and tried variations in the evolution schemes to achieve a shorter plateau.

\begin{figure}[htb!]
\resizebox{\hsize}{!}{\includegraphics{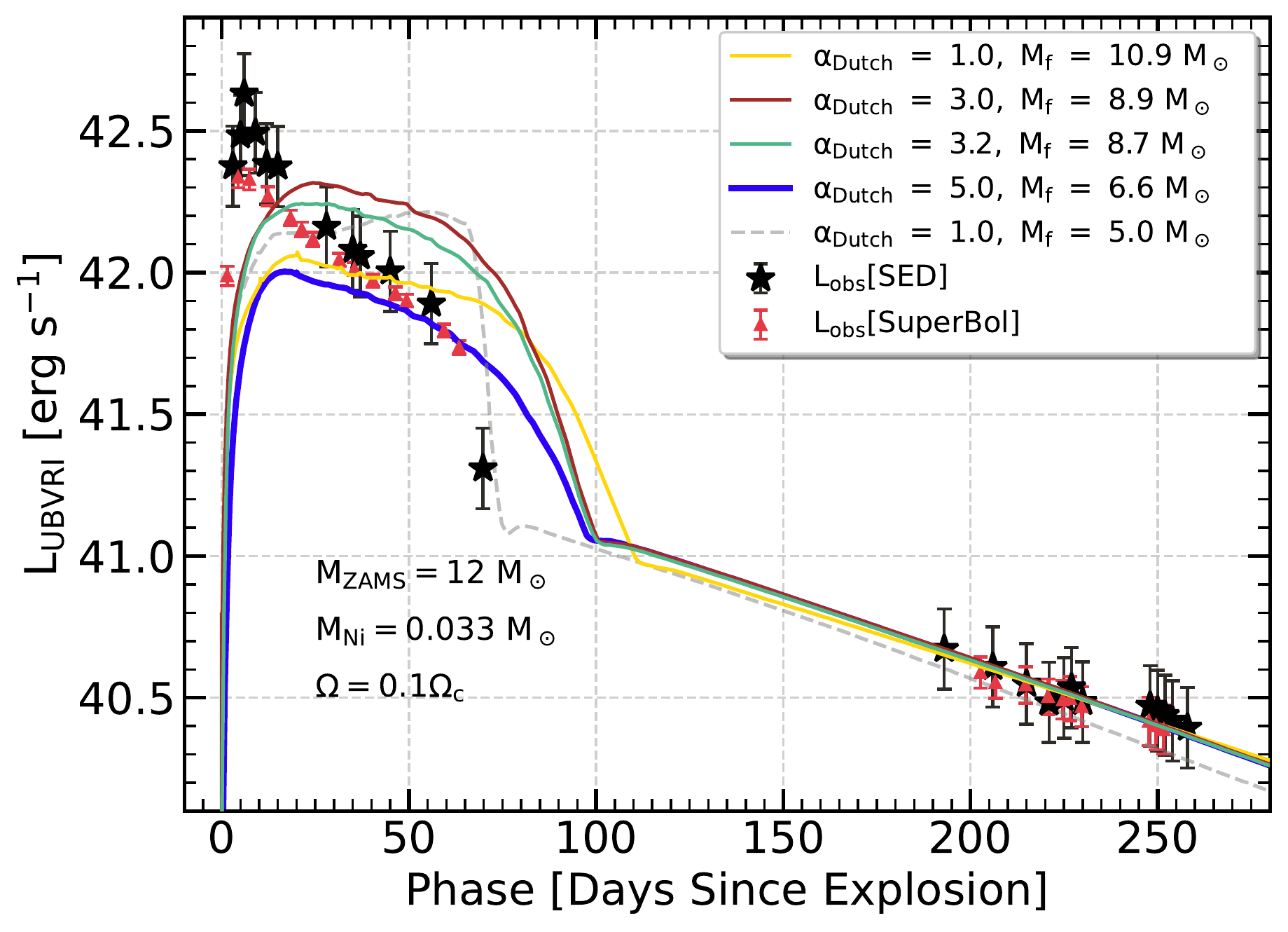}}
    \caption{Quasi-bolometric light curves obtained from \texttt{MESA\,+\,STELLA} modelling with different values of wsf (1.0, 3.0, 3.2 and 5.0) for the $\rm 12\,M_\odot$ ZAMS model. Q-bol of SN~2020jfo is over-plotted for comparison.} 
    \label{fig:StellaQuasibBol}
\end{figure}

A 12\,$\rm M_\odot$ progenitor was evolved with an initial metallicity slightly higher than Solar and with a finite amount of rotation ($\rm \Omega = 0.1\,\Omega_{critical}$). Figure \ref{fig:MesaFractions} shows the pre-Supernova mass fractions of an evolved model for `approx21' network. It was found that, for a typical mass loss rate due to winds, a short plateau was not possible as there was not enough stripping of the hydrogen envelope of the progenitor's ejecta. Hence, an enhanced mass loss due to winds was applied during the evolution, which is highly possible in a higher metallicity environment with a rotating progenitor. Mass loss was controlled by the wind scaling factor (wsf) in MESA. We tried varying this parameter from a default value of 1.0 onwards. At a value of $\rm wsf=5.0$, we could get enough material stripped off from its surface in order to achieve a short plateau with a similar period as SN~2020jfo. We also note that the sharp transition could not be produced using physical mass loss schemes.
However, a sharp transition is achieved if the mass is removed by hand to leave the final mass as 5.0\,M$_\odot$. Exploring amount of nickel mixing in layers, density structure of the progenitor, etc. is beyond the scope of this work. Some of the models did not converge as the central density was not sufficient enough for the ignition of higher masses during the course of evolution. The simulated light curves obtained for various wind scaling factors along with Q-bol for SN~2020jfo are shown in the Figure~\ref{fig:StellaQuasibBol} and corresponding pre-SN values for various models are presented in Table \ref{tab:mesa_preSN}.

\begin{table}[hbt!]
    \centering
    \caption{Pre-SN parameters for different models}
    \begin{tabular}{|c|c|c|c|c|}
    \hline \hline 
    \multicolumn{5}{|c|}{$\rm M_i\ =\ 12.0\ M_\odot,\ \Omega\ =\ 0.1\Omega_c,\ Z\ =\ 0.024$} \\
    \hline \hline
         $\rm M_f$  & Age &	$\rm \alpha_{Dutch}$ & Radius  & $\rm E_{tot} $ \\
         $\rm (M_\odot)$ & (Myr) & & $\rm (R_\odot)$ & (foe) \\
         \hline
10.9	& 18.7 &	1.0	& 470 &	--0.94 \\
8.9	 & 19.2	& 3.0	& 780 &	--1.00 \\
8.7	& 19.3	& 3.2 &	723 & 	--0.93 \\
{\bf 6.6}	&{\bf  19.7}	& {\bf 5.0} &{\bf	679}	&{\bf --0.91 }\\
\hline \hline
    \end{tabular}
    \label{tab:mesa_preSN}
\end{table}

\begin{figure}[htb!]
\resizebox{\hsize}{!}{\includegraphics{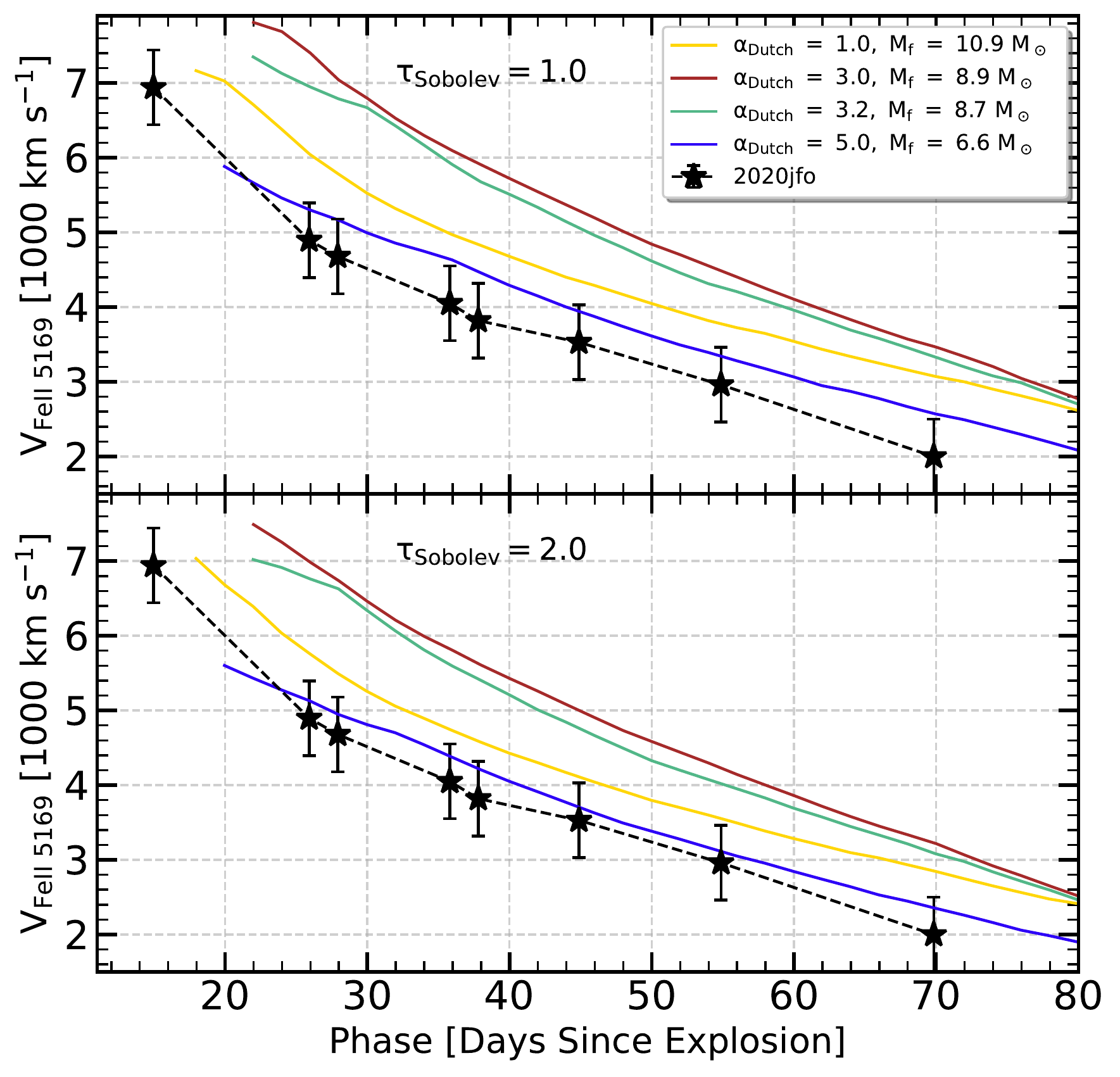}}
    \caption{Photospheric velocity evolution as obtained from \texttt{MESA\,+\,STELLA} modelling for two different values of optical depth ($\rm \tau_{Sobolev}$=1.0 and 2.0) compared with the observed photospheric velocities.} 
    \label{fig:StellaVel}
\end{figure}

It was demonstrated by comprehensive modelling that the model of ZAMS with 12\,$\rm M_\odot$ and the final mass of the progenitor as 6.6\,$\rm M_\odot$, fitted closely the decline to nebular phase and late phase evolution of the observed light curve (Figure~\ref{fig:StellaQuasibBol}). It had an ejected mass of about 5\,$\rm M_\odot$ and an excised core of about 1.6\,$\rm M_\odot$. The explosion energy for best-fitting models was $\rm E_{exp} = 0.2-0.4\,foe$. The photospheric velocity evolution for this model is in agreement with the estimates from the observed spectral sequence (see Figure~\ref{fig:StellaVel}). Slow velocity evolution could also be attributed to the low energy of the explosion as obtained from hydrodynamic modelling ($\rm 0.2-0.4\,foe$) along with the low ejecta mass as most of the mass was blown away by winds during evolution. A nickel mass of 0.033\,$\rm M_\odot$ used in models substantiate our earlier Ni mass estimates. While the modelled light curve matched with the observed plateau duration, decline to the nebular phase, and late phase light curve evolution, it failed to reproduce the early steep rise to a high luminosity observed in the quasi-bolometric light curve. As the calculations are based on normal type IIP SNe, only primary radiation sources viz. shock breakout and cooling, hydrogen recombination, and radioactive decay are considered. The inadequacy of the current model to fit the early part of the light curve indicates the need to introduce a secondary source of radiation for early times, and the best possible source could be the presence of CSM close to the progenitor.

\begin{figure}[htb!]
\resizebox{\hsize}{!}{\includegraphics{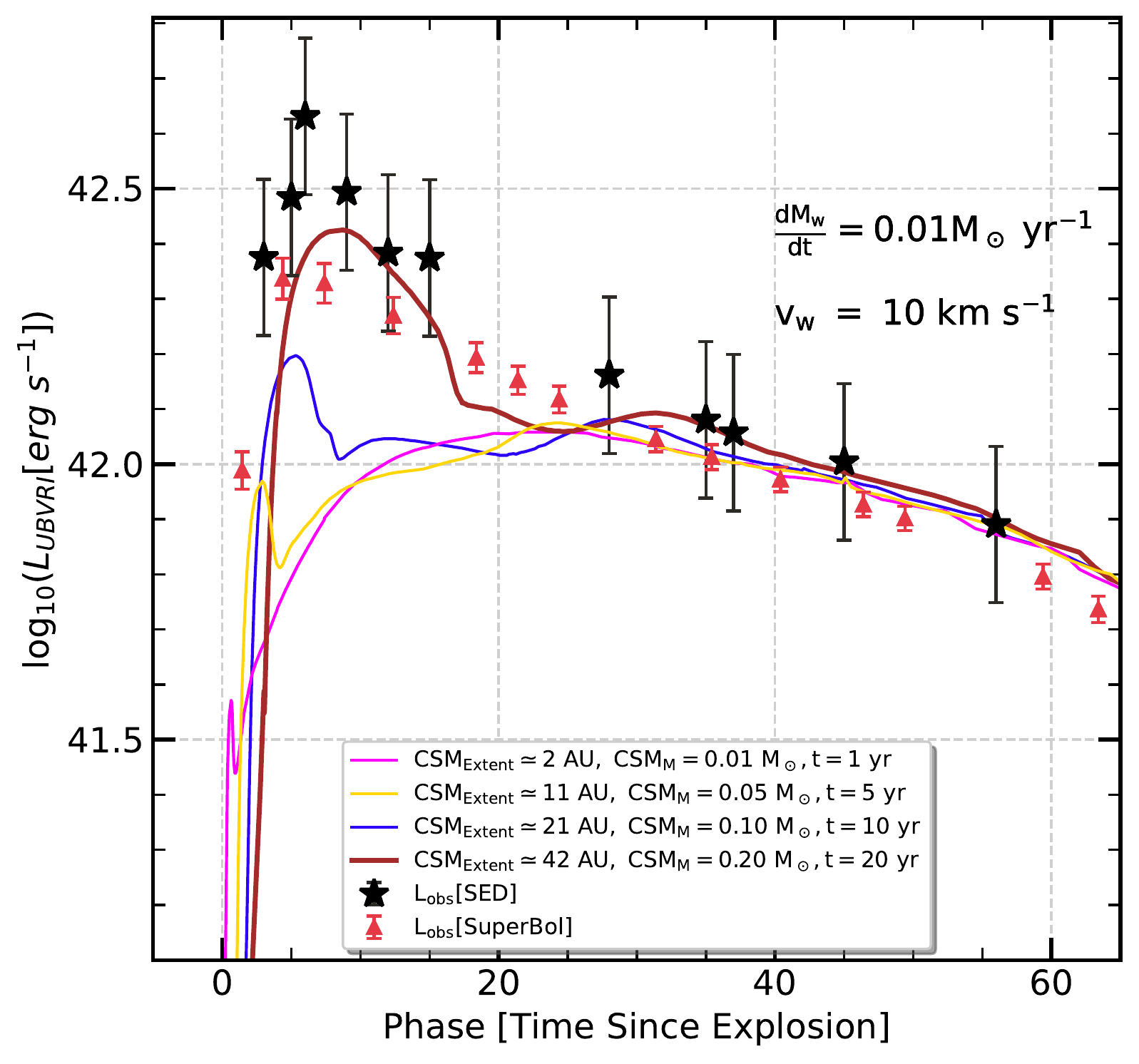}}
    \caption{Quasi-bolometric light curves as obtained from \texttt{MESA\,+\,STELLA} modelling for a $\rm 12\,M_\odot$ ZAMS progenitor ($\rm M_f=6.6\,M_\odot$) with different CSM configurations. Q-bol light curve of SN~2020jfo is over-plotted for comparison.} 
    \label{fig:StellaLbol}
\end{figure}

\begin{figure}[htb!]
\resizebox{\hsize}{!}{\includegraphics{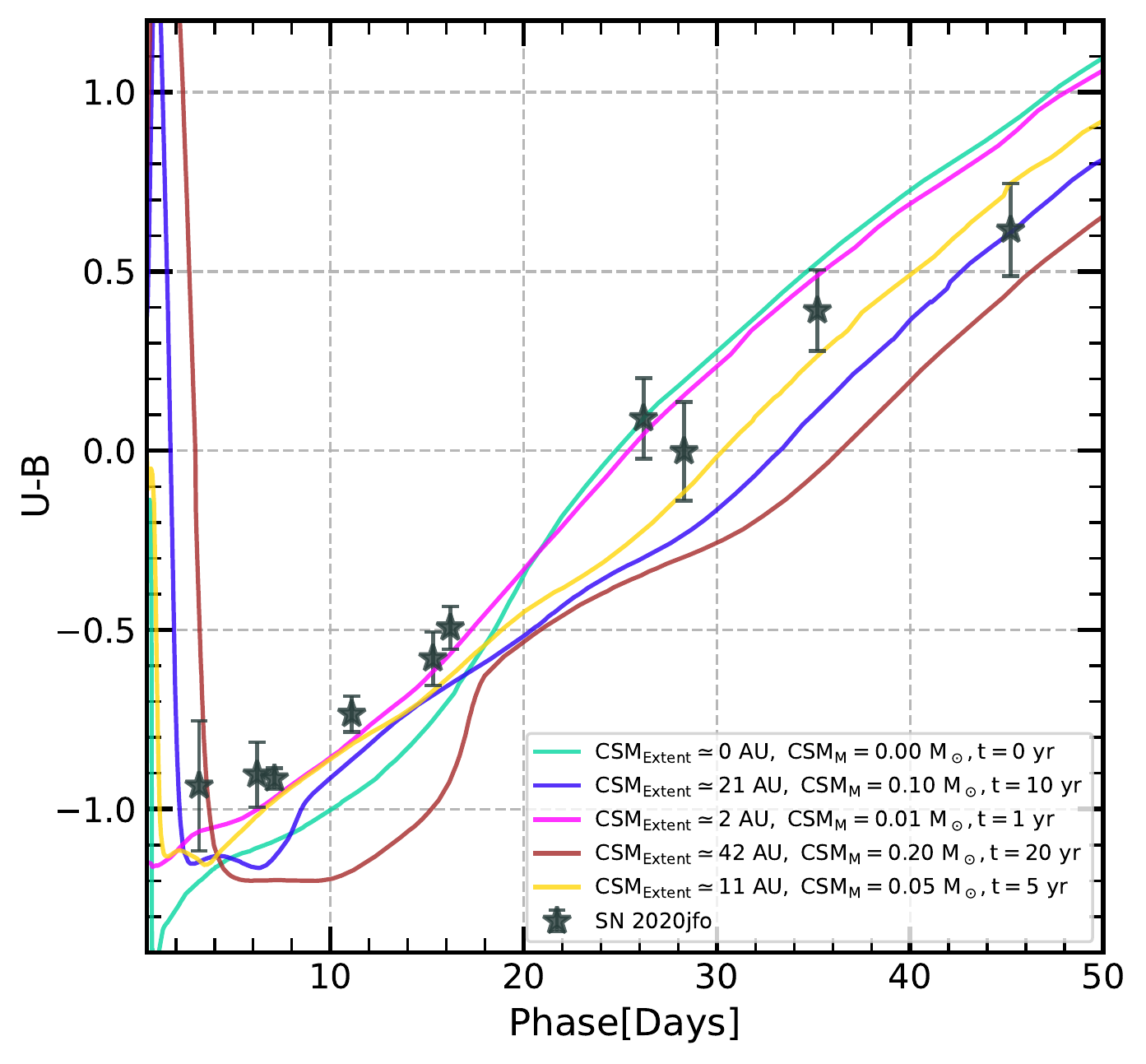}}
    \caption{Colour evolution as obtained from \texttt{MESA\,+\,STELLA} modelling for a $\rm 12\,M_\odot$ ZAMS progenitor ($\rm M_f=6.6\,M_\odot$) with different CSM configurations.} 
    \label{fig:StellaColors}
\end{figure}

Due to the lack of direct signatures of CSM interaction, we could not calculate the extent and density of the CSM. In order to estimate the same, we used \texttt{STELLA}, where it was possible to place the CSM around the progenitor with its configuration defined by the parameters wind velocity, mass loss rate, and its duration. The density profile in \texttt{STELLA} is dependent on the radius, $r$, away from the progenitor's centre as:

\begin{equation}
    \rho(r)_w = \frac{\dot{M_w}}{4\pi r^2 v_w},
\end{equation}

where $\dot{M_w}$ is mass loss rate in $\rm M_\odot\,yr^{-1}$ due to winds and v$_w$ is the wind velocity. We allocated 40 zones out of 400 for CSM configuration and the bolometric flux was obtained at four extents ($\rm ~2, 10, 20, 40\ AU$) with various mass loss rates ($\rm ~0.001, 0.005, 0.01, 0.05\,M_\odot\ yr^{-1}$). A typical wind velocity of $\rm ~10^6\,cm\,s^{-1}$ was affixed for all configurations. The modelled light curve with $\dot{M}\rm=0.01\,M_\odot\,yr^{-1},\ t=20\,yr$ (which corresponds to a CSM extent of roughly 40 AU) is found to fit the observed quasi-bolometric light curve well (see Figure~\ref{fig:StellaLbol}). The \ion{He}{2} feature considered as proxy for interaction signature is not observed beyond +10\,d. This could be due to formation and increasing strength of other lines in the spectra. Furthermore, it could also be due to decrease in CSM-ejecta interaction, giving rise to a steep decline in the quasi-bolometric light curve. Light curve modelling is suggestive of interaction up to +15\,d, which is considered as an upper bound. Further, we compare $U-B$ colour evolution (Figure~\ref{fig:StellaColors}) of models with observed $U-B$ colour. We find the colour to be flat for initial 8\,d and later it evolves towards red. This initial trend is only seen in the models with added CSM.

\subsection{CSM Interaction}

There have been numerous instances where studies have provided enough evidence for CSM surrounding Type II progenitors both in spectra and light curves. \cite{2018Forster} attributed the steeper light curve rise and delayed shock emergence to the dense CSM from their sample of 26 Type II SNe. Another study, combining light curve modelling and observations \citep{2018Morozova} summarised that $\sim 70\%$ SNe have CSM, and the estimated CSM masses ranged between $\rm 0.18-0.83\,M_\odot$. \cite{2021bruch} emphasised the appearance of narrow flash emission features, especially \ion{He}{2} 4686\,\AA, in the very early spectra, ideally taken less than 48 hrs of explosion. We do not see such narrow signatures of CSM interaction in our earlier spectra, although we do see broad ionised lines of Helium, which were likely formed at the CDS arising due to the shock ionisation of the outer layers or the CSM close to the ejecta. Along with this, the presence of HV $\rm H\alpha$ feature in the mid to late plateau phase spectra is an indication of CSM's presence.

The higher peak luminosity and steeper early phase decline seen in SN~2020jfo also strengthen the case of CSM close to its progenitor. High decline rates in the early phase of Type II SNe have been attributed to the interaction with CSM. The diagnostic in SN~2013by was the presence of asymmetric line profiles with photospheric signatures of high velocity features of hydrogen \citep{2013byValenti}. In SN~2014G, the presence of highly ionised spectroscopic features was attributed to a metal-rich CSM accumulated from the mass-loss events prior to the explosion \citep{2014GTerraran}. It is likely that the higher luminosity of SN~2020jfo during the early epoch is due to  interaction with the nearby CSM, and its density profile is such that this is not sustained for prolonged periods. We ascertained this possibility with hydro-dynamical modelling using \texttt{MESA\,+\,STELLA}.


Furthermore, the colour evolution of SN~2020jfo in the early phase is bluer as compared to other SNe, while in the late phase, it flattens out and merges with the normal Type II SN colour evolution. The bluer early phase colour evolution is similar to the CSM-interacting events. The Q-bol of SN~2020jfo during the early phase is comparable to SN~2009au and SN~2014G (see Figure~\ref{fig:bolom1}), which showed clear signs of interaction in their spectra. Though the luminosity is higher during the early phase, the steeper decline in the plateau phase leads to a luminosity comparable to normal Type IIP events such as SN~2016X towards the end of the plateau. The additional source giving rise to the higher luminosity in the early phase is likely due to CSM interaction, however, the CSM remains hidden. \citet{discCSM} showed that a CSM distributed in the form of a disc, when viewed from a polar angle, would only cause enhancement in flux due to heating effects and would not leave any signatures of interaction in the spectra. Although, as clearly shown in \citet{discCSM}, overluminous Type IIP SNe might not be powered by such disc interaction, but a slight enhancement is a likely proposition during the early phase. 

To have a complete picture, we also looked at the field of SN~2020jfo for any radio detection. The  field was observed on October 17, 2021 (JD 2459504.5) in the VLA sky survey (VLASS) (image cut-outs can be found here \footnote{\href{http://cutouts.cirada.ca/}{http://cutouts.cirada.ca/}}). No significant radio emission was detected at the SN position, and a flux density limit of 309 $\rm \mu$Jy ($\rm 3\sigma$ upper limit) at 3 GHz was obtained. Unsurprisingly, the SN being $>$ 500 days old at the time of VLASS observations, the radio flux density declined below the sensitivity limits of current radio telescopes around this period, even for radio bright Type IIP SNe. Using the expression for mass-loss rate given in \cite{1986radioWeiler}, we obtained an upper limit of $ \dot{M}<2.5 \times 10^{-5}\,\rm M_\odot\ yr^{-1}$. This value is somewhat consistent with typical Type IIP SNe but smaller for our case where we estimated a higher mass-loss rate (see Section~\ref{subsec:mesahydro}) from the MESA+STELLA modelling of the early phase bolometric light curve. This is likely due to the difference in the epochs of observations of the radio and the modelled light curve with CSM. \citet{2021Sollerman} also looked for X-ray emission post-explosion for SN~2020jfo and could only cite an upper limit based on their estimates. It might be the case that the X-ray emission from dense CSM was earlier on and could have been missed as it was absorbed by nearby CSM (similar scenario as pointed out in \citealp{2021Jacobson}).

\begin{figure}
    \centering
    \resizebox{\hsize}{!}{\includegraphics{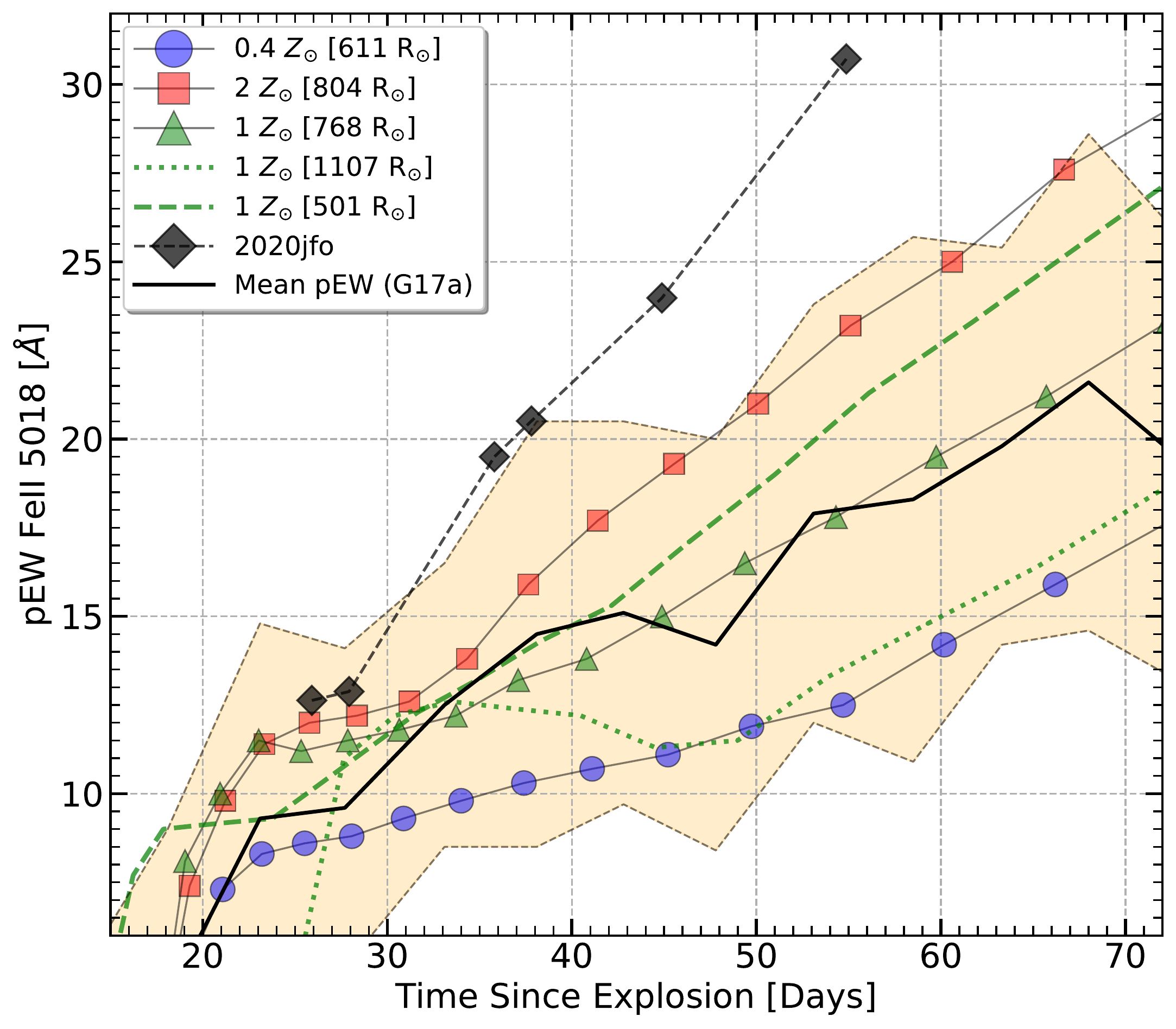}}
    \caption{Temporal evolution of pEW of \ion{Fe}{2} 5018\,\AA\ in comparison with models from \citet{2013bdessart} having different metallicities (0.4, 1 and 2\,$\rm Z_{\odot}$). The black solid line represents the mean value of the pEW of \ion{Fe}{2}\,5018\,\AA\ and shaded region shows its dispersion from the extensive sample of \citet{2017Claudia}.}
    \label{fig:pEWFe}
\end{figure}

\subsection{Case for a stripped low mass progenitor}

The metallicity estimated, was slightly higher than the Solar values and might have helped in escalating the mass-loss rate of the progenitor star. It has been argued in \cite{2013bdessart} that lower envelope mass at higher metallicity ought to produce a Type II SN with a shorter duration plateau. A direct result of this is seen in the temporal evolution of pseudo-equivalent widths (pEW) of \ion{Fe}{2} 5018\,\AA\ of SN~2020jfo in comparison to the estimates from spectral models from \citet{2013bdessart} (see Figure~\ref{fig:pEWFe}). We find that with increasing metallicities, the pEW increases. The higher pEW evolution of SN~2020jfo during the photospheric phase corroborates with the enhanced metallicity environment. However, the implied metallicity is much higher than that estimated for its host (i.e. $\sim$\,1.5 $\rm Z_{\odot}$ in Section~\ref{subsec:host}).  \cite{2013bdessart} have shown that the compactness (i.e. radius) of the progenitor affects the pEW of the metal features during the plateau phase. This is depicted in Figure~\ref{fig:pEWFe}, which shows 3 variants of 1\,$\rm Z_{\odot}$ metallicity progenitor, with different final progenitor radii. Observationally, this indicates the presence of a stripped hydrogen envelope, which would make the progenitor more compact, and probably explain the lack of a clear early decline ($s1$) phase prior to the onset of the plateau ($s2$) phase.

The semi-analytical modelling of the bolometric light curve of SN~2020jfo in Section~\ref{subsec:semi} infers a progenitor of low mass with an ejected mass, $\rm M_{ej}$ of $\rm \sim 7.3\,M_\odot$ and an RSG radius of $\rm \sim 350\ R_\odot$. Considering a typical Neutron Star (NS) remnant core of $\rm \sim 1.5\,M_\odot$, we infer a pre-SN mass of $\rm \sim\,9\,M_\odot$.

From \texttt{MESA+STELLA} modelling, we obtained a stripped progenitor with a pre-SN mass of $\sim 6.6\rm\,M_\odot$ where the initial ZAMS mass was $12\rm,M_\odot$. A similar ZAMS value for the progenitor was also obtained  using late nebular spectra (see Section~\ref{subsec:jerkstrand}).  
However, we did not implement any synthetic mass loss scheme (deliberate removal of the H envelope mass) to achieve stripping of progenitor (see Section~\ref{subsec:mesahydro}). Nevertheless, there could be numerous short time windows where enhanced mass loss is possible \citep{DecinHighMass}, which are hard to predict and hence challenging to include in modelling. Observationally, mass-loss rates have not been constrained very well and vary by over two orders of magnitudes ($10^{-4}$ to $10^{-6}\,\rm M_\odot \,yr^{-1}$; see \citealp{vanloon, mauron}). The mass loss rate adopted for the progenitor of SN~2020jfo in \texttt{MESA} is five times the typical mass-loss rates for an RSG progenitor, but is well within the observed limits. It is difficult to predict what could have caused such a high mass-loss rate, whether it was solely due to the rotation and high metallicity environment, or due to other factors such as interaction with a binary companion or multiple episodes of enhanced mass loss. 

It was shown by \citet{2018Eldridge} that the initial progenitor masses around 8-15\,$\rm M_{\sun}$ in the binary scenario possibly give light curves with shorter plateaus of the order of tens of days. However, their physical parameter space was limited, and not much could be said quantitatively about the progenitor properties. Another attempt by \cite{2021Hiramatsu} showed that the RSG progenitors with initial masses of 18-25\,$\rm M_{\sun}$ with enhanced mass-loss rates could reproduce shorter duration plateaus. However, the observed properties viz. nebular spectra, the mass of synthesised radioactive nickel, and evolution velocities of the events (SN~2006Y, SN~2006ai and SN~2016egz) were also supportive of higher mass progenitors. Both these studies had shown a continuous population of Type IIP-IIL-IIb events, wherein a higher progenitor mass leads to an increased amount of stripping of the hydrogen envelope. However, SN~2020jfo poses a question to the standard progenitor scenario. The arguments presented in our analysis and discussion weigh in on a low mass progenitor with enhanced mass loss that gave birth to the short plateau supernova SN~2020jfo.

\section{Summary}
\label{sec:summary}

In this work, we presented an extensive multi-band photometric and optical spectroscopic study of a short-plateau Type II event, namely SN~2020jfo.
 
\begin{enumerate}
    \item We estimated a plateau duration of $<65$\,d for SN~2020jfo, putting it under the category of the rare, short plateau Type IIP SNe. 
    \item The observational properties associated with SN~2020jfo are: peak V-band absolute magnitude of $\rm M_V=-17.4\pm0.4$ mag, peak optical luminosity of $\rm 4.3\pm1.4\times10^{42}\,erg\,s^{-1}$, and a synthesised $\rm ^{56}Ni$ mass of $\rm 0.033\pm0.006\,M_\odot$.
    \item Using nebular phase spectrum, we estimated a progenitor of mass $\sim 12$\,$\rm M_\odot$.
    \item We estimated the progenitor properties for SN~2020jfo from hydrodynamical modelling and concluded that the most plausible progenitor is a Red Super Giant with an initial mass around 12\,$\rm M_\odot$, radius $\rm \sim 679\,R_\odot$ and a final pre-supernova mass $\rm< 6.6\,M_\odot$. It evolved in a relatively high metallicity environment and shredded a significant amount of mass during its course of evolution.
    \item A high Ni/Fe ratio of 0.18$\pm$0.04 by mass was estimated for SN~2020jfo that is consistent with a low mass progenitor ($\rm M_{ZAMS}\leq13\,M_\odot$).
    \item The evolution of pEW values of \ion{Fe}{2} 5018\,\AA\ are much higher than the other Type II SNe, confirming the presence of a high metallicity environment and a compact progenitor.
    \item The presence of ionised \ion{He}{2} line, HV $\rm H\alpha$ feature, higher luminosity in contrast to slower velocity evolution, and the steeper decline in luminosity indicated the presence of CSM, which was confirmed by \texttt{MESA+STELLA} hydrodynamical modelling. It was deduced that a $\rm 0.2\,M_\odot$ CSM, extended to a region of $\rm \sim 40\,AU$ around progenitor was required to explain the early higher luminosity and a faster decline of the bolometric light curve.
\end{enumerate}
\vspace{5mm}
\section*{Acknowledgements}
We thank the referee for a thorough evaluation of the manuscript that helped in
improving it. We thank the staff of IAO, Hanle and CREST, Hosakote, that made these observations possible. The facilities at IAO and CREST are operated by the Indian Institute of Astrophysics, Bangalore. RST acknowledges Anirban Dutta for helpful discussions during this work. NAJ would like to acknowledge DST-INSPIRE Faculty Fellowship (IFA20-
PH-259) for supporting this research. This research has made use of the High Performance Computing (HPC) resources\footnote{\href{https://www.iiap.res.in/?q=facilities/computing/nova}{https://www.iiap.res.in/?q=facilities/computing/nova}} made available by the Computer Center of the Indian Institute of Astrophysics, Bangalore. This research made use of \textsc{RedPipe}\footnote{\url{https://github.com/sPaMFouR/RedPipe}} \citep{2021redpipe}, an assemblage of data reduction and analysis scripts written by AS. This research is also based on observations obtained at the 3.6m Devasthal Optical Telescope (DOT), which is a National Facility run and managed by Aryabhatta Research Institute of Observational Sciences (ARIES), an autonomous Institute under the Department of Science and Technology, Government of India. This research has also made use of the NASA/IPAC Extragalactic Database (NED\footnote{\url{https://ned.ipac.caltech.edu}}), which is funded by the National Aeronautics and Space Administration and operated by the California Institute of Technology. We used the Open Supernova Catalog\footnote{\url{https://sne.space}} \citep[OSC,][]{2017guillochon} to retrieve all the light curve and spectral data for comparison. We also acknowledge Wiezmann Interactive Supernova data REPository\footnote{\url{https://wiserep.weizmann.ac.il}} \citep[(WISeREP)]{2012yaron} for spectral data downloads. This research also made use of \textsc{tardis}, a community-developed software package for spectral synthesis in supernovae \citep{2014TARDIS}. The development of \textsc{tardis} received support from GitHub, the Google Summer of Code initiative, and from ESA's Summer of Code in Space program. \textsc{tardis} is a fiscally sponsored project of NumFOCUS. \textsc{tardis} makes extensive use of Astropy and Pyne.

\vspace{5mm}
\facilities{HCT: 2\,m, Swift (UVOT), Palomer-ZTF, DOT: 3.6\,m}

\software{MESA (release 15140), STELLA, PyRAF (v2.1.14), Astropy \citep{2018Astropy}, emcee \citep{2013emcee}, matplotlib \citep{2007matplotlib}, pandas \citep{pandas2010}, SciPy \citep{2020SciPy}, seaborn \citep{seaborn2021}} 
\clearpage



\bibliography{sn2020jfo}{}

\begin{thebibliography}{}
\expandafter\ifx\csname natexlab\endcsname\relax\def\natexlab#1{#1}\fi
\providecommand{\url}[1]{\href{#1}{#1}}
\providecommand{\dodoi}[1]{doi:~\href{http://doi.org/#1}{\nolinkurl{#1}}}
\providecommand{\doeprint}[1]{\href{http://ascl.net/#1}{\nolinkurl{http://ascl.net/#1}}}
\providecommand{\doarXiv}[1]{\href{https://arxiv.org/abs/#1}{\nolinkurl{https://arxiv.org/abs/#1}}}

\bibitem[{{Ahumada} {et~al.}(2020){Ahumada}, {Prieto}, {Almeida}, {Anders},
  {Anderson}, {Andrews}, {Anguiano}, {Arcodia}, {Armengaud}, {Aubert}, {Avila},
  {Avila-Reese}, {Badenes}, {Balland}, {Barger}, {Barrera-Ballesteros}, {Basu},
  {Bautista}, {Beaton}, {Beers}, {Benavides}, {Bender}, {Bernardi}, {Bershady},
  {Beutler}, {Bidin}, {Bird}, {Bizyaev}, {Blanc}, {Blanton}, {Boquien},
  {Borissova}, {Bovy}, {Brandt}, {Brinkmann}, {Brownstein}, {Bundy}, {Bureau},
  {Burgasser}, {Burtin}, {Cano-D{\'\i}az}, {Capasso}, {Cappellari}, {Carrera},
  {Chabanier}, {Chaplin}, {Chapman}, {Cherinka}, {Chiappini}, {Doohyun Choi},
  {Chojnowski}, {Chung}, {Clerc}, {Coffey}, {Comerford}, {Comparat}, {da
  Costa}, {Cousinou}, {Covey}, {Crane}, {Cunha}, {Ilha}, {Dai}, {Damsted},
  {Darling}, {Davidson}, {Davies}, {Dawson}, {De}, {de la Macorra}, {De Lee},
  {Queiroz}, {Deconto Machado}, {de la Torre}, {Dell'Agli}, {du Mas des
  Bourboux}, {Diamond-Stanic}, {Dillon}, {Donor}, {Drory}, {Duckworth},
  {Dwelly}, {Ebelke}, {Eftekharzadeh}, {Davis Eigenbrot}, {Elsworth},
  {Eracleous}, {Erfanianfar}, {Escoffier}, {Fan}, {Farr},
  {Fern{\'a}ndez-Trincado}, {Feuillet}, {Finoguenov}, {Fofie},
  {Fraser-McKelvie}, {Frinchaboy}, {Fromenteau}, {Fu}, {Galbany}, {Garcia},
  {Garc{\'\i}a-Hern{\'a}ndez}, {Oehmichen}, {Ge}, {Maia}, {Geisler}, {Gelfand},
  {Goddy}, {Gonzalez-Perez}, {Grabowski}, {Green}, {Grier}, {Guo}, {Guy},
  {Harding}, {Hasselquist}, {Hawken}, {Hayes}, {Hearty}, {Hekker}, {Hogg},
  {Holtzman}, {Horta}, {Hou}, {Hsieh}, {Huber}, {Hunt}, {Chitham}, {Imig},
  {Jaber}, {Angel}, {Johnson}, {Jones}, {J{\"o}nsson}, {Jullo}, {Kim},
  {Kinemuchi}, {Kirkpatrick}, {Kite}, {Klaene}, {Kneib}, {Kollmeier}, {Kong},
  {Kounkel}, {Krishnarao}, {Lacerna}, {Lan}, {Lane}, {Law}, {Le Goff}, {Leung},
  {Lewis}, {Li}, {Lian}, {Lin}, {Long}, {Longa-Pe{\~n}a}, {Lundgren}, {Lyke},
  {Ted Mackereth}, {MacLeod}, {Majewski}, {Manchado}, {Maraston}, {Martini},
  {Masseron}, {Masters}, {Mathur}, {McDermid}, {Merloni}, {Merrifield},
  {M{\'e}sz{\'a}ros}, {Miglio}, {Minniti}, {Minsley}, {Miyaji}, {Mohammad},
  {Mosser}, {Mueller}, {Muna}, {Mu{\~n}oz-Guti{\'e}rrez}, {Myers}, {Nadathur},
  {Nair}, {Nandra}, {do Nascimento}, {Nevin}, {Newman}, {Nidever}, {Nitschelm},
  {Noterdaeme}, {O'Connell}, {Olmstead}, {Oravetz}, {Oravetz}, {Osorio},
  {Pace}, {Padilla}, {Palanque-Delabrouille}, {Palicio}, {Pan}, {Pan},
  {Parker}, {Paviot}, {Peirani}, {Ram{\'r}ez}, {Penny}, {Percival},
  {Perez-Fournon}, {P{\'e}rez-R{\`a}fols}, {Petitjean}, {Pieri},
  {Pinsonneault}, {Poovelil}, {Povick}, {Prakash}, {Price-Whelan}, {Raddick},
  {Raichoor}, {Ray}, {Rembold}, {Rezaie}, {Riffel}, {Riffel}, {Rix}, {Robin},
  {Roman-Lopes}, {Rom{\'a}n-Z{\'u}{\~n}iga}, {Rose}, {Ross}, {Rossi},
  {Rowlands}, {Rubin}, {Salvato}, {S{\'a}nchez}, {S{\'a}nchez-Menguiano},
  {S{\'a}nchez-Gallego}, {Sayres}, {Schaefer}, {Schiavon}, {Schimoia},
  {Schlafly}, {Schlegel}, {Schneider}, {Schultheis}, {Schwope}, {Seo},
  {Serenelli}, {Shafieloo}, {Shamsi}, {Shao}, {Shen}, {Shetrone}, {Shirley},
  {Aguirre}, {Simon}, {Skrutskie}, {Slosar}, {Smethurst}, {Sobeck}, {Sodi},
  {Souto}, {Stark}, {Stassun}, {Steinmetz}, {Stello}, {Stermer},
  {Storchi-Bergmann}, {Streblyanska}, {Stringfellow}, {Stutz}, {Su{\'a}rez},
  {Sun}, {Taghizadeh-Popp}, {Talbot}, {Tayar}, {Thakar}, {Theriault}, {Thomas},
  {Thomas}, {Tinker}, {Tojeiro}, {Toledo}, {Tremonti}, {Troup}, {Tuttle},
  {Unda-Sanzana}, {Valentini}, {Vargas-Gonz{\'a}lez}, {Vargas-Maga{\~n}a},
  {V{\'a}zquez-Mata}, {Vivek}, {Wake}, {Wang}, {Weaver}, {Weijmans}, {Wild},
  {Wilson}, {Wilson}, {Wolthuis}, {Wood-Vasey}, {Yan}, {Yang}, {Y{\`e}che},
  {Zamora}, {Zarrouk}, {Zasowski}, {Zhang}, {Zhao}, {Zhao}, {Zheng}, {Zheng},
  {Zhu}, \& {Zou}}]{2020ApJS..249....3A}
{Ahumada}, R., {Prieto}, C.~A., {Almeida}, A., {et~al.} 2020, \apjs, 249, 3,
  \dodoi{10.3847/1538-4365/ab929e}

\bibitem[{{Anderson}(2019)}]{2019AndersonNi}
{Anderson}, J.~P. 2019, \aap, 628, A7, \dodoi{10.1051/0004-6361/201935027}

\bibitem[{{Anderson} {et~al.}(2014){Anderson}, {Gonz{\'a}lez-Gait{\'a}n},
  {Hamuy}, {Guti{\'e}rrez}, {Stritzinger}, {Olivares E.}, {Phillips},
  {Schulze}, {Antezana}, {Bolt}, {Campillay}, {Castell{\'o}n}, {Contreras}, {de
  Jaeger}, {Folatelli}, {F{\"o}rster}, {Freedman}, {Gonz{\'a}lez}, {Hsiao},
  {Krzemi{\'n}ski}, {Krisciunas}, {Maza}, {McCarthy}, {Morrell}, {Persson},
  {Roth}, {Salgado}, {Suntzeff}, \& {Thomas-Osip}}]{2014Anderson}
{Anderson}, J.~P., {Gonz{\'a}lez-Gait{\'a}n}, S., {Hamuy}, M., {et~al.} 2014,
  \apj, 786, 67, \dodoi{10.1088/0004-637X/786/1/67}

\bibitem[{{Andrews} {et~al.}(2019){Andrews}, {Sand}, {Valenti}, {Smith},
  {Dastidar}, {Sahu}, {Misra}, {Singh}, {Hiramatsu}, {Brown}, {Hosseinzadeh},
  {Wyatt}, {Vinko}, {Anupama}, {Arcavi}, {Ashall}, {Benetti}, {Berton},
  {Bostroem}, {Bulla}, {Burke}, {Chen}, {Chomiuk}, {Cikota}, {Congiu}, {Cseh},
  {Davis}, {Elias-Rosa}, {Faran}, {Fraser}, {Galbany}, {Gall}, {Gal-Yam},
  {Gangopadhyay}, {Gromadzki}, {Haislip}, {Howell}, {Hsiao}, {Inserra},
  {Kankare}, {Kuncarayakti}, {Kouprianov}, {Kumar}, {Li}, {Lin}, {Maguire},
  {Mazzali}, {McCully}, {Milne}, {Mo}, {Morrell}, {Nicholl}, {Ochner},
  {Olivares}, {Pastorello}, {Patat}, {Phillips}, {Pignata}, {Prentice},
  {Reguitti}, {Reichart}, {Rodr{\'\i}guez}, {Rui}, {Sanwal}, {S{\'a}rneczky},
  {Shahbandeh}, {Singh}, {Smartt}, {Strader}, {Stritzinger}, {Szak{\'a}ts},
  {Tartaglia}, {Wang}, {Wang}, {Wang}, {Wheeler}, {Xiang}, {Yaron}, {Young}, \&
  {Zhang}}]{2017gmrAndrews}
{Andrews}, J.~E., {Sand}, D.~J., {Valenti}, S., {et~al.} 2019, \apj, 885, 43,
  \dodoi{10.3847/1538-4357/ab43e3}

\bibitem[{{Angulo}(1999)}]{1999nacre}
{Angulo}, C. 1999, in American Institute of Physics Conference Series, Vol.
  495, Experimental Nuclear Physics in europe: Facing the next millennium,
  365--366, \dodoi{10.1063/1.1301821}

\bibitem[{{Arcavi}(2017)}]{HydrogenRich}
{Arcavi}, I. 2017, in Handbook of Supernovae, ed. A.~W. {Alsabti} \&
  P.~{Murdin}, 239, \dodoi{10.1007/978-3-319-21846-5\_39}

\bibitem[{{Arnett}(1980)}]{NiArnett}
{Arnett}, W.~D. 1980, \apj, 237, 541, \dodoi{10.1086/157898}

\bibitem[{{Arnett} \& {Fu}(1989)}]{ArnettAndFu}
{Arnett}, W.~D., \& {Fu}, A. 1989, \apj, 340, 396, \dodoi{10.1086/167402}

\bibitem[{{Asplund} {et~al.}(2006){Asplund}, {Grevesse}, \& {Jacques
  Sauval}}]{asplund}
{Asplund}, M., {Grevesse}, N., \& {Jacques Sauval}, A. 2006, \nphysa, 777, 1,
  \dodoi{10.1016/j.nuclphysa.2005.06.010}

\bibitem[{{Astropy Collaboration} {et~al.}(2018){Astropy Collaboration},
  {Price-Whelan}, {Sip{\H{o}}cz}, {G{\"u}nther}, {Lim}, {Crawford}, {Conseil},
  {Shupe}, {Craig}, {Dencheva}, {Ginsburg}, {VanderPlas}, {Bradley},
  {P{\'e}rez-Su{\'a}rez}, {de Val-Borro}, {Aldcroft}, {Cruz}, {Robitaille},
  {Tollerud}, {Ardelean}, {Babej}, {Bach}, {Bachetti}, {Bakanov}, {Bamford},
  {Barentsen}, {Barmby}, {Baumbach}, {Berry}, {Biscani}, {Boquien}, {Bostroem},
  {Bouma}, {Brammer}, {Bray}, {Breytenbach}, {Buddelmeijer}, {Burke},
  {Calderone}, {Cano Rodr{\'\i}guez}, {Cara}, {Cardoso}, {Cheedella}, {Copin},
  {Corrales}, {Crichton}, {D'Avella}, {Deil}, {Depagne}, {Dietrich}, {Donath},
  {Droettboom}, {Earl}, {Erben}, {Fabbro}, {Ferreira}, {Finethy}, {Fox},
  {Garrison}, {Gibbons}, {Goldstein}, {Gommers}, {Greco}, {Greenfield},
  {Groener}, {Grollier}, {Hagen}, {Hirst}, {Homeier}, {Horton}, {Hosseinzadeh},
  {Hu}, {Hunkeler}, {Ivezi{\'c}}, {Jain}, {Jenness}, {Kanarek}, {Kendrew},
  {Kern}, {Kerzendorf}, {Khvalko}, {King}, {Kirkby}, {Kulkarni}, {Kumar},
  {Lee}, {Lenz}, {Littlefair}, {Ma}, {Macleod}, {Mastropietro}, {McCully},
  {Montagnac}, {Morris}, {Mueller}, {Mumford}, {Muna}, {Murphy}, {Nelson},
  {Nguyen}, {Ninan}, {N{\"o}the}, {Ogaz}, {Oh}, {Parejko}, {Parley}, {Pascual},
  {Patil}, {Patil}, {Plunkett}, {Prochaska}, {Rastogi}, {Reddy Janga},
  {Sabater}, {Sakurikar}, {Seifert}, {Sherbert}, {Sherwood-Taylor}, {Shih},
  {Sick}, {Silbiger}, {Singanamalla}, {Singer}, {Sladen}, {Sooley},
  {Sornarajah}, {Streicher}, {Teuben}, {Thomas}, {Tremblay}, {Turner},
  {Terr{\'o}n}, {van Kerkwijk}, {de la Vega}, {Watkins}, {Weaver}, {Whitmore},
  {Woillez}, {Zabalza}, \& {Astropy Contributors}}]{2018Astropy}
{Astropy Collaboration}, {Price-Whelan}, A.~M., {Sip{\H{o}}cz}, B.~M., {et~al.}
  2018, \aj, 156, 123, \dodoi{10.3847/1538-3881/aabc4f}

\bibitem[{{Baklanov} {et~al.}(2005){Baklanov}, {Blinnikov}, \&
  {Pavlyuk}}]{Baklanov2005}
{Baklanov}, P.~V., {Blinnikov}, S.~I., \& {Pavlyuk}, N.~N. 2005, Astronomy
  Letters, 31, 429, \dodoi{10.1134/1.1958107}

\bibitem[{{Barbon} {et~al.}(1990){Barbon}, {Benetti}, {Cappellaro}, {Rosino},
  \& {Turatto}}]{1990A&A...237...79B}
{Barbon}, R., {Benetti}, S., {Cappellaro}, E., {Rosino}, L., \& {Turatto}, M.
  1990, \aap, 237, 79

\bibitem[{{Bellm} {et~al.}(2019){Bellm}, {Kulkarni}, {Graham}, {Dekany},
  {Smith}, {Riddle}, {Masci}, {Helou}, {Prince}, {Adams}, {Barbarino},
  {Barlow}, {Bauer}, {Beck}, {Belicki}, {Biswas}, {Blagorodnova}, {Bodewits},
  {Bolin}, {Brinnel}, {Brooke}, {Bue}, {Bulla}, {Burruss}, {Cenko}, {Chang},
  {Connolly}, {Coughlin}, {Cromer}, {Cunningham}, {De}, {Delacroix}, {Desai},
  {Duev}, {Eadie}, {Farnham}, {Feeney}, {Feindt}, {Flynn}, {Franckowiak},
  {Frederick}, {Fremling}, {Gal-Yam}, {Gezari}, {Giomi}, {Goldstein},
  {Golkhou}, {Goobar}, {Groom}, {Hacopians}, {Hale}, {Henning}, {Ho}, {Hover},
  {Howell}, {Hung}, {Huppenkothen}, {Imel}, {Ip}, {Ivezi{\'c}}, {Jackson},
  {Jones}, {Juric}, {Kasliwal}, {Kaspi}, {Kaye}, {Kelley}, {Kowalski},
  {Kramer}, {Kupfer}, {Landry}, {Laher}, {Lee}, {Lin}, {Lin}, {Lunnan},
  {Giomi}, {Mahabal}, {Mao}, {Miller}, {Monkewitz}, {Murphy}, {Ngeow},
  {Nordin}, {Nugent}, {Ofek}, {Patterson}, {Penprase}, {Porter}, {Rauch},
  {Rebbapragada}, {Reiley}, {Rigault}, {Rodriguez}, {van Roestel}, {Rusholme},
  {van Santen}, {Schulze}, {Shupe}, {Singer}, {Soumagnac}, {Stein}, {Surace},
  {Sollerman}, {Szkody}, {Taddia}, {Terek}, {Van Sistine}, {van Velzen},
  {Vestrand}, {Walters}, {Ward}, {Ye}, {Yu}, {Yan}, \& {Zolkower}}]{2019bellm}
{Bellm}, E.~C., {Kulkarni}, S.~R., {Graham}, M.~J., {et~al.} 2019, \pasp, 131,
  018002, \dodoi{10.1088/1538-3873/aaecbe}

\bibitem[{{Bessell} {et~al.}(1998){Bessell}, {Castelli}, \& {Plez}}]{bessell}
{Bessell}, M.~S., {Castelli}, F., \& {Plez}, B. 1998, \aap, 333, 231

\bibitem[{{Blinnikov} \& {Sorokina}(2004)}]{Blinnikov2004}
{Blinnikov}, S., \& {Sorokina}, E. 2004, \apss, 290, 13,
  \dodoi{10.1023/B:ASTR.0000022161.03559.42}

\bibitem[{{Blinnikov} {et~al.}(2006){Blinnikov}, {R{\"o}pke}, {Sorokina},
  {Gieseler}, {Reinecke}, {Travaglio}, {Hillebrandt}, \&
  {Stritzinger}}]{Blinnikov2006}
{Blinnikov}, S.~I., {R{\"o}pke}, F.~K., {Sorokina}, E.~I., {et~al.} 2006, \aap,
  453, 229, \dodoi{10.1051/0004-6361:20054594}

\bibitem[{{Blondin} \& {Tonry}(2007)}]{2007ApJ...666.1024B}
{Blondin}, S., \& {Tonry}, J.~L. 2007, \apj, 666, 1024, \dodoi{10.1086/520494}

\bibitem[{{Bose} \& {Kumar}(2014)}]{2014ApJ...782...98B}
{Bose}, S., \& {Kumar}, B. 2014, \apj, 782, 98,
  \dodoi{10.1088/0004-637X/782/2/98}

\bibitem[{{Bottinelli} {et~al.}(1984){Bottinelli}, {Gouguenheim}, {Paturel}, \&
  {de Vaucouleurs}}]{1984A&AS...56..381B}
{Bottinelli}, L., {Gouguenheim}, L., {Paturel}, G., \& {de Vaucouleurs}, G.
  1984, \aaps, 56, 381

\bibitem[{Branch \& Wheeler(2017)}]{Branch2017TypeIIP}
Branch, D., \& Wheeler, J.~C. 2017, Type IIP Supernovae (Berlin, Heidelberg:
  Springer Berlin Heidelberg), 245--265, \dodoi{10.1007/978-3-662-55054-0_12}

\bibitem[{{Brown} {et~al.}(2009){Brown}, {Holland}, {Immler}, {Milne},
  {Roming}, {Gehrels}, {Nousek}, {Panagia}, {Still}, \& {Vanden
  Berk}}]{brownAJ....137.4517B}
{Brown}, P.~J., {Holland}, S.~T., {Immler}, S., {et~al.} 2009, \aj, 137, 4517,
  \dodoi{10.1088/0004-6256/137/5/4517}

\bibitem[{{Bruch} {et~al.}(2021){Bruch}, {Gal-Yam}, {Schulze}, {Yaron}, {Yang},
  {Soumagnac}, {Rigault}, {Strotjohann}, {Ofek}, {Sollerman}, {Masci},
  {Barbarino}, {Ho}, {Fremling}, {Perley}, {Nordin}, {Cenko}, {Adams},
  {Adreoni}, {Bellm}, {Blagorodnova}, {Bulla}, {Burdge}, {De}, {Dhawan},
  {Drake}, {Duev}, {Dugas}, {Graham}, {Graham}, {Irani}, {Jencson},
  {Karamehmetoglu}, {Kasliwal}, {Kim}, {Kulkarni}, {Kupfer}, {Liang},
  {Mahabal}, {Miller}, {Prince}, {Riddle}, {Sharma}, {Smith}, {Taddia},
  {Taggart}, {Walters}, \& {Yan}}]{2021bruch}
{Bruch}, R.~J., {Gal-Yam}, A., {Schulze}, S., {et~al.} 2021, \apj, 912, 46,
  \dodoi{10.3847/1538-4357/abef05}

\bibitem[{{Bullivant} {et~al.}(2018{\natexlab{a}}){Bullivant}, {Smith},
  {Williams}, {Mauerhan}, {Andrews}, {Fong}, {Bilinski}, {Kilpatrick}, {Milne},
  {Fox}, {Cenko}, {Filippenko}, {Zheng}, {Kelly}, \& {Clubb}}]{2013fsBullivant}
{Bullivant}, C., {Smith}, N., {Williams}, G.~G., {et~al.} 2018{\natexlab{a}},
  \mnras, 476, 1497, \dodoi{10.1093/mnras/sty045}

\bibitem[{{Bullivant} {et~al.}(2018{\natexlab{b}}){Bullivant}, {Smith},
  {Williams}, {Mauerhan}, {Andrews}, {Fong}, {Bilinski}, {Kilpatrick}, {Milne},
  {Fox}, {Cenko}, {Filippenko}, {Zheng}, {Kelly}, \& {Clubb}}]{2018bullivant}
---. 2018{\natexlab{b}}, \mnras, 476, 1497, \dodoi{10.1093/mnras/sty045}

\bibitem[{{Burrows} \& {Vartanyan}(2021)}]{BurrowsINtro}
{Burrows}, A., \& {Vartanyan}, D. 2021, \nat, 589, 29,
  \dodoi{10.1038/s41586-020-03059-w}

\bibitem[{{Cardelli} {et~al.}(1989){Cardelli}, {Clayton}, \&
  {Mathis}}]{Cardelli}
{Cardelli}, J.~A., {Clayton}, G.~C., \& {Mathis}, J.~S. 1989, \apj, 345, 245,
  \dodoi{10.1086/167900}

\bibitem[{{Chugai}(2020)}]{2020chugai}
{Chugai}, N.~N. 2020, \mnras, 494, L86, \dodoi{10.1093/mnrasl/slaa042}

\bibitem[{{Chugai} {et~al.}(2007){Chugai}, {Chevalier}, \&
  {Utrobin}}]{2007Chugai}
{Chugai}, N.~N., {Chevalier}, R.~A., \& {Utrobin}, V.~P. 2007, \apj, 662, 1136,
  \dodoi{10.1086/518160}

\bibitem[{{Curtis} {et~al.}(2020){Curtis}, {Wolfe}, {Fr{\"o}hlich}, {Miller},
  {Wollaeger}, \& {Ebinger}}]{NSP1}
{Curtis}, S., {Wolfe}, N., {Fr{\"o}hlich}, C., {et~al.} 2020, arXiv e-prints,
  arXiv:2008.05498.
\newblock \doarXiv{2008.05498}

\bibitem[{{Cyburt} {et~al.}(2010){Cyburt}, {Amthor}, {Ferguson}, {Meisel},
  {Smith}, {Warren}, {Heger}, {Hoffman}, {Rauscher}, {Sakharuk}, {Schatz},
  {Thielemann}, \& {Wiescher}}]{2010JINA}
{Cyburt}, R.~H., {Amthor}, A.~M., {Ferguson}, R., {et~al.} 2010, \apjs, 189,
  240, \dodoi{10.1088/0067-0049/189/1/240}

\bibitem[{{Decin} {et~al.}(2006){Decin}, {Hony}, {de Koter}, {Justtanont},
  {Tielens}, \& {Waters}}]{DecinHighMass}
{Decin}, L., {Hony}, S., {de Koter}, A., {et~al.} 2006, \aap, 456, 549,
  \dodoi{10.1051/0004-6361:20065230}

\bibitem[{{Dessart} {et~al.}(2013){Dessart}, {Hillier}, {Waldman}, \&
  {Livne}}]{2013bdessart}
{Dessart}, L., {Hillier}, D.~J., {Waldman}, R., \& {Livne}, E. 2013, \mnras,
  433, 1745, \dodoi{10.1093/mnras/stt861}

\bibitem[{{Dessart} {et~al.}(2010){Dessart}, {Livne}, \&
  {Waldman}}]{dessart2010}
{Dessart}, L., {Livne}, E., \& {Waldman}, R. 2010, \mnras, 408, 827,
  \dodoi{10.1111/j.1365-2966.2010.17190.x}

\bibitem[{{Dom{\'\i}nguez} {et~al.}(2013){Dom{\'\i}nguez}, {Siana}, {Henry},
  {Scarlata}, {Bedregal}, {Malkan}, {Atek}, {Ross}, {Colbert}, {Teplitz},
  {Rafelski}, {McCarthy}, {Bunker}, {Hathi}, {Dressler}, {Martin}, \&
  {Masters}}]{2013ApJ...763..145D}
{Dom{\'\i}nguez}, A., {Siana}, B., {Henry}, A.~L., {et~al.} 2013, \apj, 763,
  145, \dodoi{10.1088/0004-637X/763/2/145}

\bibitem[{{Dong} {et~al.}(2021){Dong}, {Valenti}, {Bostroem}, {Sand},
  {Andrews}, {Galbany}, {Jha}, {Eweis}, {Kwok}, {Hsiao}, {Davis}, {Brown},
  {Kuncarayakti}, {Maeda}, {Rho}, {Amaro}, {Anderson}, {Arcavi}, {Burke},
  {Dastidar}, {Folatelli}, {Haislip}, {Hiramatsu}, {Hosseinzadeh}, {Howell},
  {Jencson}, {Kouprianov}, {Lundquist}, {Lyman}, {McCully}, {Misra},
  {Reichart}, {S{\'a}nchez}, {Smith}, {Wang}, {Wang}, \& {Wyatt}}]{2018cufDong}
{Dong}, Y., {Valenti}, S., {Bostroem}, K.~A., {et~al.} 2021, \apj, 906, 56,
  \dodoi{10.3847/1538-4357/abc417}

\bibitem[{{Eldridge} {et~al.}(2018){Eldridge}, {Xiao}, {Stanway}, {Rodrigues},
  \& {Guo}}]{2018Eldridge}
{Eldridge}, J.~J., {Xiao}, L., {Stanway}, E.~R., {Rodrigues}, N., \& {Guo},
  N.~Y. 2018, \pasa, 35, e049, \dodoi{10.1017/pasa.2018.47}

\bibitem[{{Elmhamdi} {et~al.}(2003{\natexlab{a}}){Elmhamdi}, {Chugai}, \&
  {Danziger}}]{1999emElmhamdi}
{Elmhamdi}, A., {Chugai}, N.~N., \& {Danziger}, I.~J. 2003{\natexlab{a}}, \aap,
  404, 1077, \dodoi{10.1051/0004-6361:20030522}

\bibitem[{{Elmhamdi} {et~al.}(2003{\natexlab{b}}){Elmhamdi}, {Danziger},
  {Chugai}, {Pastorello}, {Turatto}, {Cappellaro}, {Altavilla}, {Benetti},
  {Patat}, \& {Salvo}}]{ElmhamdiIR2003MNRAS.338..939E}
{Elmhamdi}, A., {Danziger}, I.~J., {Chugai}, N., {et~al.} 2003{\natexlab{b}},
  \mnras, 338, 939, \dodoi{10.1046/j.1365-8711.2003.06150.x}

\bibitem[{{Farmer} {et~al.}(2016){Farmer}, {Fields}, {Petermann}, {Dessart},
  {Cantiello}, {Paxton}, \& {Timmes}}]{Farmer}
{Farmer}, R., {Fields}, C.~E., {Petermann}, I., {et~al.} 2016, \apjs, 227, 22,
  \dodoi{10.3847/1538-4365/227/2/22}

\bibitem[{{Filippenko}(1997)}]{1997AVFillippenko}
{Filippenko}, A.~V. 1997, \araa, 35, 309,
  \dodoi{10.1146/annurev.astro.35.1.309}

\bibitem[{{Foreman-Mackey} {et~al.}(2013){Foreman-Mackey}, {Hogg}, {Lang}, \&
  {Goodman}}]{2013emcee}
{Foreman-Mackey}, D., {Hogg}, D.~W., {Lang}, D., \& {Goodman}, J. 2013, \pasp,
  125, 306, \dodoi{10.1086/670067}

\bibitem[{{F{\"o}rster} {et~al.}(2018){F{\"o}rster}, {Moriya}, {Maureira},
  {Anderson}, {Blinnikov}, {Bufano}, {Cabrera-Vives}, {Clocchiatti}, {de
  Jaeger}, {Est{\'e}vez}, {Galbany}, {Gonz{\'a}lez-Gait{\'a}n}, {Gr{\"a}fener},
  {Hamuy}, {Hsiao}, {Huentelemu}, {Huijse}, {Kuncarayakti}, {Mart{\'\i}nez},
  {Medina}, {Olivares E.}, {Pignata}, {Razza}, {Reyes}, {San Mart{\'\i}n},
  {Smith}, {Vera}, {Vivas}, {de Ugarte Postigo}, {Yoon}, {Ashall}, {Fraser},
  {Gal-Yam}, {Kankare}, {Le Guillou}, {Mazzali}, {Walton}, \&
  {Young}}]{2018Forster}
{F{\"o}rster}, F., {Moriya}, T.~J., {Maureira}, J.~C., {et~al.} 2018, Nature
  Astronomy, 2, 808, \dodoi{10.1038/s41550-018-0563-4}

\bibitem[{{Fransson} \& {Chevalier}(1989)}]{1989fransson}
{Fransson}, C., \& {Chevalier}, R.~A. 1989, \apj, 343, 323,
  \dodoi{10.1086/167707}

\bibitem[{{Fraser} {et~al.}(2012){Fraser}, {Maund}, {Smartt}, {Botticella},
  {Dall'Ora}, {Inserra}, {Tomasella}, {Benetti}, {Ciroi}, {Eldridge}, {Ergon},
  {Kotak}, {Mattila}, {Ochner}, {Pastorello}, {Reilly}, {Sollerman},
  {Stephens}, {Taddia}, \& {Valenti}}]{2012awFraser}
{Fraser}, M., {Maund}, J.~R., {Smartt}, S.~J., {et~al.} 2012, \apjl, 759, L13,
  \dodoi{10.1088/2041-8205/759/1/L13}

\bibitem[{{Gehrels} {et~al.}(2004){Gehrels}, {Chincarini}, {Giommi}, {Mason},
  {Nousek}, {Wells}, {White}, {Barthelmy}, {Burrows}, {Cominsky}, {Hurley},
  {Marshall}, {M{\'e}sz{\'a}ros}, {Roming}, {Angelini}, {Barbier}, {Belloni},
  {Campana}, {Caraveo}, {Chester}, {Citterio}, {Cline}, {Cropper}, {Cummings},
  {Dean}, {Feigelson}, {Fenimore}, {Frail}, {Fruchter}, {Garmire}, {Gendreau},
  {Ghisellini}, {Greiner}, {Hill}, {Hunsberger}, {Krimm}, {Kulkarni}, {Kumar},
  {Lebrun}, {Lloyd-Ronning}, {Markwardt}, {Mattson}, {Mushotzky}, {Norris},
  {Osborne}, {Paczynski}, {Palmer}, {Park}, {Parsons}, {Paul}, {Rees},
  {Reynolds}, {Rhoads}, {Sasseen}, {Schaefer}, {Short}, {Smale}, {Smith},
  {Stella}, {Tagliaferri}, {Takahashi}, {Tashiro}, {Townsley}, {Tueller},
  {Turner}, {Vietri}, {Voges}, {Ward}, {Willingale}, {Zerbi}, \&
  {Zhang}}]{2004gehrels}
{Gehrels}, N., {Chincarini}, G., {Giommi}, P., {et~al.} 2004, \apj, 611, 1005,
  \dodoi{10.1086/422091}

\bibitem[{{Glebbeek} {et~al.}(2009){Glebbeek}, {Gaburov}, {de Mink}, {Pols}, \&
  {Portegies Zwart}}]{dutch1}
{Glebbeek}, E., {Gaburov}, E., {de Mink}, S.~E., {Pols}, O.~R., \& {Portegies
  Zwart}, S.~F. 2009, \aap, 497, 255, \dodoi{10.1051/0004-6361/200810425}

\bibitem[{{Goldberg} {et~al.}(2019){Goldberg}, {Bildsten}, \&
  {Paxton}}]{2019Goldberg}
{Goldberg}, J.~A., {Bildsten}, L., \& {Paxton}, B. 2019, \apj, 879, 3,
  \dodoi{10.3847/1538-4357/ab22b6}

\bibitem[{{Guillochon} {et~al.}(2017){Guillochon}, {Parrent}, {Kelley}, \&
  {Margutti}}]{2017guillochon}
{Guillochon}, J., {Parrent}, J., {Kelley}, L.~Z., \& {Margutti}, R. 2017, \apj,
  835, 64, \dodoi{10.3847/1538-4357/835/1/64}

\bibitem[{Guti{\'{e}}rrez {et~al.}(2014)Guti{\'{e}}rrez, Anderson, Hamuy,
  Gonz{\'{a}}lez-Gait{\'{a}}n, Folatelli, Morrell, Stritzinger, Phillips,
  McCarthy, Suntzeff, \& Thomas-Osip}]{Guti_rrez_2014}
Guti{\'{e}}rrez, C.~P., Anderson, J.~P., Hamuy, M., {et~al.} 2014, The
  Astrophysical Journal, 786, L15, \dodoi{10.1088/2041-8205/786/2/l15}

\bibitem[{{Guti{\'e}rrez} {et~al.}(2017{\natexlab{a}}){Guti{\'e}rrez},
  {Anderson}, {Hamuy}, {Gonz{\'a}lez-Gaitan}, {Galbany}, {Dessart},
  {Stritzinger}, {Phillips}, {Morrell}, \& {Folatelli}}]{2017Claudia}
{Guti{\'e}rrez}, C.~P., {Anderson}, J.~P., {Hamuy}, M., {et~al.}
  2017{\natexlab{a}}, \apj, 850, 90, \dodoi{10.3847/1538-4357/aa8f42}

\bibitem[{{Guti{\'e}rrez} {et~al.}(2017{\natexlab{b}}){Guti{\'e}rrez},
  {Anderson}, {Hamuy}, {Morrell}, {Gonz{\'a}lez-Gaitan}, {Stritzinger},
  {Phillips}, {Galbany}, {Folatelli}, {Dessart}, {Contreras}, {Della Valle},
  {Freedman}, {Hsiao}, {Krisciunas}, {Madore}, {Maza}, {Suntzeff}, {Prieto},
  {Gonz{\'a}lez}, {Cappellaro}, {Navarrete}, {Pizzella}, {Ruiz}, {Smith}, \&
  {Turatto}}]{2017gutierrez}
---. 2017{\natexlab{b}}, \apj, 850, 89, \dodoi{10.3847/1538-4357/aa8f52}

\bibitem[{Gutiérrez {et~al.}(2018)Gutiérrez, Anderson, Sullivan, Dessart,
  González-Gaitan, Galbany, Dimitriadis, Arcavi, Bufano, Chen, Dennefeld,
  Gromadzki, Haislip, Hosseinzadeh, Howell, Inserra, Kankare, Leloudas,
  Maguire, McCully, Morrell, Olivares E, Pignata, Reichart, Reynolds, Smartt,
  Sollerman, Taddia, Takáts, Terreran, Valenti, \& Young}]{2018guiterez}
Gutiérrez, C.~P., Anderson, J.~P., Sullivan, M., {et~al.} 2018, Monthly
  Notices of the Royal Astronomical Society, 479, 3232,
  \dodoi{10.1093/mnras/sty1581}

\bibitem[{{Hamuy}(2003)}]{hamuy2003ApJ...582..905H}
{Hamuy}, M. 2003, \apj, 582, 905, \dodoi{10.1086/344689}

\bibitem[{{Haynie} \& {Piro}(2021)}]{PS1-13arp}
{Haynie}, A., \& {Piro}, A.~L. 2021, \apj, 910, 128,
  \dodoi{10.3847/1538-4357/abe938}

\bibitem[{{Henyey} {et~al.}(1965){Henyey}, {Vardya}, \&
  {Bodenheimer}}]{henyey1965}
{Henyey}, L., {Vardya}, M.~S., \& {Bodenheimer}, P. 1965, \apj, 142, 841,
  \dodoi{10.1086/148357}

\bibitem[{{Hiramatsu} {et~al.}(2021{\natexlab{a}}){Hiramatsu}, {Howell},
  {Moriya}, {Goldberg}, {Hosseinzadeh}, {Arcavi}, {Anderson}, {Guti{\'e}rrez},
  {Burke}, {McCully}, {Valenti}, {Galbany}, {Fang}, {Maeda}, {Folatelli},
  {Hsiao}, {Morrell}, {Phillips}, {Stritzinger}, {Suntzeff}, {Gromadzki},
  {Maguire}, {M{\"u}ller-Bravo}, \& {Young}}]{2021Hiramatsu}
{Hiramatsu}, D., {Howell}, D.~A., {Moriya}, T.~J., {et~al.} 2021{\natexlab{a}},
  \apj, 913, 55, \dodoi{10.3847/1538-4357/abf6d6}

\bibitem[{{Hiramatsu} {et~al.}(2021{\natexlab{b}}){Hiramatsu}, {Howell}, {Van
  Dyk}, {Goldberg}, {Maeda}, {Moriya}, {Tominaga}, {Nomoto}, {Hosseinzadeh},
  {Arcavi}, {McCully}, {Burke}, {Bostroem}, {Valenti}, {Dong}, {Brown},
  {Andrews}, {Bilinski}, {Williams}, {Smith}, {Smith}, {Sand}, {Anand}, {Xu},
  {Filippenko}, {Bersten}, {Folatelli}, {Kelly}, {Noguchi}, \&
  {Itagaki}}]{2018zdHiramatsu}
{Hiramatsu}, D., {Howell}, D.~A., {Van Dyk}, S.~D., {et~al.}
  2021{\natexlab{b}}, Nature Astronomy, 5, 903,
  \dodoi{10.1038/s41550-021-01384-2}

\bibitem[{{Huang} {et~al.}(2018){Huang}, {Wang}, {Hosseinzadeh}, {Brown}, {Mo},
  {Zhang}, {Zhang}, {Zhang}, {Howell}, {Arcavi}, {McCully}, {Valenti}, {Rui},
  {Song}, {Xiang}, {Li}, {Lin}, \& {Wang}}]{2016XHuang}
{Huang}, F., {Wang}, X.~F., {Hosseinzadeh}, G., {et~al.} 2018, \mnras, 475,
  3959, \dodoi{10.1093/mnras/sty066}

\bibitem[{Hunter(2007)}]{2007matplotlib}
Hunter, J.~D. 2007, Computing in Science \& Engineering, 9, 90,
  \dodoi{10.1109/MCSE.2007.55}

\bibitem[{{Jacobson-Gal{\'a}n} {et~al.}(2021){Jacobson-Gal{\'a}n}, {Dessart},
  {Jones}, {Margutti}, {Coppejans}, {Dimitriadis}, {Foley}, {Kilpatrick},
  {Matthews}, {Rest}, {Terreran}, {Aleo}, {Auchettl}, {Blanchard}, {Coulter},
  {Davis}, {de Boer}, {DeMarchi}, {Drout}, {Earl}, {Gagliano}, {Gall},
  {Hjorth}, {Huber}, {Ibik}, {Milisavljevic}, {Pan}, {Rest}, {Ridden-Harper},
  {Rojas-Bravo}, {Siebert}, {Smith}, {Taggart}, {Tinyanont}, {Wang}, \&
  {Zenati}}]{2021Jacobson}
{Jacobson-Gal{\'a}n}, W., {Dessart}, L., {Jones}, D., {et~al.} 2021, arXiv
  e-prints, arXiv:2109.12136.
\newblock \doarXiv{2109.12136}

\bibitem[{{Jacobson-Gal{\'a}n} {et~al.}(2022){Jacobson-Gal{\'a}n}, {Dessart},
  {Jones}, {Margutti}, {Coppejans}, {Dimitriadis}, {Foley}, {Kilpatrick},
  {Matthews}, {Rest}, {Terreran}, {Aleo}, {Auchettl}, {Blanchard}, {Coulter},
  {Davis}, {de Boer}, {DeMarchi}, {Drout}, {Earl}, {Gagliano}, {Gall},
  {Hjorth}, {Huber}, {Ibik}, {Milisavljevic}, {Pan}, {Rest}, {Ridden-Harper},
  {Rojas-Bravo}, {Siebert}, {Smith}, {Taggart}, {Tinyanont}, {Wang}, \&
  {Zenati}}]{2020tlfJacobson}
{Jacobson-Gal{\'a}n}, W.~V., {Dessart}, L., {Jones}, D.~O., {et~al.} 2022,
  \apj, 924, 15, \dodoi{10.3847/1538-4357/ac3f3a}

\bibitem[{{Janka}(2012)}]{2012Janka}
{Janka}, H.-T. 2012, Annual Review of Nuclear and Particle Science, 62, 407,
  \dodoi{10.1146/annurev-nucl-102711-094901}

\bibitem[{{Jerkstrand} {et~al.}(2014){Jerkstrand}, {Smartt}, {Fraser},
  {Fransson}, {Sollerman}, {Taddia}, \& {Kotak}}]{2014jerkstrand}
{Jerkstrand}, A., {Smartt}, S.~J., {Fraser}, M., {et~al.} 2014, \mnras, 439,
  3694, \dodoi{10.1093/mnras/stu221}

\bibitem[{{Jerkstrand} {et~al.}(2015{\natexlab{a}}){Jerkstrand}, {Timmes},
  {Magkotsios}, {Sim}, {Fransson}, {Spyromilio}, {M{\"u}ller}, {Heger},
  {Sollerman}, \& {Smartt}}]{2015NiFe}
{Jerkstrand}, A., {Timmes}, F.~X., {Magkotsios}, G., {et~al.}
  2015{\natexlab{a}}, \apj, 807, 110, \dodoi{10.1088/0004-637X/807/1/110}

\bibitem[{{Jerkstrand} {et~al.}(2015{\natexlab{b}}){Jerkstrand}, {Smartt},
  {Sollerman}, {Inserra}, {Fraser}, {Spyromilio}, {Fransson}, {Chen},
  {Barbarino}, {Dall'Ora}, {Botticella}, {Della Valle}, {Gal-Yam}, {Valenti},
  {Maguire}, {Mazzali}, \& {Tomasella}}]{2015FeNiMethodology}
{Jerkstrand}, A., {Smartt}, S.~J., {Sollerman}, J., {et~al.}
  2015{\natexlab{b}}, \mnras, 448, 2482, \dodoi{10.1093/mnras/stv087}

\bibitem[{{Jester} {et~al.}(2005){Jester}, {Schneider}, {Richards}, {Green},
  {Schmidt}, {Hall}, {Strauss}, {Vanden Berk}, {Stoughton}, {Gunn},
  {Brinkmann}, {Kent}, {Smith}, {Tucker}, \& {Yanny}}]{Jester2005}
{Jester}, S., {Schneider}, D.~P., {Richards}, G.~T., {et~al.} 2005, \aj, 130,
  873, \dodoi{10.1086/432466}

\bibitem[{{Kerzendorf} \& {Sim}(2014)}]{2014TARDIS}
{Kerzendorf}, W.~E., \& {Sim}, S.~A. 2014, \mnras, 440, 387,
  \dodoi{10.1093/mnras/stu055}

\bibitem[{{Maguire} {et~al.}(2012){Maguire}, {Jerkstrand}, {Smartt},
  {Fransson}, {Pastorello}, {Benetti}, {Valenti}, {Bufano}, \&
  {Leloudas}}]{maguire2012}
{Maguire}, K., {Jerkstrand}, A., {Smartt}, S.~J., {et~al.} 2012, \mnras, 420,
  3451, \dodoi{10.1111/j.1365-2966.2011.20276.x}

\bibitem[{{Maund} {et~al.}(2013){Maund}, {Fraser}, {Smartt}, {Botticella},
  {Barbarino}, {Childress}, {Gal-Yam}, {Inserra}, {Pignata}, {Reichart},
  {Schmidt}, {Sollerman}, {Taddia}, {Tomasella}, {Valenti}, \&
  {Yaron}}]{2012ecMaund}
{Maund}, J.~R., {Fraser}, M., {Smartt}, S.~J., {et~al.} 2013, \mnras, 431,
  L102, \dodoi{10.1093/mnrasl/slt017}

\bibitem[{{Mauron} \& {Josselin}(2011)}]{mauron}
{Mauron}, N., \& {Josselin}, E. 2011, \aap, 526, A156,
  \dodoi{10.1051/0004-6361/201013993}

\bibitem[{{Minkowski}(1941)}]{minkwoski1941}
{Minkowski}, R. 1941, \pasp, 53, 224, \dodoi{10.1086/125315}

\bibitem[{{Moriya} {et~al.}(2011){Moriya}, {Tominaga}, {Blinnikov}, {Baklanov},
  \& {Sorokina}}]{2011Moriya}
{Moriya}, T., {Tominaga}, N., {Blinnikov}, S.~I., {Baklanov}, P.~V., \&
  {Sorokina}, E.~I. 2011, \mnras, 415, 199,
  \dodoi{10.1111/j.1365-2966.2011.18689.x}

\bibitem[{{Morozova} {et~al.}(2018){Morozova}, {Piro}, \&
  {Valenti}}]{2018Morozova}
{Morozova}, V., {Piro}, A.~L., \& {Valenti}, S. 2018, \apj, 858, 15,
  \dodoi{10.3847/1538-4357/aab9a6}

\bibitem[{{Nagao} {et~al.}(2020){Nagao}, {Maeda}, \& {Ouchi}}]{discCSM}
{Nagao}, T., {Maeda}, K., \& {Ouchi}, R. 2020, \mnras, 497, 5395,
  \dodoi{10.1093/mnras/staa2360}

\bibitem[{{Nagy} \& {Vink{\'o}}(2016)}]{NagyandVinko}
{Nagy}, A.~P., \& {Vink{\'o}}, J. 2016, \aap, 589, A53,
  \dodoi{10.1051/0004-6361/201527931}

\bibitem[{{Nicholl}(2018)}]{SuperBol}
{Nicholl}, M. 2018, Research Notes of the American Astronomical Society, 2,
  230, \dodoi{10.3847/2515-5172/aaf799}

\bibitem[{{Nordin} {et~al.}(2020){Nordin}, {Brinnel}, {Giomi}, {Santen},
  {Gal-Yam}, {Yaron}, \& {Schulze}}]{2020TNSTR1248....1N}
{Nordin}, J., {Brinnel}, V., {Giomi}, M., {et~al.} 2020, Transient Name Server
  Discovery Report, 2020-1248, 1

\bibitem[{{Nugis} \& {Lamers}(2000)}]{nugis1}
{Nugis}, T., \& {Lamers}, H.~J.~G.~L.~M. 2000, \aap, 360, 227

\bibitem[{{Omar} {et~al.}(2019){Omar}, {Kumar}, {Krishna Reddy}, {Pant}, \&
  {Mahto}}]{2019DOTOmar}
{Omar}, A., {Kumar}, T.~S., {Krishna Reddy}, B., {Pant}, J., \& {Mahto}, M.
  2019, arXiv e-prints, arXiv:1902.05857.
\newblock \doarXiv{1902.05857}

\bibitem[{{Patat} {et~al.}(2001){Patat}, {Cappellaro}, {Danziger}, {Mazzali},
  {Sollerman}, {Augusteijn}, {Brewer}, {Doublier}, {Gonzalez}, {Hainaut},
  {Lidman}, {Leibundgut}, {Nomoto}, {Nakamura}, {Spyromilio}, {Rizzi},
  {Turatto}, {Walsh}, {Galama}, {van Paradijs}, {Kouveliotou}, {Vreeswijk},
  {Frontera}, {Masetti}, {Palazzi}, \& {Pian}}]{patat2001ApJ...555..900P}
{Patat}, F., {Cappellaro}, E., {Danziger}, J., {et~al.} 2001, \apj, 555, 900,
  \dodoi{10.1086/321526}

\bibitem[{{Paxton} {et~al.}(2011){Paxton}, {Bildsten}, {Dotter}, {Herwig},
  {Lesaffre}, \& {Timmes}}]{Paxton2011}
{Paxton}, B., {Bildsten}, L., {Dotter}, A., {et~al.} 2011, \apjs, 192, 3,
  \dodoi{10.1088/0067-0049/192/1/3}

\bibitem[{{Paxton} {et~al.}(2013){Paxton}, {Cantiello}, {Arras}, {Bildsten},
  {Brown}, {Dotter}, {Mankovich}, {Montgomery}, {Stello}, {Timmes}, \&
  {Townsend}}]{Paxton2013}
{Paxton}, B., {Cantiello}, M., {Arras}, P., {et~al.} 2013, \apjs, 208, 4,
  \dodoi{10.1088/0067-0049/208/1/4}

\bibitem[{{Paxton} {et~al.}(2015){Paxton}, {Marchant}, {Schwab}, {Bauer},
  {Bildsten}, {Cantiello}, {Dessart}, {Farmer}, {Hu}, {Langer}, {Townsend},
  {Townsley}, \& {Timmes}}]{Paxton2015}
{Paxton}, B., {Marchant}, P., {Schwab}, J., {et~al.} 2015, \apjs, 220, 15,
  \dodoi{10.1088/0067-0049/220/1/15}

\bibitem[{{Paxton} {et~al.}(2018){Paxton}, {Schwab}, {Bauer}, {Bildsten},
  {Blinnikov}, {Duffell}, {Farmer}, {Goldberg}, {Marchant}, {Sorokina},
  {Thoul}, {Townsend}, \& {Timmes}}]{Paxton2018}
{Paxton}, B., {Schwab}, J., {Bauer}, E.~B., {et~al.} 2018, \apjs, 234, 34,
  \dodoi{10.3847/1538-4365/aaa5a8}

\bibitem[{{Paxton} {et~al.}(2019){Paxton}, {Smolec}, {Schwab}, {Gautschy},
  {Bildsten}, {Cantiello}, {Dotter}, {Farmer}, {Goldberg}, {Jermyn}, {Kanbur},
  {Marchant}, {Thoul}, {Townsend}, {Wolf}, {Zhang}, \& {Timmes}}]{Paxton2019}
{Paxton}, B., {Smolec}, R., {Schwab}, J., {et~al.} 2019, \apjs, 243, 10,
  \dodoi{10.3847/1538-4365/ab2241}

\bibitem[{{Perley} {et~al.}(2020){Perley}, {Barbarino}, {Sollerman},
  {Schweyer}, {Schulze}, \& {Yang}}]{2020TNSCR1259....1P}
{Perley}, D., {Barbarino}, C., {Sollerman}, J., {et~al.} 2020, Transient Name
  Server Classification Report, 2020-1259, 1

\bibitem[{{Pettini} \& {Pagel}(2004)}]{o3n2}
{Pettini}, M., \& {Pagel}, B. E.~J. 2004, \mnras, 348, L59,
  \dodoi{10.1111/j.1365-2966.2004.07591.x}

\bibitem[{{Poole} {et~al.}(2008){Poole}, {Breeveld}, {Page}, {Landsman},
  {Holland}, {Roming}, {Kuin}, {Brown}, {Gronwall}, {Hunsberger}, {Koch},
  {Mason}, {Schady}, {vanden Berk}, {Blustin}, {Boyd}, {Broos}, {Carter},
  {Chester}, {Cucchiara}, {Hancock}, {Huckle}, {Immler}, {Ivanushkina},
  {Kennedy}, {Marshall}, {Morgan}, {Pandey}, {de Pasquale}, {Smith}, \&
  {Still}}]{poole2008MNRAS.383..627P}
{Poole}, T.~S., {Breeveld}, A.~A., {Page}, M.~J., {et~al.} 2008, \mnras, 383,
  627, \dodoi{10.1111/j.1365-2966.2007.12563.x}

\bibitem[{{Poznanski} {et~al.}(2012){Poznanski}, {Prochaska}, \&
  {Bloom}}]{2012MNRAS.426.1465P}
{Poznanski}, D., {Prochaska}, J.~X., \& {Bloom}, J.~S. 2012, \mnras, 426, 1465,
  \dodoi{10.1111/j.1365-2966.2012.21796.x}

\bibitem[{{Prabhu}(2014)}]{2014Prabhu}
{Prabhu}, T.~P. 2014, Proceedings of the Indian National Science Academy Part
  A, 80, 887, \dodoi{10.16943/ptinsa/2014/v80i4/55174}

\bibitem[{Quimby {et~al.}(2007)Quimby, Wheeler, Hoflich, Akerlof, Brown, \&
  Rykoff}]{Quimby_2007}
Quimby, R.~M., Wheeler, J.~C., Hoflich, P., {et~al.} 2007, The Astrophysical
  Journal, 666, 1093, \dodoi{10.1086/520532}

\bibitem[{{Rodr{\'i}guez} {et~al.}(2014){Rodr{\'i}guez}, {Clocchiatti}, \&
  {Hamuy}}]{2014AJ....148..107R}
{Rodr{\'i}guez}, {\'O}., {Clocchiatti}, A., \& {Hamuy}, M. 2014, \aj, 148, 107,
  \dodoi{10.1088/0004-6256/148/6/107}

\bibitem[{{Rodr{\'\i}guez} {et~al.}(2020){Rodr{\'\i}guez}, {Pignata},
  {Anderson}, {Moriya}, {Clocchiatti}, {F{\"o}rster}, {Prieto}, {Phillips},
  {Burns}, {Contreras}, {Folatelli}, {Guti{\'e}rrez}, {Hamuy}, {Morrell},
  {Stritzinger}, {Suntzeff}, {Benetti}, {Cappellaro}, {Elias-Rosa},
  {Pastorello}, {Turatto}, {Maza}, {Antezana}, {Cartier}, {Gonz{\'a}lez},
  {Haislip}, {Kouprianov}, {L{\'o}pez}, {Marchi-Lasch}, \&
  {Reichart}}]{2009auRodriguez}
{Rodr{\'\i}guez}, {\'O}., {Pignata}, G., {Anderson}, J.~P., {et~al.} 2020,
  \mnras, 494, 5882, \dodoi{10.1093/mnras/staa1133}

\bibitem[{{Roming} {et~al.}(2005){Roming}, {Kennedy}, {Mason}, {Nousek}, {Ahr},
  {Bingham}, {Broos}, {Carter}, {Hancock}, {Huckle}, {Hunsberger}, {Kawakami},
  {Killough}, {Koch}, {McLelland}, {Smith}, {Smith}, {Soto}, {Boyd},
  {Breeveld}, {Holland}, {Ivanushkina}, {Pryzby}, {Still}, \&
  {Stock}}]{2005roming}
{Roming}, P.~W.~A., {Kennedy}, T.~E., {Mason}, K.~O., {et~al.} 2005, Space
  Science Reviews, 120, 95, \dodoi{10.1007/s11214-005-5095-4}

\bibitem[{{Rui} {et~al.}(2019){Rui}, {Wang}, {Mo}, {Xiang}, {Zhang}, {Maund},
  {Gal-Yam}, {Wang}, \& {Zhang}}]{2017eawRui}
{Rui}, L., {Wang}, X., {Mo}, J., {et~al.} 2019, \mnras, 485, 1990,
  \dodoi{10.1093/mnras/stz503}

\bibitem[{{Sagar} {et~al.}(2019){Sagar}, {Kumar}, \& {Omar}}]{Sagar2019}
{Sagar}, R., {Kumar}, B., \& {Omar}, A. 2019, Current Science, 117, 365.
\newblock \doarXiv{1905.12896}

\bibitem[{{Sahu} {et~al.}(2006){Sahu}, {Anupama}, {Srividya}, \&
  {Muneer}}]{2004etSahu}
{Sahu}, D.~K., {Anupama}, G.~C., {Srividya}, S., \& {Muneer}, S. 2006, \mnras,
  372, 1315, \dodoi{10.1111/j.1365-2966.2006.10937.x}

\bibitem[{{S{\'a}nchez-S{\'a}ez} {et~al.}(2021){S{\'a}nchez-S{\'a}ez}, {Reyes},
  {Valenzuela}, {F{\"o}rster}, {Eyheramendy}, {Elorrieta}, {Bauer},
  {Cabrera-Vives}, {Est{\'e}vez}, {Catelan}, {Pignata}, {Huijse}, {De Cicco},
  {Ar{\'e}valo}, {Carrasco-Davis}, {Abril}, {Kurtev}, {Borissova}, {Arredondo},
  {Castillo-Navarrete}, {Rodriguez}, {Ruz-Mieres}, {Moya},
  {Sabatini-Gacit{\'u}a}, {Sep{\'u}lveda-Cobo}, \&
  {Camacho-I{\~n}iguez}}]{2021Alerce}
{S{\'a}nchez-S{\'a}ez}, P., {Reyes}, I., {Valenzuela}, C., {et~al.} 2021, \aj,
  161, 141, \dodoi{10.3847/1538-3881/abd5c1}

\bibitem[{{Sanders} {et~al.}(2015){Sanders}, {Soderberg}, {Gezari},
  {Betancourt}, {Chornock}, {Berger}, {Foley}, {Challis}, {Drout}, {Kirshner},
  {Lunnan}, {Marion}, {Margutti}, {McKinnon}, {Milisavljevic}, {Narayan},
  {Rest}, {Kankare}, {Mattila}, {Smartt}, {Huber}, {Burgett}, {Draper},
  {Hodapp}, {Kaiser}, {Kudritzki}, {Magnier}, {Metcalfe}, {Morgan}, {Price},
  {Tonry}, {Wainscoat}, \& {Waters}}]{2015Sanders}
{Sanders}, N.~E., {Soderberg}, A.~M., {Gezari}, S., {et~al.} 2015, \apj, 799,
  208, \dodoi{10.1088/0004-637X/799/2/208}

\bibitem[{{Schlafly} \& {Finkbeiner}(2011)}]{2011ApJ...737..103S}
{Schlafly}, E.~F., \& {Finkbeiner}, D.~P. 2011, \apj, 737, 103,
  \dodoi{10.1088/0004-637X/737/2/103}

\bibitem[{{Singh}(2021)}]{2021redpipe}
{Singh}, A. 2021, {RedPipe: Reduction Pipeline}.
\newblock \doeprint{2106.024}

\bibitem[{{Singh} {et~al.}(2019){Singh}, {Kumar}, {Moriya}, {Anupama}, {Sahu},
  {Brown}, {Andrews}, \& {Smith}}]{2016gfy}
{Singh}, A., {Kumar}, B., {Moriya}, T.~J., {et~al.} 2019, \apj, 882, 68,
  \dodoi{10.3847/1538-4357/ab3050}

\bibitem[{{Singh} {et~al.}(2018{\natexlab{a}}){Singh}, {Srivastav}, {Kumar},
  {Anupama}, \& {Sahu}}]{2018avinash}
{Singh}, A., {Srivastav}, S., {Kumar}, B., {Anupama}, G.~C., \& {Sahu}, D.~K.
  2018{\natexlab{a}}, \mnras, 480, 2475, \dodoi{10.1093/mnras/sty1957}

\bibitem[{{Singh} {et~al.}(2018{\natexlab{b}}){Singh}, {Misra}, {Sahu},
  {Dastidar}, {Gangopadhyay}, {Bose}, {Srivastav}, {Anupama}, {Chakradhari},
  {Kumar}, {Kumar}, \& {Pandey}}]{2014dt}
{Singh}, M., {Misra}, K., {Sahu}, D.~K., {et~al.} 2018{\natexlab{b}}, \mnras,
  474, 2551, \dodoi{10.1093/mnras/stx2916}

\bibitem[{{Sollerman} {et~al.}(2021){Sollerman}, {Yang}, {Schulze},
  {Strotjohann}, {Jerkstrand}, {Van Dyk}, {Kool}, {Barbarino}, {Brink},
  {Bruch}, {De}, {Filippenko}, {Fremling}, {Patra}, {Perley}, {Yan}, {Yang},
  {Andreoni}, {Campbell}, {Coughlin}, {Kasliwal}, {Kim}, {Rigault}, {Shin},
  {Tzanidakis}, {Ashley}, {Moore}, \& {Travouillon}}]{2021Sollerman}
{Sollerman}, J., {Yang}, S., {Schulze}, S., {et~al.} 2021, \aap, 655, A105,
  \dodoi{10.1051/0004-6361/202141374}

\bibitem[{{Sparks}(1994)}]{1994ApJ...433...19S}
{Sparks}, W.~B. 1994, \apj, 433, 19, \dodoi{10.1086/174621}

\bibitem[{{Srinivasaragavan} {et~al.}(2021){Srinivasaragavan}, {Sfaradi},
  {Jencson}, {De}, {Horesh}, {Kasliwal}, {Tinyanont}, {Hankins}, {Schulze},
  {Ashley}, {Graham}, {Karambelkar}, {Lau}, {Mahabal}, {Moore}, {Ofek},
  {Sharma}, {Sollerman}, {Soon}, {Soria}, {Travouillon}, \&
  {Walters}}]{2020qmp_nebular}
{Srinivasaragavan}, G.~P., {Sfaradi}, I., {Jencson}, J., {et~al.} 2021, arXiv
  e-prints, arXiv:2109.02159.
\newblock \doarXiv{2109.02159}

\bibitem[{Steer(2020)}]{Steer_2020}
Steer, I. 2020, The Astronomical Journal, 160, 199,
  \dodoi{10.3847/1538-3881/abafba}

\bibitem[{{Sukhbold} {et~al.}(2016){Sukhbold}, {Ertl}, {Woosley}, {Brown}, \&
  {Janka}}]{sukhbold}
{Sukhbold}, T., {Ertl}, T., {Woosley}, S.~E., {Brown}, J.~M., \& {Janka}, H.~T.
  2016, \apj, 821, 38, \dodoi{10.3847/0004-637X/821/1/38}

\bibitem[{{Szalai} {et~al.}(2019){Szalai}, {Vink{\'o}}, {K{\"o}nyves-T{\'o}th},
  {Nagy}, {Bostroem}, {S{\'a}rneczky}, {Brown}, {Pejcha}, {B{\'o}di}, {Cseh},
  {Cs{\"o}rnyei}, {Dencs}, {Hanyecz}, {Ign{\'a}cz}, {Kalup}, {Kriskovics},
  {Ordasi}, {P{\'a}l}, {Seli}, {S{\'o}dor}, {Szak{\'a}ts}, {Vida}, {Zsidi},
  {Konkoly Team}, {Arcavi}, {Ashall}, {Burke}, {Galbany}, {Hiramatsu},
  {Hosseinzadeh}, {Hsiao}, {Howell}, {McCully}, {Moran}, {Rho}, {Sand},
  {Shahbandeh}, {Valenti}, {Wang}, {Wheeler}, \& {Supernova
  Project}}]{2019sazlai}
{Szalai}, T., {Vink{\'o}}, J., {K{\"o}nyves-T{\'o}th}, R., {et~al.} 2019, \apj,
  876, 19, \dodoi{10.3847/1538-4357/ab12d0}

\bibitem[{{Terreran} {et~al.}(2016){Terreran}, {Jerkstrand}, {Benetti},
  {Smartt}, {Ochner}, {Tomasella}, {Howell}, {Morales-Garoffolo},
  {Harutyunyan}, {Kankare}, {Arcavi}, {Cappellaro}, {Elias-Rosa},
  {Hosseinzadeh}, {Kangas}, {Pastorello}, {Tartaglia}, {Turatto}, {Valenti},
  {Wiggins}, \& {Yuan}}]{2014GTerraran}
{Terreran}, G., {Jerkstrand}, A., {Benetti}, S., {et~al.} 2016, \mnras, 462,
  137, \dodoi{10.1093/mnras/stw1591}

\bibitem[{{Turatto} {et~al.}(1998){Turatto}, {Mazzali}, {Young}, {Nomoto},
  {Iwamoto}, {Benetti}, {Cappellaro}, {Danziger}, {de Mello}, {Phillips},
  {Suntzeff}, {Clocchiatti}, {Piemonte}, {Leibundgut}, {Covarrubias}, {Maza},
  \& {Sollerman}}]{56NIMass1987A}
{Turatto}, M., {Mazzali}, P.~A., {Young}, T.~R., {et~al.} 1998, \apjl, 498,
  L129, \dodoi{10.1086/311324}

\bibitem[{{Valenti} {et~al.}(2014){Valenti}, {Sand}, {Pastorello}, {Graham},
  {Howell}, {Parrent}, {Tomasella}, {Ochner}, {Fraser}, {Benetti}, {Yuan},
  {Smartt}, {Maund}, {Arcavi}, {Gal-Yam}, {Inserra}, \& {Young}}]{2014Valenti}
{Valenti}, S., {Sand}, D., {Pastorello}, A., {et~al.} 2014, \mnras, 438, L101,
  \dodoi{10.1093/mnrasl/slt171}

\bibitem[{Valenti {et~al.}(2015)Valenti, Sand, Stritzinger, Howell, Arcavi,
  McCully, Childress, Hsiao, Contreras, Morrell, Phillips, Gromadzki, Kirshner,
  \& Marion}]{2013byValenti}
Valenti, S., Sand, D., Stritzinger, M., {et~al.} 2015, Monthly Notices of the
  Royal Astronomical Society, 448, 2608, \dodoi{10.1093/mnras/stv208}

\bibitem[{{Valenti} {et~al.}(2016){Valenti}, {Howell}, {Stritzinger}, {Graham},
  {Hosseinzadeh}, {Arcavi}, {Bildsten}, {Jerkstrand}, {McCully}, {Pastorello},
  {Piro}, {Sand}, {Smartt}, {Terreran}, {Baltay}, {Benetti}, {Brown},
  {Filippenko}, {Fraser}, {Rabinowitz}, {Sullivan}, \& {Yuan}}]{2016Valenti}
{Valenti}, S., {Howell}, D.~A., {Stritzinger}, M.~D., {et~al.} 2016, \mnras,
  459, 3939, \dodoi{10.1093/mnras/stw870}

\bibitem[{{Van Dyk}(2017)}]{2017VanDyk}
{Van Dyk}, S.~D. 2017, Philosophical Transactions of the Royal Society of
  London Series A, 375, 20160277, \dodoi{10.1098/rsta.2016.0277}

\bibitem[{{Van Dyk} {et~al.}(2012){Van Dyk}, {Cenko}, {Poznanski}, {Arcavi},
  {Gal-Yam}, {Filippenko}, {Silverio}, {Stockton}, {Cuillandre}, {Marcy},
  {Howard}, \& {Isaacson}}]{2012awVanDyk}
{Van Dyk}, S.~D., {Cenko}, S.~B., {Poznanski}, D., {et~al.} 2012, \apj, 756,
  131, \dodoi{10.1088/0004-637X/756/2/131}

\bibitem[{{Van Dyk} {et~al.}(2019){Van Dyk}, {Zheng}, {Maund}, {Brink},
  {Srinivasan}, {Andrews}, {Smith}, {Leonard}, {Morozova}, {Filippenko},
  {Conner}, {Milisavljevic}, {de Jaeger}, {Long}, {Isaacson}, {Crossfield},
  {Kosiarek}, {Howard}, {Fox}, {Kelly}, {Piro}, {Littlefair}, {Dhillon},
  {Wilson}, {Butterley}, {Yunus}, {Channa}, {Jeffers}, {Falcon}, {Ross},
  {Hestenes}, {Stegman}, {Zhang}, \& {Kumar}}]{VanDyk2019}
{Van Dyk}, S.~D., {Zheng}, W., {Maund}, J.~R., {et~al.} 2019, \apj, 875, 136,
  \dodoi{10.3847/1538-4357/ab1136}

\bibitem[{{van Loon} {et~al.}(2005){van Loon}, {Marshall}, \&
  {Zijlstra}}]{vanloon}
{van Loon}, J.~T., {Marshall}, J.~R., \& {Zijlstra}, A.~A. 2005, \aap, 442,
  597, \dodoi{10.1051/0004-6361:20053528}

\bibitem[{{Vink} {et~al.}(2001){Vink}, {de Koter}, \& {Lamers}}]{vink1}
{Vink}, J.~S., {de Koter}, A., \& {Lamers}, H.~J.~G.~L.~M. 2001, \aap, 369,
  574, \dodoi{10.1051/0004-6361:20010127}

\bibitem[{Virtanen {et~al.}(2020)Virtanen, Gommers, Oliphant, Haberland, Reddy,
  Cournapeau, Burovski, Peterson, Weckesser, Bright, {van der Walt}, Brett,
  Wilson, Millman, Mayorov, Nelson, Jones, Kern, Larson, Carey, Polat, Feng,
  Moore, {VanderPlas}, Laxalde, Perktold, Cimrman, Henriksen, Quintero, Harris,
  Archibald, Ribeiro, Pedregosa, {van Mulbregt}, \& {SciPy 1.0
  Contributors}}]{2020SciPy}
Virtanen, P., Gommers, R., Oliphant, T.~E., {et~al.} 2020, Nature Methods, 17,
  261, \dodoi{10.1038/s41592-019-0686-2}

\bibitem[{{Vogl} {et~al.}(2019){Vogl}, {Sim}, {Noebauer}, {Kerzendorf}, \&
  {Hillebrandt}}]{2019Vogl}
{Vogl}, C., {Sim}, S.~A., {Noebauer}, U.~M., {Kerzendorf}, W.~E., \&
  {Hillebrandt}, W. 2019, \aap, 621, A29, \dodoi{10.1051/0004-6361/201833701}

\bibitem[{Waskom(2021)}]{seaborn2021}
Waskom, M.~L. 2021, Journal of Open Source Software, 6, 3021,
  \dodoi{10.21105/joss.03021}

\bibitem[{{Weiler} {et~al.}(1986){Weiler}, {Sramek}, {Panagia}, {van der
  Hulst}, \& {Salvati}}]{1986radioWeiler}
{Weiler}, K.~W., {Sramek}, R.~A., {Panagia}, N., {van der Hulst}, J.~M., \&
  {Salvati}, M. 1986, \apj, 301, 790, \dodoi{10.1086/163944}

\bibitem[{{W}es {M}c{K}inney(2010)}]{pandas2010}
{W}es {M}c{K}inney. 2010, in {P}roceedings of the 9th {P}ython in {S}cience
  {C}onference, ed. {S}t\'efan van~der {W}alt \& {J}arrod {M}illman, 56 -- 61,
  \dodoi{10.25080/Majora-92bf1922-00a}

\bibitem[{{Yaron} \& {Gal-Yam}(2012)}]{2012yaron}
{Yaron}, O., \& {Gal-Yam}, A. 2012, \pasp, 124, 668, \dodoi{10.1086/666656}

\bibitem[{{Zhang} {et~al.}(2022){Zhang}, {Wang}, {Sai}, {Niculescu-Duvaz},
  {Filippenko}, {Zheng}, {Brink}, {Lin}, {Zhang}, {Cai}, {Mo}, {Zhang},
  {Baron}, {DerKacy}, {Huang}, \& {Zhang}}]{2018hfmZhang}
{Zhang}, X., {Wang}, X., {Sai}, H., {et~al.} 2022, \mnras, 509, 2013,
  \dodoi{10.1093/mnras/stab3007}

\end{thebibliography}
\bibliographystyle{aasjournal}



\end{document}